\documentclass{aa}

\usepackage{graphicx}
\usepackage{txfonts}
\usepackage{placeins}
\usepackage{natbib}
\bibpunct{(}{)}{;}{a}{}{,}

\usepackage[whole]{bxcjkjatype}

\usepackage{amsmath}
\usepackage{algpseudocode}
\usepackage{algorithm}
\usepackage[bookmarks=false]{hyperref}

\begin{document}
\title{A fast tree algorithm for multi-component coagulation equation}

\author{
    Taichi K. Watanabe (渡邊K.太一)\inst{\ref{inst1}}\fnmsep\inst{\ref{inst2}}\fnmsep \thanks{Corresponding author: taichi.astrocat@gmail.com}
    \and Akimasa Kataoka (片岡章雅)\inst{\ref{inst1}}\fnmsep\inst{\ref{inst2}}
}

\institute{
    Astronomical Science Program, The Graduate University for Advanced Studies (SOKENDAI), 2-21-1 Osawa, Mitaka, Tokyo 181-8588, Japan\label{inst1}
    \and National Astronomical Observatory of Japan, 2-21-1 Osawa, Mitaka, Tokyo 181-8588, Japan\label{inst2}
}

\date{Received 21 December 2025 / Accepted 14 May 2026}

\abstract
{ Dust properties, such as mass, porosity, electric charge, and chemical composition, impact planet formation directly. Understanding the time evolution of dust distribution across multiple properties requires numerical computation. However, available ways to calculate the multi-component coagulation-fragmentation are highly time-consuming.}
{ This study aims to develop a fast and accurate algorithm for multi-component coagulation. We assumed that two pairs of colliding aggregates reproduce a similar outcome if the dust properties are similar, and that the ratio of dust properties in logarithmic space gives the similarity as a "distance". These assumptions enable us to apply the tree algorithm, which groups distant bins and calculates interactions together, to coagulation. The algorithm reduces the computational complexity from $\mathcal{O} (N^{2d})$ to $\mathcal{O} (d N^d \log N)$, considering $N$ bins per $d$ components.}
{ We tested the tree algorithm by comparing it with the conventional direct method for cases where analytic solutions are known. We measured the dependencies of the wall-clock time, $L_2$ error in the distribution, and relative error of the total mass, on the $d, N$, opening angle $\theta_c$, and maximum dust distribution width after coagulation $k_c$.}
{ The algorithms are found to calculate coagulation consistently. For $d=1$, the tree method is faster than the direct method for a specific range of parameters. For $d=2$, however, the tree method is faster for all parameter regions surveyed, speeding it up by tens to a hundred times. Increasing $N$ and decreasing $\theta_c$ or $k_c$ made it slower and more accurate. Additionally, using a small $k_c$ performs worse than when using a large $k_c$, suggesting that limiting $k_c$ is unnecessary. }
{ We present a fast tree algorithm for the multi-component coagulation equation. It will enable us to evolve the multi-component dust distribution, such as in mass-porosity space, in protoplanetary disks. }

\keywords{
    Methods: numerical
    -- Planets and satellites: formation
    -- Protoplanetary disks
    -- dust, extinction
}

\authorrunning{T. K. Watanabe \& A. Kataoka}

\maketitle
\nolinenumbers

\section{Introduction} \label{sec-1-introduction}

Multiple characteristics of a dust aggregate other than mass, including porosity, electrical charge, and chemical composition, are highlighted in planet formation. Planets form in protoplanetary disks through the collisional growth of micron-size dust grains \cite[e.g.][]{ safronovEvolutionProtoplanetaryCloud1972, hayashiStructureSolarNebula1981}. Planet formation theory has a major problem where the coagulation of dust particles is hindered by several barriers: radial drift \citep{adachiGasDragEffect1976, weidenschillingAerodynamicsSolidBodies1977}, bouncing \cite[e.g.][]{ zsomOutcomeProtoplanetaryDust2010, zsomOutcomeProtoplanetaryDust2011, windmarkPlanetesimalFormationSweepup2012}, electrostatic repulsion \citep{okuzumiElectricChargingDust2009, okuzumiELECTROSTATICBARRIERDUST2011, akimkinInhibitedCoagulationMicronsize2020}, fragmentation \cite[e.g.][]{ blumGrowthMechanismsMacroscopic2008}, and erosion\cite[e.g.][]{ krijtErosionLimitsPlanetesimal2015}. Recent findings of exoplanets \citep[e.g.][]{winnOccurrenceArchitectureExoplanetary2015}, comets and asteroids, which are remnants of planetesimals \citep[e.g.][]{ahearnCometsBuildingBlocks2011}, and certainly the existence of our eight planets require the overcoming of the barriers. While several mechanisms have been proposed to overcome these barriers, such as the streaming instability \citep{johansenRapidPlanetesimalFormation2007}, one group of the proposed mechanisms is dust microphysics arising from properties other than mass.
Porous dust aggregates follow gas flow more closely, weakening the radial drift barrier \citep{okuzumiRapidCoagulationPorous2012, kataokaFluffyDustForms2013, kataokaStaticCompressionPorous2013}.
The electric charge of dust aggregates can either stop or promote the coagulation, since aggregates with the same sign of charge repel \citep{okuzumiElectricChargingDust2009, okuzumiELECTROSTATICBARRIERDUST2011}, but those with different signs attract \citep{steinpilzElectricalChargingOvercomes2020, teiserGrowthSuperlargePreplanetary2025}. Chemical composition is also a key to understanding planet formation, as it can characterize disk environments and determine the elemental and isotopic ratios of planets, asteroids, and comets as the final product \citep[e.g.][]{obergProtoplanetaryDiskChemistry2023}. Further understanding of dust aggregates with multiple characteristics, or multi-component dust aggregates, through theoretical, simulation, observational, and laboratory-experimental perspectives will surely refine the current view of planet formation.

Although numerical coagulation simulations of multi-component dust aggregates are in high demand, conventional methods are highly time-consuming.
The simple, direct method of discretizing the multi-dimensional distribution of aggregate properties (e.g., mass and volume) and calculating by summation has computational complexity of $\mathcal{O} (N^{2d})$, where $d$ is the number of components and $N$ is the number of bins per component. The direct method is widely used when only mass is considered \citep[e.g.][]{stammlerDustPyPythonPackage2022}, but it is too costly when other dust characteristics are taken into account.
Schemes that utilize the conservation form of the equation \citep[e.g.][]{maHighResolutionSimulationMultidimensional2002, gunawanHighResolutionAlgorithms2004, liuHighOrderPositivity2019, lombartGrainGrowthAstrophysics2021, laibeCourantFriedrichsLewy2022, lombartFragmentationDiscontinuousGalerkin2022, lombartGeneralNonlinearFragmentation2024}, cannot handle basic aggregate parameters that do not conserve, such as volume, easily.
Some other fast multi-component coagulation schemes developed in different fields \citep[e.g.][]{matveevTensorTrainMonte2016, smirnovFastAccurateFinitedifference2016, dyachenkoMosaicskeletonApproximationAll2025} assume a linear gridding for the dust properties, and cannot be applied to planet formation, where the scale ranges from micron-size to km-size and a logarithmic gridding is required.
The methods that average the dust distribution, such as methods of moments \citep{hulburtProblemsParticleTechnology1964, estradaSolvingCoagulationEquation2008, estradaGlobalModelingNebulae2022}, volume-averaging methods \citep{okuzumiRapidCoagulationPorous2012, kataokaFluffyDustForms2013}, and mono-disperse models, \citep[e.g.][]{stepinskiGlobalEvolutionSolid1996, michoulierCompactionFragmentationBouncing2024}, cannot be used for situations where the complete information on the multi-dimensional dust parameter distribution is important. For example, we cannot judge whether a planet on the high-mass end of the distribution forms when tracking just the average of the mass distribution. Even if we tracked the mass distribution carefully but averaged the porosity distribution, it would not be accurate since it is known that bouncing and fragmentation are affected by porosity \citep[e.g.][]{blumExperimentsStickingRestructuring2000, guttlerOutcomeProtoplanetaryDust2010, shimakiLowvelocityCollisionsCentimetersized2012, shimakiTensileStrengthElastic2021, planesCollisionsMicrosizedAggregates2021, oshiroInvestigatingBouncingBarrier2025}. This implies that a coagulation calculation with a full two-dimensional distribution is required to understand these interactions.
Monte Carlo methods require many particles to achieve sufficient accuracy, making them slower than direct methods by orders of magnitude \citep{drazkowskaModelingDustGrowth2014}.
Therefore, to fully simulate planet formation and verify the numerical results from various methods, we need a new algorithm to solve the coagulation of multi-component dust aggregates.

\begin{figure}
    \resizebox{\hsize}{!}{\includegraphics[width=17cm]{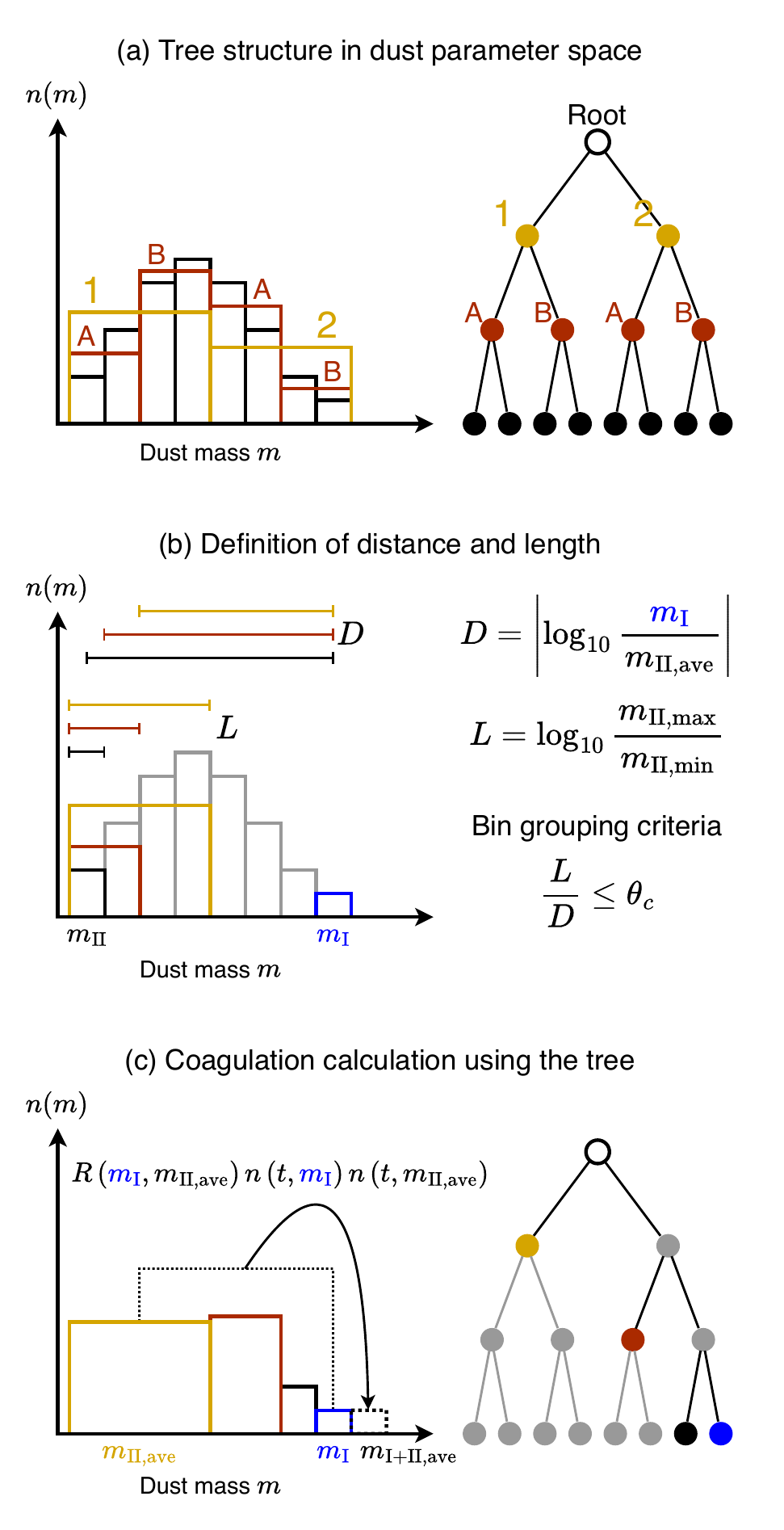}}
    \caption{Diagram of tree algorithm for one-component Smoluchowski coagulation calculations. (a) The dust property (here, mass) distribution is discretized into bins, and the bitree corresponds to the hierarchical grouping of the bins. (b) The tree algorithm assumes that the dust parameter space is a virtual metric space, where a distance is defined. To calculate coagulation using the tree, the algorithm first fixes one of the colliding dust aggregates: I-dust, and iterates through the other dust aggregate: II-dust using the tree. For each level of grouping, the distance between the average of the II-dust bin and the I-dust bin $D$, the length of the grouped bin $L$, and the ratio of these two values $\theta$ are calculated. (c) The algorithm computes the coagulation (Eq. (\ref{eq-1D-continuous-SCE-discretized})) using the averaged II-dust bin with the largest bin width under the condition of $\theta < \theta_c$.}
    \label{fig-1-scheme}
\end{figure}

Here, we present a novel $\mathcal{O} (dN^d \log N)$ multi-component coagulation tree algorithm inspired by the tree algorithm in $N$-body gravity simulations. The tree method for $N$-body gravity simulations groups together the gravitational force from particles far away and reduces the number of interactions from $\mathcal{O} (N^2)$ to $\mathcal{O} (N \log N)$ \citep[e.g.][]{barnesHierarchicalLogForcecalculation1986}. A tree data structure represents the geometry or closeness of particles, as the algorithm's name suggests. We applied this tree method to the coagulation. Our tree algorithm for coagulation assumes that two similar pairs of colliding aggregates will result in a similar outcome, and that the ratio of dust parameters, such as mass and volume, in logarithmic space, is a "distance" (Fig. \ref{fig-1-scheme} b). With this, we can impose a tree structure on the dust property distribution bins (Fig. \ref{fig-1-scheme} a). The tree structure holds data of the average dust parameter and the sum of the number density for each grouped bin. The algorithm groups bins far away using the tree and the opening angle criteria (Fig. \ref{fig-1-scheme} b), and computes coagulation using the grouped bins (Fig. \ref{fig-1-scheme} c). By introducing this algorithm, we can reduce the number of interactions and computational complexity from $\mathcal{O} (N^{2d})$ to $\mathcal{O} (dN^d \log N)$ (see Section \ref{sec-3-4-procon}). This tree method can be used with logarithmic axes and does not assume the conservation of dust properties upon coagulation. Thus, our tree method can be applied to planet formation simulations, such as porosity evolution in protoplanetary disks.

This paper is organized as follows. In Section \ref{sec-2-background}, we review the basic equation: Smoluchowski coagulation equation (SCE) and its numerical calculations as a background. In Section \ref{sec-3-treealgorithm}, we explain the details of the tree algorithm proposed in our work. In Section \ref{sec-4-methods}, we explain the numerical experiment conditions for comparing the tree method and the direct method, which uses simple kernels for which analytic solutions are known. In Section \ref{sec-5-results}, we show the results on how numerical parameters affect calculation speed and numerical accuracy. In Section \ref{sec-6-discussion}, we discuss the interpretations of results, techniques for a faster and more accurate scheme, and applications to planet formation. We summarize in Section \ref{sec-7-summary}.

\section{Background: the Smoluchowski coagulation equation and its numerical methods}\label{sec-2-background}

This section reviews the governing equation for dust coagulation: the SCE and its numerical method. In Section \ref{sec-2-onecomponent}, we first cover the one-component version of the equation, where only mass is considered, and in Section \ref{sec-2-review-onecomponent}, we review the numerical methods to calculate the one-component equation. In Section \ref{sec-2-multicomponent}, we cover the multi-component version, and in Section \ref{sec-2-review-multicomponent}, we review the numerical method for the multi-component equation. In Section \ref{sec-2-directmethod-onecomponent}, we explain the conventional and simple direct method, a method that is the basis of our tree method, for the one-component. Finally, in Section \ref{sec-2-directmethod-multicomponent}, we extend the direct method to the multi-component.

\subsection{The one-component Smoluchowski coagulation equation} \label{sec-2-onecomponent}

First, we introduce the coagulation equation of one-component dust aggregates, which have mass as their sole characteristic. The coagulation can be expressed using the following one-component continuous SCE \citep{smoluchowskiDreiVortrageUber1916, muellerZurAllgemeinenTheorie1928, schumannTheoreticalAspectsSize1940, safronovEvolutionProtoplanetaryCloud1972}:
\begin{align}
  \frac{\partial n (t, m)}{\partial t} &= \frac{1}{2}\int_0^m \mathrm{d}m' R (m', m-m') n (t, m') n (t, m-m')\nonumber\\
  & \quad - n (t, m) \int_0^\infty \mathrm{d}m' R (m, m') n (t, m'). \label{eq-1D-continuous-SCE}
\end{align}
If the particle mass can be assumed to be integral multiples of a unit mass $m_0$, the mass distribution $n(m)$ can be expressed with a discrete distribution $n_i$, where $i$ is the index corresponding to the mass $m = m_0 i$. For the discrete mass distribution, the following one-component discrete SCE is used \citep{smoluchowskiDreiVortrageUber1916, ohtsukiArtificialAccelerationAccumulation1990}:
\begin{align}
    \frac{\mathrm{d}n_i (t)}{\mathrm{d}t} &= \frac{1}{2} \sum_{j=1}^{i - 1} R(i - j, j) n_{i - j} (t) n_j (t) \nonumber\\
    &\quad - n_i (t) \sum_{j=1}^\infty R(i, j) n_j (t). \label{eq-1D-discrete-SCE}
\end{align}

These differential-integro equations describe the time evolution of the dust mass distribution. For the continuous equation (Eq. \ref{eq-1D-continuous-SCE}), $n(t, m)\ \mathrm{d} m$ is the number density of dust aggregates with mass from $m$ to $m + \mathrm{d} m$. For the discrete equation (Eq. \ref{eq-1D-discrete-SCE}), $n(t, m)$ is the number density of dust aggregates with mass $m$. A kernel $R (m, m')=R (m', m)$ is a coagulation frequency between mass $m$ and $m'$ aggregates. The first term on the right-hand side of the equation increases $n(t, m)$ due to the coagulation of mass $m'$ and $m-m'$ dust aggregates. The second term decreases $n(t, m)$ due to the coagulation of mass $m'$ and $m$ dust aggregates. The existence and uniqueness of the solution to the equation is shown by \citet{melzakScalarTransportEquation1957}.

One of the major differences between the continuous equation and the discrete equation is the existence of dust aggregates with mass smaller than the monomer $m=m_0$. The continuous equation allows infinitesimal dust aggregates, but the discrete one only treats the integer multiples of the monomer. This paper focuses on the continuous version of the equation because the multi-component analytical solutions only exist for the continuous equation (Section \ref{sec-2-multicomponent}). Infinitely small dust aggregates are non-physical and purely mathematical, implying that a discrete equation should be preferred in physical applications. However, the numerical simulations for the continuous equation discretize it, which essentially becomes the same as the original discrete equation. The difference explained above turns into the difference in the minimum mass $m_\mathrm{min}$ of the grid (and the maximum mass $m_\mathrm{max}$). This slightly affects the numerical setup, which is detailed in Section \ref{sec-2-directmethod-onecomponent} and \ref{sec-2-directmethod-multicomponent}. 

The SCE calculates the time evolution of dust mass distribution at only one spatial point, where the distribution can be assumed to be homogeneous. The equation must be calculated at all spatial grids and implemented with advection-diffusion to consider the nonhomogeneity in dust spatial distribution. This paper focuses on the coagulation at only one point in space.

The SCE is used not only for planet formation but also for the coagulation of giant molecular clouds \citep{kobayashiEvolutionaryDescriptionGiant2017}, dust coagulation in protostellar collapse \citep{lebreuillyProtostellarCollapseSimulations2023} and interstellar medium \citep{hirashitaEvolutionDustPorosity2021}, polymer synthesis \citep{ziffKineticsPolymerization1980}, nano-particles \citep{zhangNanoparticleAggregationPrinciples2014}, and aerosol dynamics \citep{seigneurSimulationAerosolDynamics1986}. The SCE is also known as the kinetic rate equation \cite[e.g.][]{ ziffKineticsPolymerization1980}, the kinetic collection equation \cite[e.g.][]{ alfonsoValidityKineticCollection2008}, and the population balance equation\cite[e.g.][]{ solsvikFoundationPopulationBalance2015}, especially in fields other than astronomy.

For the one-component continuous SCE (Eq. \ref{eq-1D-continuous-SCE}), the following three analytic solutions corresponding to three main simple kernels are known. The three kernels are constant kernel $R (m_i, m_j)=\alpha$ (also known as “size-independent kernel” in chemical engineering), additive kernel $R (m_i, m_j) = \alpha (m_i + m_j)$ (also known as “size-dependent kernel” in chemical engineering), and multiplicative kernel $R (m_i, m_j) = \alpha m_i m_j$. Here, $\alpha$ is a constant.

The analytic solution for the constant kernel with the initial condition of $n(0, m) = N_0/m_0 \exp (-m/m_0)$ is
\begin{align}
  \tau (t) &= \alpha N_0 t,\\
  f(t) &= \left(1 + \frac{\tau (t)}{2}\right)^{-1}, \\
  x(m) &= \frac{m}{m_0},\\
  n(t, m) &= \frac{N_0}{m_0} f(t)^2 \exp \left(- f(t) x \right)
\end{align}
\citep{muellerZurAllgemeinenTheorie1928, schumannTheoreticalAspectsSize1940, scottAnalyticStudiesCloud1968, silkStatisticalModelInitial1979}. Here, $m_0$ is the total mass in the system (constant), and $N_0$ is the initial total number of particles.

The analytic solution for the additive kernel with the initial condition of $n(0, m) = N_0/m_0 \exp (-m/m_0)$ is
\begin{align}
  \tau (t) &= \alpha N_0 m_0 t,\\
  f(t) &= \exp (-\tau (t)),\\
  g(t) &= 1 - f(t),\\
  x(m) &= \frac{m}{m_0},\\
  n(t, m) &= \frac{N_0}{m_0} \frac{f(t)}{x \sqrt{g(t)}} e^{-(2 - f(t))x} I_1 \left(2x\sqrt{g(t)}\right),
\end{align}
where $I_\nu(x)$ is the modified Bessel function of the first kind \citep{safronovParticularCaseSolution1962, golovinSolutionCoagulationEquation1963, scottAnalyticStudiesCloud1968, wetherillComparisonAnalyticalPhysical1990}.

And the analytic solution for the multiplicative kernel with the initial condition of $n(0, m) = N_0/m_0 \cdot m_0/m \exp (-m/m_0) = N_0/m \exp (-m/m_0)$ is
\begin{align}
  \tau (t) &= \alpha N_0 m_0^2 t,\\
  T (t) &= \begin{cases}
      \tau (t) & (\tau \leq 1)\\
      2\sqrt{\tau (t)} - 1 & (\tau \geq 1)
  \end{cases}\ ,\\
  x(m) &= \frac{m}{m_0},\\
  n(t, m) &= \frac{N_0}{m_0} \frac{1}{x^2 \tau} e^{-(1 +T(t))x} I_1 \left(2 x \sqrt{\tau} \right)
\end{align}
\citep{mcleodScalarTransportEquation1964, ernstCoagulationProcessesPhase1984}.

For the analytic solutions for the delta function initial condition and ones for the discrete equation, see Appendix \ref{sec-a-additionalkernels}. The analytic solutions for a linear combination of three main kernels\citep{spougeAnalyticSolutionsSmoluchowskis1985} and ones for more complicated initial conditions \citep{scottAnalyticStudiesCloud1968} are known, but we do not cover these.

The continuous equation can be rewritten using a double integral as
\begin{align}
  \frac{\partial n (t, m)}{\partial t} &= \frac{1}{2}\iint_0^m \mathrm{d} m_\mathrm{I} \mathrm{d}m_\mathrm{II} R (m_\mathrm{I}, m_\mathrm{II}) n (t, m_\mathrm{I}) n (t, m_\mathrm{II})\nonumber\\
  & \qquad \times \delta (m - (m_\mathrm{I} + m_\mathrm{II}))\nonumber\\
  & \quad - n (t, m) \int_0^\infty \mathrm{d} m_\mathrm{II} R (m, m_\mathrm{II}) n (t, m_\mathrm{II}), \label{eq-1D-continuous-SCE-var}
\end{align}
where $\delta(x)$ is the Dirac delta function that satisfies $\int \mathrm{d} x \delta(x - x_0) f(x) = f(x_0)$.

The first term in the right-hand side of the SCE (Eq. \ref{eq-1D-continuous-SCE-var}) expresses the increase in the I+II-dust ($m$) from the coagulation of the I-dust ($m_\mathrm{I}$) and the II-dust ($m_\mathrm{II}$). The second term expresses the decrease in the I-dust ($m$) from the coagulation of the I-dust ($m$) and the II-dust ($m_\mathrm{II}$). Instead of two loops with $m$ and $m'$ as in the equation \ref{eq-1D-continuous-SCE}, the loops in this equation can be regarded as the loops of I-dust and II-dust that are colliding. The subscripts in the first term are as they are, and for the second term, $m$ should be interpreted as $m_\mathrm{I}$ in the first term. Moreover, this formulation directly connects to the multi-component version of the equation, which is explained in Section \ref{sec-2-multicomponent}, and the discretized form for the numerical calculation, as explained in Section \ref{sec-2-directmethod-onecomponent}. 

\subsection{Review of numerical calculations of the one-component Smoluchowski coagulation equation} \label{sec-2-review-onecomponent}

Since we do not know the analytic solution for the kernel in planet formation, which is much more complex than the ones shown in Section \ref{sec-2-onecomponent}, we need to calculate the SCE numerically. The SCE calculation can be grouped mainly into two ways. One of them is deterministic methods, which discretize the dust distribution and solve the equation deterministically. The other group of methods is Monte Carlo methods, which randomly sample the dust distribution and evaluate the integrals. Furthermore, the deterministic methods can be classified into three types: the pair-interaction method, the finite element method (FEM), and the method of moments. Section 2.5 of \citet{lombartGrainGrowthAstrophysics2021} reviews these methods comprehensively.

The pair-interaction method discretizes the dust parameter distribution into histograms and integrates by summing up the coagulation of all bin pairs. This method is widely used in many preceding studies of planet formation \citep[e.g.][]{nakagawaGrowthSedimentationDust1981, dullemondDustCoagulationProtoplanetary2005, tanakaDustGrowthSettling2005, nomuraDustSizeGrowth2006, brauerCoagulationFragmentationRadial2008, birnstielGasDustEvolution2010, kobayashiRapidFormationGasgiant2021, stammlerDustPyPythonPackage2022}, and we refer to this scheme as the “direct method” hereafter. Even limiting to planet formation, its validity has been well studied by \citet{wetherillComparisonAnalyticalPhysical1990, ohtsukiArtificialAccelerationAccumulation1990, leeValidityCoagulationEquation2000, drazkowskaModelingDustGrowth2014}. Our novel “tree method”, which we explain later in Section \ref{sec-3-treealgorithm}, also uses the same discretization but approximates part of the summation.

The finite element method discretizes the dust parameter distribution by referencing a conservative form of the equation. In planet formation, \citet{liuHighOrderPositivity2019, lombartGrainGrowthAstrophysics2021} proposed the discontinuous Galerkin schemes for coagulation based on the conservative form proposed by \citet{tanakaSteadyStateSizeDistribution1996}. When applying these methods with an explicit time integration, the Courant--Friedrichs--Lewy conditions for coagulation should be considered for a valid simulation since these methods use a hyperbolic formulation \citep{laibeCourantFriedrichsLewy2022}. Outside planet formation, the population balance equations, which are variations of SCEs, are widely used with FEM. \citet{osullivanConservativeFiniteVolume2022} gives further reviews on those.

The method of moments discretizes the equation using the moments $M_k = \int_0^\infty X^k n(t, X) \mathrm{d}X$ of the parameter distribution \citep{hulburtProblemsParticleTechnology1964, estradaSolvingCoagulationEquation2008}. In a broader sense, the mono-disperse model can also be included in the method of moments, since it uses the mean mass to rewrite the coagulation. The mono-disperse model is computationally less demanding and fits better if the model needs to be coupled with fluid dynamics or chemical network computation \cite[e.g.][]{ stepinskiGlobalEvolutionSolid1996, stepinskiGlobalEvolutionSolid1997, kornetDiversityPlanetarySystems2001, cieslaEvolutionWaterDistribution2006, garaudGrowthMigrationSolids2007, johansenCoagulationfragmentationModelTurbulent2008, laibeSPHSimulationsGrain2008, satoWaterDeliveryTerrestrial2016, krijtTransportCOProtoplanetary2018, garciaEvolutionPorousDust2020, michoulierDustGrainShattering2022, michoulierCompactionFragmentationBouncing2024}. Several methods that extend the mono-disperse model have been developed, such as the two-population model \citep{birnstielSimpleModelEvolution2012}, the tri-population model \citep{pfeilTriPoDTriPopulationSize2024}, and the batch method \citep{krijtPanopticModelPlanetesimal2016}.

Monte Carlo methods use random numbers to sample (super-)particles from the distribution. These methods are also used widely in planet formation \citep[e.g.][]{ormelDustCoagulationProtoplanetary2007, zsomRepresentativeParticleApproach2008, zsomOutcomeProtoplanetaryDust2011, drazkowskaPlanetesimalFormationSweepup2013, krijtErosionLimitsPlanetesimal2015, beutelEfficientSimulationStochastic2024, vaikundaramanMcdust2DMonte2025, gurrutxagaMonteCarloMethod2026}. The comparison with the direct method on numerical accuracy and speed was conducted by \citet{drazkowskaModelingDustGrowth2014}.

\subsection{The multi-component Smoluchowski coagulation equation} \label{sec-2-multicomponent}

Now, we extend the SCE into a multi-component system. The coagulation of dust aggregates with multiple components, or parameters, is expressed by the multi-component SCE \citep{ossenkopfDustCoagulationDense1993, kostoglouEvolutionAggregateSize2001, okuzumiNumericalModelingCoagulation2009}:
\begin{align}
  \frac{\partial n (t,\boldsymbol{X})}{\partial t} &= \frac{1}{2} \int \mathrm{d}\boldsymbol{X}_\mathrm{I} \mathrm{d} \boldsymbol{X}_\mathrm{II} R (\boldsymbol{X}_\mathrm{I}, \boldsymbol{X}_\mathrm{II}) n (t, \boldsymbol{X}_\mathrm{I}) n (t, \boldsymbol{X}_\mathrm{II})\nonumber\\
  &\qquad \times \delta (\boldsymbol{X} - \boldsymbol{X}_\mathrm{I+II} (\boldsymbol{X}_\mathrm{I}, \boldsymbol{X}_\mathrm{II}))\nonumber\\
  &\quad - n (t, \boldsymbol{X}) \int \mathrm{d}\boldsymbol{X}_\mathrm{II} R (\boldsymbol{X}, \boldsymbol{X}_\mathrm{II}) n (t, \boldsymbol{X}_\mathrm{II}), \label{eq-multiD-continuous-SCE}
\end{align}
where $\boldsymbol{X} = (X^{(1)}, X^{(2)}, \dots, X^{(d)})$ is the dust aggregate parameter vector with $d$ components. The multi-dimensional Dirac delta function $\delta (\boldsymbol{X}, \boldsymbol{X}')$ satisfies $\int \mathrm{d} \boldsymbol{X} \delta(\boldsymbol{X} - \boldsymbol{X}_0) f(\boldsymbol{X}) = f(\boldsymbol{X}_0)$. $\boldsymbol{X}_\mathrm{I+II} (\boldsymbol{X}_\mathrm{I}, \boldsymbol{X}_\mathrm{II})$ is a function of $\boldsymbol{X}_\mathrm{I}$ and $\boldsymbol{X}_\mathrm{II}$ that returns the dust parameter after the coagulation of the two dust aggregates. This formulation is a natural extension of the one-component coagulation equation (Eq. \ref{eq-1D-continuous-SCE-var}). The variables subscripted with I and II correspond to single-dashed and double-dashed variables in the preceding studies. Here, instead of "single-dashed-dust" or "double-dashed-bin", we use the terminology of "I-dust" or "II-bins" for simplicity.

For example, to calculate porosity distribution evolution as in \citet{kostoglouEvolutionAggregateSize2001, okuzumiNumericalModelingCoagulation2009}, one would use mass and volume as the dust aggregate parameter $\boldsymbol{X} = (X^{(1)}, X^{(2)}) = (m, v)$, and the two-component coagulation equation is expressed as
\begin{align}
    \frac{\partial n (t, m, v)}{\partial t} &= \frac{1}{2} \iiiint_0^\infty \mathrm{d}m_\mathrm{I} \mathrm{d}v_\mathrm{I} \mathrm{d}m_\mathrm{II} \mathrm{d}v_\mathrm{II}\nonumber\\
    &\qquad \times \delta (m - (m_\mathrm{I} + m_\mathrm{II})) \delta (v - v_\mathrm{I+II} (m_\mathrm{I}, v_\mathrm{I}; m_\mathrm{II}, v_\mathrm{II}))\nonumber\\
    &\qquad \times R(m_\mathrm{I}, v_\mathrm{I}; m_\mathrm{II}, v_\mathrm{II}) n(t, m_\mathrm{I}, v_\mathrm{I}) n(t, m_\mathrm{II}, v_\mathrm{II})\nonumber\\
    &\quad -  n(t, m, v) \iint_0^\infty \mathrm{d}m_\mathrm{II} \mathrm{d}v_\mathrm{II}\nonumber\\
    &\qquad \times R(m, v; m_\mathrm{II}, v_\mathrm{II}) n(t, m_\mathrm{II}, v_\mathrm{II}).
\end{align}
Here, $v_\mathrm{I+II} (m_\mathrm{I}, v_\mathrm{I}; m_\mathrm{II}, v_\mathrm{II})$ is a function that returns the volume after coagulation with void creation and compression included. The maximum of the integration interval in the first term is extended to infinity since the compression may reduce the volume.

One of the challenges with the multi-component SCE is that it is generally $\boldsymbol{X}_\mathrm{I+II} \neq \boldsymbol{X}_\mathrm{I} + \boldsymbol{X}_\mathrm{II}$, which we call the "non-convervative dust property". Taking dust porosity evolution as an example, dust undergoes fractal growth and compression, leading to non-conservation of volume: $v_\mathrm{I+II} \neq v_\mathrm{I} + v_\mathrm{II}$. The non-conservation dust property only refers to the dust parameters after coagulation; other fundamental physical quantities, such as the total mass of the distribution, still conserve even in the SCE with non-conservative dust property.

The multi-component continuous SCE (Eq. \ref{eq-multiD-continuous-SCE}) with conservative dust properties (i.e., $\boldsymbol{X}_\mathrm{I+II} = \boldsymbol{X}_\mathrm{I} + \boldsymbol{X}_\mathrm{II}$) has analytic solutions, but only for the constant kernel $R(\boldsymbol{X}_\mathrm{I}, \boldsymbol{X}_\mathrm{II}) = \alpha$ and the additive kernel $R(\boldsymbol{X}_\mathrm{I}, \boldsymbol{X}_\mathrm{II}) = \alpha \left(\sum_{i=1}^d X_\mathrm{I}^{(i)} + X_\mathrm{II}^{(i)}\right)$, where $\alpha$ is a constant. We only show here the solution for the two-component case, where the dust property vector is $\boldsymbol{X} = (m, v)$, because we only tested the algorithms with one or two components in later sections. The extensions to three or more components are described in the cited works.

The analytic solution for the constant kernel with the initial condition of $n(0, m, v) = N_0/(m_0 v_0) \exp (-m/m_0 - v/v_0)$ is
\begin{align}
    \tau(t) &= \alpha N_0 t,\\
    f(t) &= \left(1 + \frac{\tau(t)}{2}\right)^{-1},\\
    x(m) &= \frac{m}{m_0},\qquad y(v) = \frac{v}{v_0},\\
    n(t, m, v) &= \frac{N_0}{m_0 v_0} f(t)^2 \exp\left(- x - y\right) I_0 \left(2 \sqrt{(1 - f(t)) x y}\right)
\end{align}
\citep{lushnikovEvolutionCoagulatingSystems1976, gelbardCoagulationGrowthMulticomponent1978, alexopoulosPartDynamicEvolution2009}.

The analytic solution for the additive kernel with the initial condition of $n(0, m, v) = N_0/(m_0 v_0) \exp (-m/m_0 - v/v_0)$ is
\begin{align}
    \tau(t) &= \alpha N_0 (m_0 + v_0) t,\\
    f(t) &= \exp(-\tau(t)),\\
    g(t) &= 1 - f(t),\\
    x(m) &= \frac{m}{m_0},\qquad y(v) = \frac{v}{v_0},\\
    S(t, m, v) &= \sum_{k=0}^\infty \frac{\left[x y \frac{m + v}{m_0 + v_0} g(t)\right]^k}{(k+1)! (k!)^2},\\
    n(t, m, v) &= \frac{N_0}{m_0 v_0} f(t) \exp\left(- x - y - \frac{m + v}{m_0 + v_0} g(t)\right) S
\end{align}
\citep{fernandez-diazExactSolutionSmoluchowskis2007, singhNewFiniteVolume2021}. 

\subsection{Review of numerical calculations of the multi-component Smoluchowski coagulation equation} \label{sec-2-review-multicomponent}

The multi-component SCE calculation has several computational difficulties. The pair-interaction method is computationally intensive, and the finite element method cannot be applied easily to planet formation due to the intrinsic non-conservation of the dust properties. Monte Carlo methods require a large number of particles and substantial computational time to achieve sufficient accuracy. The methods of moment can calculate the multi-component SCE at a reasonable computational cost and can be applied to planet formation. However, it cannot capture behaviors at the tails of the dust parameter distribution (i.e., the maximum dust size cannot be resolved), and is invalid if the dust parameter distribution is complex, such as a bimodal distribution.

The high computational cost of computing the SCE is a universal characteristic. For the pair-interaction method and the finite element method, two loops, I and II, are required, leading to a computational complexity of $\mathcal{O} (N^{2d})$ \citep{gelbardSimulationMulticomponentAerosol1980}. Here, $d$ is the number of parameters or components, and $N$ is the number of bins per parameter. This high computational cost makes it unrealistic to compute the multi-component ($d > 1$) SCE directly. This is also true for the finite element methods, which can reduce the number of bins $N$ compared to the pair-interaction methods thanks to their high-resolution scheme.

The acceleration of multi-component SCE calculations has been studied widely. Like the one-component SCE computation, it has been studied using FEM, methods of moments, tensor decompositions, Monte Carlo methods, and parallelization.

One approach to accelerate multi-component SCE calculations is to reduce the number of bins $N$ through (high-order) conservative forms or FEM. The papers that described the discrete Galerkin scheme for the one-component SCE \citep{liuHighOrderPositivity2019, lombartGrainGrowthAstrophysics2021, laibeCourantFriedrichsLewy2022} also discuss the extension to the multi-component SCE. \citet{maHighResolutionSimulationMultidimensional2002, gunawanHighResolutionAlgorithms2004} proposed high-resolution schemes for the multi-component population balance equations, a conservative variant of multi-component SCE. However, applying the conservative forms and FEM to the SCE equations with non-conservative dust properties, such as porosity, is not straightforward \citep[see Section 5.6 of ][]{lombartGeneralNonlinearFragmentation2024}. \citet{singhSolutionBivariateAggregation2018} reviews FEM approaches for the coagulation.

Another orientation is to rewrite the equation using the moments of the dust parameter distribution. \citet{okuzumiRapidCoagulationPorous2012, kataokaFluffyDustForms2013} averaged the volume of dust aggregates for each mass and reduced the mass-volume multi-component SCE into the one-component SCE. This model is adopted widely, and has been applied to dust growth in the interstellar medium \citep{hirashitaEvolutionDustPorosity2021} and extended to include other physical effects such as evaporation \citep{estradaGlobalModelingNebulae2022}. Similar to this, \citet{kostoglouEvolutionAggregateSize2001, kostoglouBivariatePopulationDynamics2006, gruyPopulationBalanceAggregation2011} constructed coagulation models where the fractal dimension of dust aggregates can be written as a time-dependent function, and \citet{ottoPopulationBalanceModeling2024} constructed a coagulation model that averages porosity for each mass. These “averaging” methods are generally of the methods of moments. The methods of moments are relatively low-cost and valid for planet formation, but they cannot, in principle, derive all the information on the multi-dimensional dust parameter distribution. For example, the information on the maximum size is required to understand if a planet can form or not. The effects of dust porosity on fragmentation and bouncing, combined with two or more distinct populations of dust aggregates, can lead to inaccurate calculations. These difficulties imply that a coagulation calculation with a full two-dimensional distribution is required to understand these interactions.

Recent advances in data science have been applied to the multi-component SCE calculation, creating new approximation methods. \citet{matveevTensorTrainMonte2016, smirnovFastAccurateFinitedifference2016} proposed a method that applies Tensor-Train Decomposition (TT-Decomposition) \citep{oseledetsTensorTrainDecomposition2011}, which is a type of tensor decomposition. The SCE in a linear gridding can be rewritten using a lower-triangular convolution \citep{smirnovFastAccurateFinitedifference2016}, and the convolution can be approximated using TT-Decomposition \citep{kazeevMultilevelToeplitzMatrices2013}. This reduces the computational complexity of multi-component SCE to $\mathcal{O} (d^2 R^4 N \log N)$, where $R$ is the maximum TT-rank. Another tensor approximation method is Mosaic-Skeleton Approximation \citep{dyachenkoMosaicskeletonApproximationAll2025}, which can reduce the computational complexity to $\mathcal{O} (N (\log N)^2)$. However, assuming a linear gridding makes it difficult to apply these methods to planet formation, where scales range from micrometers to kilometers. 

Monte Carlo methods for calculating multi-component SCE are well studied in and out of astronomy \citep[e.g.][]{laurenziGeneralAlgorithmExact2002, zhaoDifferentiallyWeightedMonte2010}, since they have the advantage that their computational complexity does not increase with the number of components $d$. However, the fundamental disadvantage of the methods is the same: they require a large number of particles and computational time.

Not only algorithmic improvements but also parallelizations are used to accelerate the SCE computation. As an extension to the TT-decomposition method, 
\citet{zagidullinSupercomputerModellingSpatiallyheterogeneous2019} parallelized coagulation and advection using MPI and CUDA, resulting in 2--4 times faster. In Monte Carlo methods, \citet{weiGPUacceleratedMonteCarlo2013} used GPU parallelization to accelerate coagulation computation by 10--100 times. Moreover, \citet{xuFastMonteCarlo2014} implemented the differentially-weighted Monte Carlo, and \citet{xuAcceleratingPopulationBalanceMonte2015} implemented the Markov jump model to accelerate the calculation further.

\subsection{The direct method for the one-component Smoluchowski coagulation equation} \label{sec-2-directmethod-onecomponent}

Here, we explain the conventional direct method for the one-component SCE, which is the basis of our tree method. The direct method discretizes the dust property distribution and the SCE and calculates the integral by summing up all the bin pairs. We mainly followed \citet{leeValidityCoagulationEquation2000, brauerCoagulationFragmentationRadial2008} for this.

First, the program constructs the bin axes for each dimension of the dust parameters. The axes are in units of the monomer property (i.e., monomer mass $m_0 = 1$, monomer volume $v_0 = 1$). The representative values $m_i$ of the first $N_\mathrm{bd} + 1$ bins are in integer scale from 0 to $N_\mathrm{bd}$. The next $N_\mathrm{dec}$ decades with $N_\mathrm{bd}$ bins per decade are on the logarithmic grid. The total number of bins is
\begin{align}
    N &= N_\mathrm{bd} (N_\mathrm{dec} + 1) + 1. \label{eq-total-number-bin}
\end{align}
Combining these constructions into one equation, the representative values of bins are defined by 
\begin{align}
    m_i &= \begin{cases}
        i & \quad (i = 0, 1, \dots, N_\mathrm{bd}),\\
        N_\mathrm{bd} \times 10^{(i - N_\mathrm{bd})/N_\mathrm{bd}} & \quad (i = N_\mathrm{bd} + 1, \dots, N - 1).
    \end{cases} \label{eq-ax-representative}
\end{align}
Next, bin boundaries $m_{i-1/2}$ are defined by the arithmetic average of the representative values of the two neighboring bins 
\begin{align}
    m_{i-1/2} &= \displaystyle\frac{m_{i-1} + m_{i}}{2} & \quad (i = 0, 1, \dots, N). \label{eq-ax-boundary}
\end{align}
The representative value of the ghost bin before the smallest bin is defined as $m_{-1} = 0$ i.e., $m_{-1/2} = 0$, and the that of the ghost bin after the largest bin is defined as $m_{N} = r m_{N-1} = m_{N - 1}^2/m_{N - 2}$, assuming a geometric sequence of ratio $r = m_{N-1}/m_{N-2}$. Lastly, bin widths are defined by the difference between the adjacent boundaries 
\begin{align}
    \Delta m_i &= m_{i+1/2} - m_{i-1/2} & (i = 0, 1, \dots, N - 1).
\end{align} \label{eq-ax-width}

This integer-then-logarithmic grid, first used in \citet{leeValidityCoagulationEquation2000}, enables the coagulation calculation with a large scale difference from micron-size to kilometer-size with high precision. The former integer part allows the precise treatment of integer multiples of monomers, especially for the delta function initial conditions or for numerically calculating the discrete version of the equation. The latter logarithmic part treats the large size effectively. Our grid definition is slightly different from the ones in \citet{leeValidityCoagulationEquation2000}, where they first defined the bin boundary by the “integer-then-logarithmic” grid and then defined the representative values by the arithmetic average. Our simple tests showed slightly better mass conservation with our axes definition.

Our axis definition also includes the zero-bin, a bin with the representative value of $m=0$. This is unnecessary for the discrete equation or the one-component continuous equation, since mass $m=0$ added to any other bin has no effect. However, it is necessary for the multi-component continuous equation, which is described in Section \ref{sec-2-directmethod-multicomponent}.

We then discretize the continuous dust number density distribution as 
\begin{align}
  \omega_i (t) &= \int_{m_{i-1/2}}^{m_{i+1/2}} n(t, m) \mathrm{d}m,
\end{align}
where $m_{i+1/2}$ and $m_{i-1/2}$ are the bin boundaries defined in the equation \ref{eq-ax-boundary}. Using the defined axis and this discretized distribution, we can discretize the continuous SCE \ref{eq-1D-continuous-SCE-var} by a simple rectangle method as
\begin{align}
  \frac{\partial \omega_k (t)}{\partial t} &= \frac{1}{2} \sum_{i=0}^{k} \sum_{j=0}^{k} K (m_i + m_j, m_k) R (m_i, m_j) \omega_i (t) \omega_j (t) \nonumber\\
  &\quad - \omega_k (t) \sum_{j=0}^N R (m_k, m_j) \omega_j (t). \label{eq-1D-continuous-SCE-discretized}
\end{align}
Here, $K (m_\mathrm{I+II}, m_k)$ is an outcome function of a collision. The variables with indices $i$, $j$, and $k$ in this discretized equation correspond to the ones with I, II, and no indices in the original continuous equation (Eq. \ref{eq-1D-continuous-SCE-var}), respectively.

To construct the outcome function $K (m_\mathrm{I+II}, m_k)$, a simple 1st order scheme is used. This was first proposed in \citet{kovetzEffectCoalescenceCondensation1969} and is called the "Podolak algorithm" in astronomy \citep[e.g.][]{brauerCoagulationFragmentationRadial2008, stammlerDustPyPythonPackage2022} or the "fixed pivot technique" in chemical engineering \cite[e.g.][]{ kumarSolutionPopulationBalance1996}. Since we are using a nonlinear bin axis, the resulting property of coagulation $m_\mathrm{I+II}$ does not necessarily fall into the representative value of another bin $m_n$, but instead falls in between $m_{n-1}$ and $m_n$. This scheme distributes the coagulation term $R (m_\mathrm{I}, m_\mathrm{II}) n(m_\mathrm{I}) n(m_\mathrm{II})$ into the two bins. To determine how much of the term should be distributed into each bin, this Podolak algorithm sets the value so that it conserves the total number of particles (i.e., the 0th moment of the dust distribution) and total mass (i.e., the 1st moment of the dust distribution) before and after the distribution. Then, the proportion 
\begin{align}
\epsilon (m_\mathrm{I+II}, n) &= \frac{m_\mathrm{I+II} - m_{n-1}}{m_n - m_{n-1}} \label{eq-podolak-epsilon}
\end{align}
of the term is put into bin $m_n$ and $1 - \epsilon$ of the term is put into bin $m_{n-1}$. Thus, the outcome function $K (m_\mathrm{I+II}, m_k)$ becomes 
\begin{align}
K (m_\mathrm{I+II}, m_k) &= \begin{cases}
  \epsilon_m (m_\mathrm{I+II}, n) & (\text{if}\ m_k=m_n),\\
  1 - \epsilon_m (m_\mathrm{I+II}, n) & (\text{if}\ m_k=m_{n-1}),\\
  0 & (\text{otherwise}).
\end{cases}
\end{align}

The indices of the two bins $m_{n-1}$ and $m_n$ of which the dust parameter value after coagulation $m_\mathrm{I+II}$ falls in between can be calculated by the inverse function of the equation \ref{eq-ax-representative} as 
\begin{align}
  n(m_\mathrm{I+II}) &= \begin{cases}
      \displaystyle \lceil m_\mathrm{I+II} \rceil & \quad (\mathrm{if}\ m_\mathrm{I+II} \leq N_\mathrm{bd}),\\
      \displaystyle N_\mathrm{bd} + \left\lceil N_\mathrm{bd} \log_{10}\frac{m_\mathrm{I+II}}{N_\mathrm{bd}} \right\rceil & \quad (\mathrm{otherwise}).
  \end{cases} \label{eq-podolak-n}
\end{align}

The pseudocode for the one-component direct method is shown in the Appendix \ref{sec-a-pseudocode}.

The outcome function $K$ only has non-zero values at $m_n$ and $m_{n-1}$, so we can perform all the operations of the first term using the two loops of $i$ and $j$. The original loop indices for the second term are $j$ and $k$ in the discretized equation, but since the loop range is the same as the first term, we can rename $k$ in the second term as $i$ and do the operations for the second term in the same loop as the first term. This is essentially the same as using two loops, corresponding to the two colliding aggregates of I and II, to calculate both the increase in the I+II-dust and the decrease in the I-dust.

The "empty" nodes and bins, which are bins with zero dust number density, are skipped (\texttt{continue} statement in C++). This reduces computational time considerably in both the direct and tree methods.

At the end of each timestep, the program forces the number density to zero if it is smaller than $10^{-40}$. This stabilizes and accelerates the calculation slightly. This is also applied to both the direct method and the tree method.

\subsection{The direct method for the multi-component Smoluchowski coagulation equation} \label{sec-2-directmethod-multicomponent}

The direct method can be naturally extended to the multi-component. The axis is defined in the same manner for each dust property dimension as the mass axis in the one-component direct method. Using these axes, the $d$-dimensional array representing the Cartesian dust property grid $\boldsymbol{X}_{\boldsymbol{i}} = \left(X_{i^{(1)}}^{(1)}, X_{i^{(2)}}^{(2)}, \dots, X_{i^{(d)}}^{(d)}\right)$, where $\boldsymbol{i} = \left(i^{(1)}, i^{(2)}, \dots, i^{(d)}\right)$ is the indice vector and $\boldsymbol{X} = \left( X^{(1)}, X^{(2)}, \dots, X^{(d)} \right)$ is the actual dust property corresponding to the indice vector, is prepared. In actual implementation, preparing $d$ copies of one-dimensional axis is enough (i.e., \texttt{m[mi]} and \texttt{v[vi]}).

We used a Cartesian coordinate grid for the performance evaluation in later sections, but different types of grids are available. For example, a two-dimensional polar coordinate \citep{nandanwarNewDiscretizationSpace2008} or the X-discretization technique that adopts a diagonal grid \citep{chauhanSolutionBivariatePopulation2012}, can be used for the two-component SCE.

For the multi-component continuous equation with exponentially decreasing initial condition, the small tweak of the zero-bin greatly improves the numerical accuracy. This is because the exponentially decreasing initial condition implies that the infinitesimal aggregates dominate in terms of the number of particles, and the multi-component distribution can have aggregates with parameters such as $(m, v)=(1, 0), (0, 1)$. The former contributes to diffusing the dust distribution to a larger mass, and the latter contributes to diffusing the dust distribution to a larger volume. These aggregates are non-physical, but they have a non-negligible effect of making the numerical solution closer to the mathematical analytic solutions. This zero-bin increases the calculation time by a factor of 5--10, so it should be removed for the physical applications that use the initial condition where everything is a monomer.

The same discretization is applied to the multi-component equation. The $d$-dimensional dust parameter distribution is discretized as
\begin{align}
  \omega_{\boldsymbol{i}} (t) &= \int_{X^{(1)}_{i^{(1)} - 1/2}}^{X^{(1)}_{i^{(1)} + 1/2}} \mathrm{d} X^{(1)} \cdots \int_{X^{(d)}_{i^{(d)} - 1/2}}^{X^{(d)}_{i^{(d)} + 1/2}} \mathrm{d} X^{(d)} n (t, \boldsymbol{X}),
\end{align}
where the total number of bins is
\begin{align}
    N_\mathrm{total} &= N^d = (N_\mathrm{bd} (N_\mathrm{dec} + 1) + 1)^d.
\end{align}
Then, the discretized multi-component continuous SCE is written as:
\begin{align}
  \frac{\partial \omega_{\boldsymbol{k}} (t)}{\partial t} &= \frac{1}{2} \sum_{\boldsymbol{i}}^{} \sum_{\boldsymbol{j}} K (\boldsymbol{X}_\mathrm{I+II} ( \boldsymbol{X}_{\boldsymbol{i}}, \boldsymbol{X}_{\boldsymbol{j}} ), \boldsymbol{X}_{\boldsymbol{k}}) R (\boldsymbol{X}_{\boldsymbol{i}}, \boldsymbol{X}_{\boldsymbol{j}}) \omega_{\boldsymbol{i}} (t) \omega_{\boldsymbol{j}} (t) \nonumber\\
  &\quad - \omega_{\boldsymbol{k}} (t) \sum_{\boldsymbol{j}}^{} R (\boldsymbol{X}_{\boldsymbol{k}}, \boldsymbol{X}_{\boldsymbol{j}}) \omega_{\boldsymbol{j}} (t). \label{eq-multiD-continuous-SCE-discretized}
\end{align}

The Podolak algorithm is also extended to the multi-component. The bin that the outcome aggregate falls in $n_p$ (Eq. \ref{eq-podolak-n}) and the portion of the term $\epsilon_p$ (Eq. \ref{eq-podolak-epsilon}) is calculated for each parameter $p$ using the $p$-th element of the dust property vector $X_\mathrm{I+II}^{(p)}$. The multi-component output function $K (\boldsymbol{X}_\mathrm{I+II}, \boldsymbol{X}_{\boldsymbol{k}})$ is defined by the product of output functions for all the axes
\begin{align}
  K (\boldsymbol{X}_\mathrm{I+II}, \boldsymbol{X}_{\boldsymbol{k}}) &= K\left(X_\mathrm{I+II}^{(1)}, X_{k^{(1)}}^{(1)}\right) \cdots K\left(X_\mathrm{I+II}^{(d)}, X_{k^{(d)}}^{(d)}\right).
\end{align}
Taking mass-volume space as an example, the pseudocode for the two-component direct method is shown in the Appendix \ref{sec-a-pseudocode}.

More intricate schemes to conserve quantities before and after the distribution can be constructed, such as the modified Podolak algorithm in \citet{brauerCoagulationFragmentationRadial2008} and the three-point framework by \citet{chakrabortyNewFrameworkSolution2007}. Since the difference in choice of these does not affect the major part of our algorithm, we do not cover these.

\section{A tree algorithm for the multi-component Smoluchowski coagulation equation}\label{sec-3-treealgorithm}

This section explains our novel tree algorithm for the multi-component SCE. The key to our tree algorithm is the assumption that property ratios (e.g., mass ratio, volume ratio, etc.) serve as a distance. By defining the distance between bins, we can apply the tree algorithm, which groups bins that are far from the interacting bin partner. Our new coagulation method can reduce the computational cost from $\mathcal{O} (N^{2d})$ to $\mathcal{O} (dN^d \log N)$ (see Section \ref{sec-3-4-procon}). Our method can also consistently and easily handle the non-conservative coagulation for the first time.

This section is organized as follows. Section \ref{sec-3-1-treenbodygravity} reviews the concept of the tree algorithm in $N$-body simulations. Section \ref{sec-3-2-treeSCE} introduces our tree algorithm for the SCE, and Section \ref{sec-3-3-treeSCEdetail} explains it in detail. Section \ref{sec-3-4-procon} describes the strengths and limitations of the tree algorithm.

\subsection{Key concept of the tree algorithm in $N$-body simulations} \label{sec-3-1-treenbodygravity}

Before explaining our tree method for SCE, we briefly explain the original tree algorithm in $N$-body simulations \citep{barnesHierarchicalLogForcecalculation1986}. The simple $N$-body direct method calculates all the (gravitational) interactions between the $N$ particles. The number of these interactions is the number of all combinations of the particles receiving the force ($i$-particles) and the particles exerting the force ($j$-particles), so the computational complexity becomes $\mathcal{O} (N^2)$. The $N$-body Tree Algorithm groups $j$-particles far away from the $i$-particle, whose forces are smaller. This reduces the number of interactions and the computational complexity to $\mathcal{O} (N \log N)$. A tree data structure, used to group particles, stores the geometric proximity data of the particles. The oct-tree (if 3D), or quad-tree (if 2D), recursively divides the particles into groups and the space into cells, until the cell only contains one particle.

The $N$-body tree algorithm consists of two parts: tree construction and force iteration using the tree. The tree is constructed by recursively partitioning real space into octo-cells, half the length for each dimension. The partitioning repeats until there is only one particle in the cell. This tree construction is taken every timestep, every several timesteps, or only during initialization and adjusted every timestep, depending on the extent of optimizations.

After the tree is constructed, the program calculates the gravitational interaction between the particles. The $i$-particle (receiving the force) is iterated for all particles, and inside the $i$-loop, $j$-particles (giving the force) are iterated recursively, starting from the root of the tree. The force from the $j$-cell is added to the $i$-particle if the $j$-cell is sufficiently resolved seen from the $i$-particle. If the $j$-cell is not sufficiently resolved, forces from each child cell inside the $j$-cell are recursively calculated.

To determine if the $j$-cell is sufficiently resolved, or the $j$-cell can be grouped, the opening angle $\theta$ is used. The opening angle is defined as follows:
\begin{align}
  \theta &= \frac{L}{D},
\end{align}
where $L$ is the length of the $j$-cell and $D$ is the distance from the $i$-particle to the mass center of the $j$-cell. The resolution is enough if $\theta < \theta_c$, where $\theta_c$ is the critical opening angle.

\subsection{Introducing the tree algorithm for the Smoluchowski coagulation equation} \label{sec-3-2-treeSCE}

Here, we introduce a new coagulation method that groups bins, inspired by the $N$-body tree algorithm. The grouping is possible when the grouped particles or bins ($j$) give almost the same interactions to the particle or bin of interest ($i$). In the case of gravitational interactions, only the relative position of particles affects. The $j$-particles close together apply similar gravitational forces to the $i$-particle, due to the smoothness of the gravitational potential. The hierarchical grouping of $j$-particles by direction and distance, implemented by the gridded space and the opening angle, thus works well with $N$-body gravity simulations. In the case of the SCE, the interaction is the coagulation of a pair of bins (I and II in the original equation (Eq. \ref{eq-1D-continuous-SCE-var}), or $i$ and $j$ in the discretized equation (Eq. \ref{eq-1D-continuous-SCE-discretized})), and thus the grouped bins (II or $j$) must have similar coagulation results against the bin to coagulate (I or $i$). For example, the coagulation of mass $i=1000$ and $j=1$ aggregates can be practically treated the same as that of mass $i=1000$ and $j=2$ aggregates.

To convert this idea into a mathematical expression, we assumed that the ratio of dust properties (i.e., mass ratio, volume ratio) of colliding aggregates is a virtual distance (Fig. \ref{fig-1-scheme} (b) and Eq. \ref{eq-tree-distance}). This essential assumption enables the application of the tree method to coagulation calculation.

The assumption of the distance raises two problems: (1) This assumption only validates the equipment of the distance but does not validate the grouping of $j$-bins far from the $i$-bin. How can the bin grouping be validated? (2) Which dust property other than mass is valid for distance calculation, and which is not?

To discuss the validation of the grouping of bins far away, we need to introduce here the major difference between $N$-body simulations and coagulation calculations: the outcome $k$-bin. For gravity, only $i$ and $j$-particles are affected by the interaction between $i$ and $j$-particles, but for coagulation, $k$-bins are also affected by the coagulation of $i$ and $j$-bins. For this additional effect, we need to consider how the grouping of $j$-bins affects $k$-bins. Specifically, we need to consider which $k$-bin is increased and by how much. Another thing to consider in the multi-component SCE is that it is difficult to analytically calculate the $j$ from $k$ and $i$. For example, calculating the inverse function of Equation 15 in \citet{okuzumiRapidCoagulationPorous2012}, a nonlinear model that calculates the volume after coagulation $v_k$ from $i$ and $j$, is difficult. A conservative law for the porosity in this model is not determined, either. This hinders the use of some conservative quantities as a dust parameter other than volume or porosity. This complication also forces the loop indices to be $i$ and $j$, not $k$.

\begin{figure}
  \resizebox{\hsize}{!}{\includegraphics{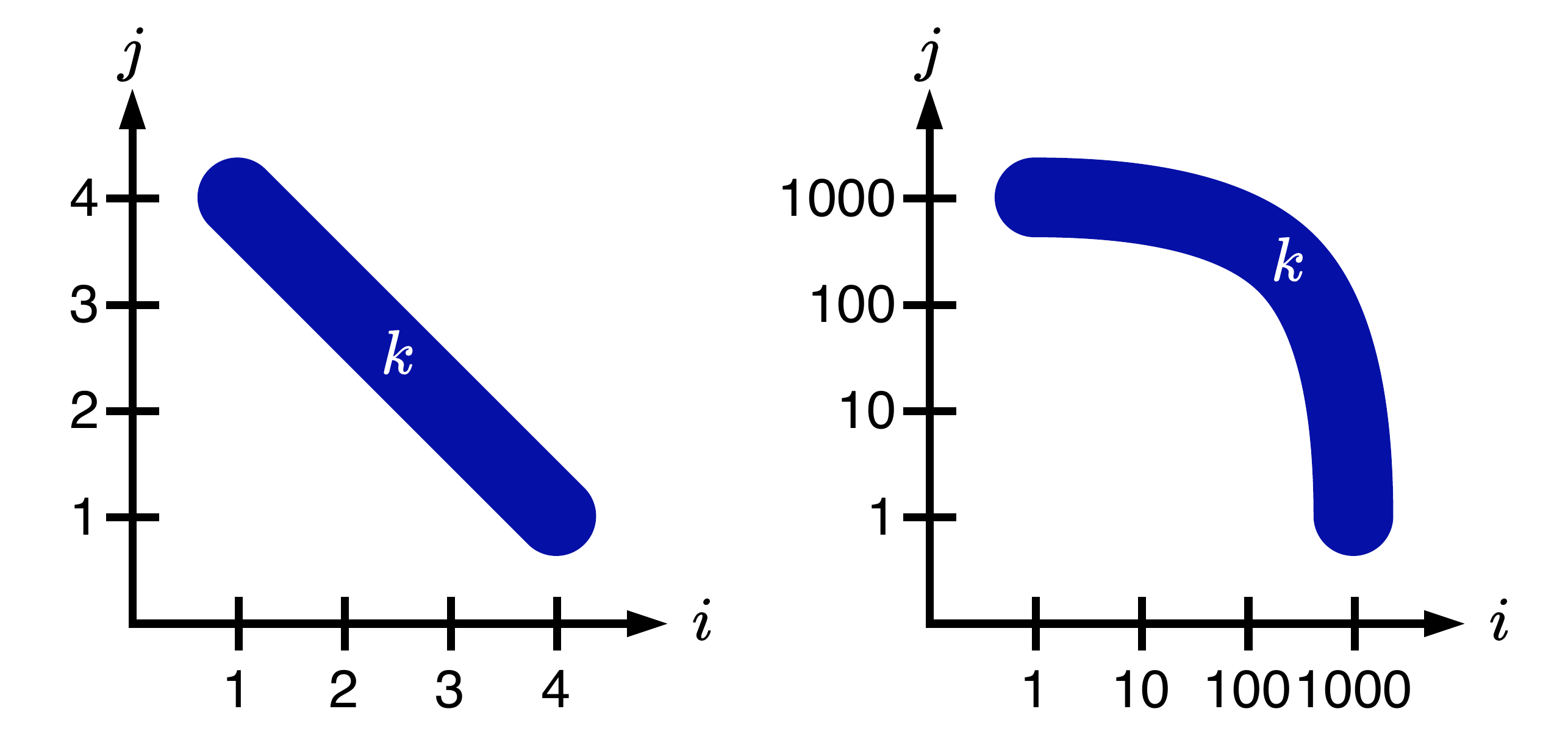}}
  \caption{Positions corresponding to $k$-dust aggregates (outcome of the coagulation) in the $(i, j)$ aggregates (two colliding aggregates) matrix. (Left) When the dust properties are on the linear axes, the $k$-bin after the coagulation spans diagonally. (Right) When the dust properties are on the logarithmic axes, the $k$-bin after the coagulation spans along the outwardly convex curve.}
  \label{fig-3-matrix-k}
\end{figure}

The grouping of the $j$-bins far away in the tree algorithm for the SCE is justified by using the logarithmic grid for the basic dust properties. Let us consider how the grouping of $j$-bins affects the selection of the $k$-bin, using mass as a first example. This validation is only important in the first term of the SCE, which describes the increase in the $k$-bin due to the coagulation of $i$ and $j$-dust aggregates. If we use the linear grid for the mass, we cannot group the $j$-bins because the loop index $i$ is fixed within the loop, and the $k$-bin, calculated by $m_i + m_j = m_k$, differs for every $j$.
However, by using the logarithmic grid, we can group the $j$-bins effectively. Since the logarithmic grid partitions bins finer in smaller quantity regions, there are more $j$-bins in $j < i$ that satisfy $m_{k-1/2} \leq m_i + m_j < m_{k+1/2}$, where $i$ and $k$ are fixed. Inside the $i$-loop, we can group these $j$-bins and calculate the coagulation of these pairs at once. For example, the increase in mass 1000 bin can be calculated at once by grouping the coagulation of mass $m_i = 990$ and $m_j = 1$ aggregates, that of mass $m_i = 990$ and $m_j = 2$ aggregates, etc. This is graphically illustrated using the positions of matrix elements corresponding to $k$, where the axes of the matrix are $i$ and $j$ (Fig. \ref{fig-3-matrix-k}). The diagonal elements in the matrix for the linear grid become bent for the logarithmic grid. Then, we can group the $j$-bins where the curve is vertical. The farther the $j$-bins are from the $i$-bin, the smaller the differences are and the coarser the grouping can be. This supports hierarchical bin grouping, where more distant bins are grouped coarsely.

If the dust property is conservative, then the same justification as mass can be made. Each of the dust parameters is equivalent, and no dust parameter is mathematically special in the multi-component coagulation equation. This naturally leads to using the same gridding of the bins and extending the bin grouping to multi-component. This paper evaluates the algorithms with the equations with the conservative dust properties, so the justification above is sufficient.

\begin{figure*}
  \sidecaption
  \includegraphics[width=12cm]{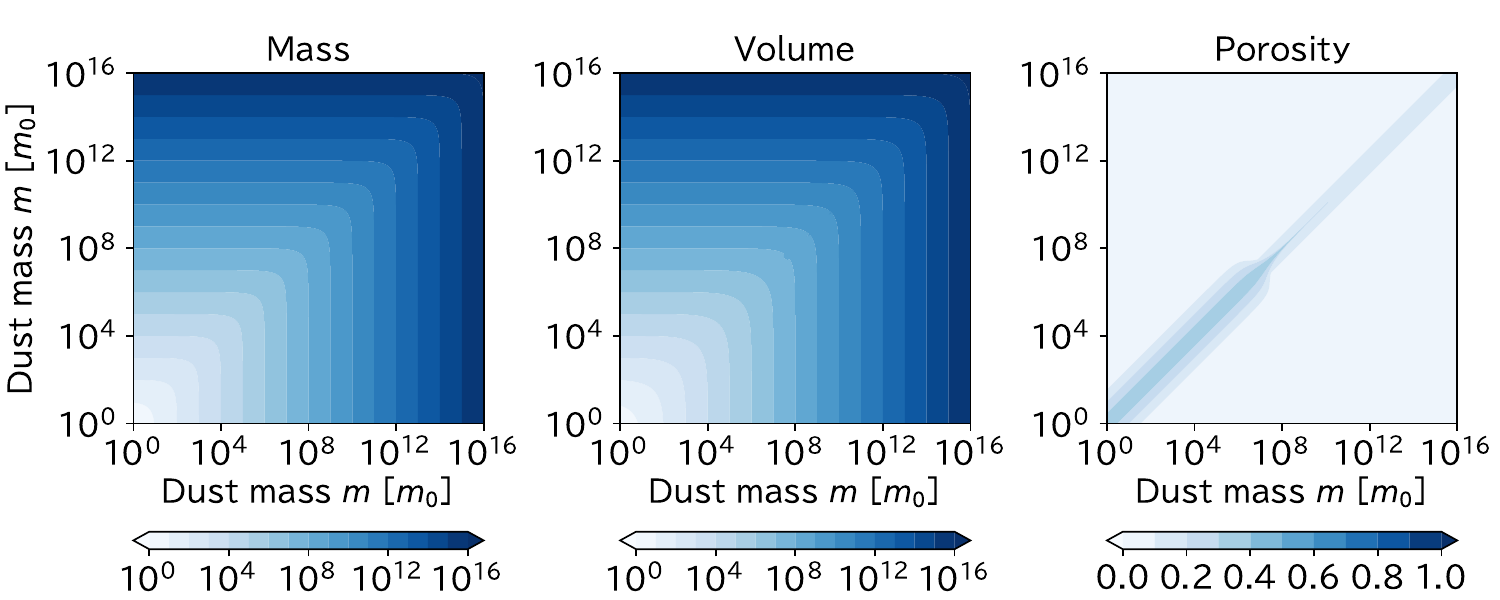}
  \caption{Contour plot of the dust property value after coagulation on $i$ and $j$ grid. This is the same as Fig. \ref{fig-3-matrix-k}, but with concrete numbers. The mass and volume are normalized by using monomer units, i.e., $m_0 = v_0 = 1$. (Left) Contour plot of $m_i + m_j$. (Middle) Contour plot of effective volume evolution with \cite{okuzumiRapidCoagulationPorous2012} model, assuming that the original dust aggregate is compact. (Right) Contour plot of porosity after the coagulation, using the mass and volume from the left and middle panels.}
  \label{fig-3-matrix-k-concrete}
\end{figure*}

For the non-conservative dust properties, the justification needs to be made on a case-by-case basis. For example, the effective volume after the coagulation $v_\mathrm{I+II}$ can be approximated as $v_\mathrm{I} + v_\mathrm{II}$, as long as the logarithmic axis and appropriate porosity model are used (Fig. \ref{fig-3-matrix-k-concrete} middle panel). However, for the value of the porosity itself, we cannot conclude the same because the shape of the outcome function of the porosity (Fig. \ref{fig-3-matrix-k-concrete} right panel) differs greatly from that of the mass (Fig. \ref{fig-3-matrix-k-concrete} left panel). We can conclude that we need to use extensive variables, such as mass, or almost-extensive variables, such as effective volume, for dust parameters in the tree algorithm.

We note that for the simple one-component SCE with mass as the dust property and the linear grid, the mass conservation allows us to use $k$ as the loop index. The bin pairs $(i, j)$ that give the same $k$ can be grouped, and this idea leads to the diagonal matrix or tensor decomposition methods for the SCE using the linear grid \citep{matveevTensorTrainMonte2016, smirnovFastAccurateFinitedifference2016, dyachenkoMosaicskeletonApproximationAll2025}. However for our matter: the multi-component SCE in planet formation, we cannot do this because (1) the models for components other than mass (e.g., porosity) are often complex and do not have conservation laws, which forbid the grouping of bins by $k$ (as aforementioned), and (2) planet formation needs a much wider range of scales, essentially requiring the logarithmic grid. Paradoxically, we solved these problems by using the logarithmic grid as above.

\begin{figure}
  \resizebox{\hsize}{!}{\includegraphics{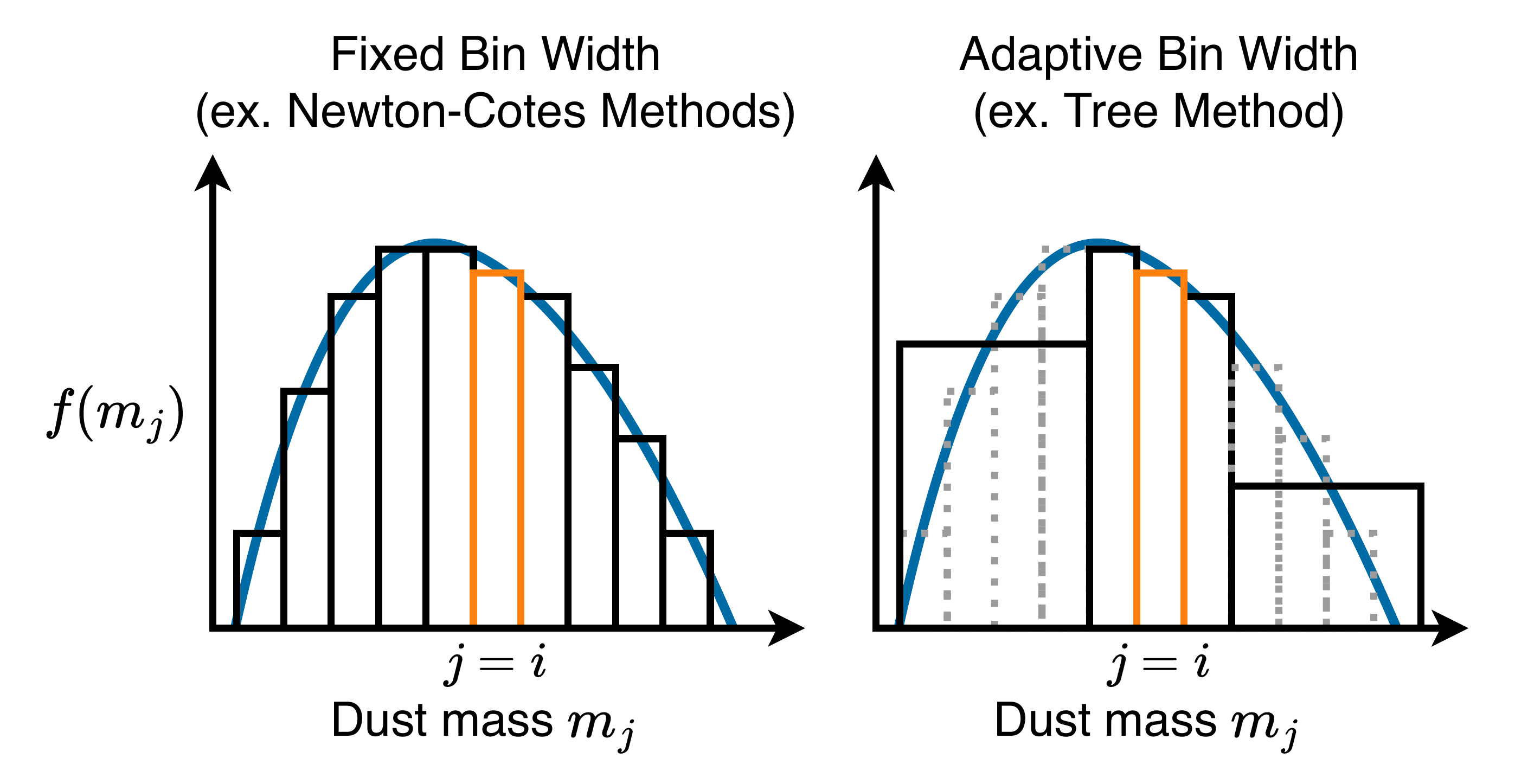}}
  \caption{Illustration of numerical integration using the fixed bin width (ex. Newton-Cotes rectangle method) (left) and the adaptive bin width (ex. our tree method) (right). In the adaptive bin width methods, the bin width becomes larger in the farther region, where the virtual distance is the mass ratio of $i$-dust and $j$-dust.}
  \label{fig-3-treequadrature}
\end{figure}

After the grouping, we can approximate the integral or summation in both terms of the SCE using the average $j$-bin value. The direct method used the rectangle method with the fixed bin width to discretize the continuous equation. Here, the tree method uses the rectangle method with adaptive bin width (i.e., the grouped bins have a larger bin width than the non-grouped ones) to discretize the equation (Fig. \ref{fig-3-treequadrature}). The farther the $j$-bins are from the $i$-bins, the more aggressively they will be grouped. The condition on how much the $j$-bins are grouped will be discussed in the next section, and here we explain what happens after the grouping of the bins. The total number of dust aggregates within the grouped $j$-bin is
\begin{align}
  \Omega_{\boldsymbol{j}} = \sum_{{\boldsymbol{j}}' \in {\boldsymbol{j}}} \omega_{\boldsymbol{j}'} (t),
\end{align}
and the average representative value of the grouped $j$-bin is
\begin{align}
  \overline{\boldsymbol{X}}_{\boldsymbol{j}} = \frac{1}{\Omega_{\boldsymbol{j}}} \sum_{\boldsymbol{j}' \in \boldsymbol{j}} \omega_{\boldsymbol{j}'} (t) \boldsymbol{X}_{\boldsymbol{j}'},
\end{align}
where $\boldsymbol{j}' \in \boldsymbol{j}$ denotes all the child nodes within the grouped $j$-bin. $\Omega_{\boldsymbol{j}}$ corresponds to the total mass within a cell in the gravitational tree algorithm, and $\overline{\boldsymbol{X}}_{\boldsymbol{j}}$ corresponds to the center of mass of the cell, respectively. Using this, we can approximate the first term of the SCE as 
\begin{align}
  & \sum_{\boldsymbol{j}} K (\boldsymbol{X}_\mathrm{I+II}(\boldsymbol{X}_{\boldsymbol{i}}, \boldsymbol{X}_{\boldsymbol{j}}), \boldsymbol{X}_{\boldsymbol{k}}) R (\boldsymbol{X}_{\boldsymbol{i}}, \boldsymbol{X}_{\boldsymbol{j}}) \omega_{\boldsymbol{i}} (t) \omega_{\boldsymbol{j}} (t)\nonumber\\
  \approx & K (\boldsymbol{X}_\mathrm{I+II}(\boldsymbol{X}_{\boldsymbol{i}}, \overline{\boldsymbol{X}}_{\boldsymbol{j}}), \boldsymbol{X}_{\boldsymbol{k}}) R (\boldsymbol{X}_{\boldsymbol{i}}, \overline{\boldsymbol{X}}_{\boldsymbol{j}}) \omega_{\boldsymbol{i}} (t) \Omega_{\boldsymbol{j}}.
\end{align}
The second term of the SCE, which describes the decrease in the $k$-bin from the coagulation of $k$ and $j$-dust aggregates, can be similarly approximated as follows.
\begin{align}
  & - \omega_{\boldsymbol{k}} (t) \sum_{\boldsymbol{X}_{\boldsymbol{j}}} R (\boldsymbol{X}_{\boldsymbol{k}}, \boldsymbol{X}_{\boldsymbol{j}}) \omega_{\boldsymbol{j}} (t)\nonumber\\
  \approx & - R (\boldsymbol{X}_{\boldsymbol{k}}, \overline{\boldsymbol{X}}_{\boldsymbol{j}}) \omega_{\boldsymbol{k}} (t) \Omega_{\boldsymbol{j}}.
\end{align}
Note that the $k$-bin here corresponds to the $i$-bin in the first term in the sense of the non-grouped bin of the colliding pair. In both terms, the $j$-bin is the virtual bin that is created by bin grouping. For this, we need to be careful about the integral interval. The first term of the SCE only needs to be integrated over $j < i$. However, this second term needs to remove the dust aggregates from the coagulation with all the aggregates, resulting in an integral over a half-infinite region. This means that we have to group not only $j$-bins that are smaller than $k$-bins but also larger ones in the second term.

Introduction of the distance into the dust property space and verification of the bin grouping enables us to apply the tree algorithm to the SCE, similarly to $N$-body simulations (Fig. \ref{fig-1-scheme}). The tree is constructed bottom-up recursively by creating nodes with $2^d$ adjacent bins as child nodes. Since the geometry of the bins is fixed, this tree construction only takes place in the initialization. The tree needs to be balanced; i.e, the height must be the same (or only differ by one) for all leaves, or it is inefficient. For each timestep, the coagulation, corresponding to the force in the $N$-body, is calculated by looping over all $i$-bins and looping over $j$-bins using the tree. If $i > j$, the addition to the $k$-bin (the first term in the SCE) and the subtraction from the $i$-bin (the second term in the SCE) are done, and if $i < j$, only the subtraction from the $i$-bin (the second term in the SCE) is done. The $j$-bins or nodes are read-only because updating them increases the computational complexity, going back to $\mathcal{O} (N^{2d})$.

\subsection{Detailed explanation of the tree algorithm for the Smoluchowski coagulation equation} \label{sec-3-3-treeSCEdetail}

This subsection explains the details of the tree algorithm code. We implemented the tree code for the SCE in C++. 

In initialization, the program first constructs the bin grid in the same way as the ones in the direct methods previously explained in Section \ref{sec-2-directmethod-onecomponent} and \ref{sec-2-directmethod-multicomponent}. Next, the program instantiates the tree structure. In our code, the tree nodes are represented as structs with the mean dust properties $\overline{\boldsymbol{X}}$, the maximum and minimum dust properties, the dust number density $\Omega$, and pointers to the next node (\texttt{next}) and the first child node (\texttt{more}). The \texttt{next} points to the next sibling node if there is one, or else the aunt-uncle node (the next sibling of the parent node). The grid and bin geometry are the same throughout the simulation, but the mean dust properties and the dust number density must be updated every timestep.

At every timestep, the program recursively updates the dust property values in the tree and calculates coagulation using the tree. The program recursively updates the values using the depth-first-search iteration. This update costs $\mathcal{O} (dN^d \log N)$. After the update, the program calculates the coagulation. The program iterates over all the $i$-bins, then over the $j$-bins and nodes using the tree. The $j$-node is assumed to be a singular bin for calculating the coagulation (i.e., the term $R(i, j) n(j)$, the $k$-bin, and $\epsilon$), with the total mass and the total number density being the sum of the bin values within the node. If the $j$-node is resolved enough, as defined later, the first term of the SCE is added to the $k$-bins, and the second term of the SCE is removed from the $i$-bins. Otherwise, the program iterates over all child nodes of the $j$-node, treating them as new $j$-nodes. The $j$-bins selected using the tree are read-only and are not updated. The "empty" nodes and bins, which are those with zero dust number density, are skipped, similarly to the direct method.

Next, we explain three conditions that determine whether the $j$-nodes are resolved sufficiently. If all three conditions are satisfied or the $j$-node is a leaf node (i.e., the bottommost node), the $j$-node is adequately resolved, and the coagulation is calculated using that $j$-node; otherwise, it is not.

The first condition is the opening angle $\theta$, which is also used in the original $N$-body gravity tree algorithms. We defined the "distance" between two dust property vectors $\boldsymbol{X}_1$ and $\boldsymbol{X}_2$ in the dust property space with logarithmic axes using the $L_2$ norm, i.e.,
\begin{align}
  || \boldsymbol{X}_1 - \boldsymbol{X}_2 || &= \sqrt{ \sum_{p=1}^d \left (\log_{10} X_1^{(p)} - \log_{10} X_2^{(p)} \right)^2 }. \label{eq-tree-distance}
\end{align}
This "distance" can be seen as a logarithm of the ratio of dust aggregate properties. The length of the cell $L$ is defined using the "distance" between the maximum and the minimum of the dust property value within the $j$-node, and the distance $D$ is defined using the "distance" between the average property values of the $i$-bin and $j$-node (see Fig. \ref{fig-1-scheme} b). Using these values and the critical opening angle constant $\theta_c$, we defined it as sufficiently resolved if 
\begin{align}
  \theta &= \frac{L}{D} < \theta_c. \label{eq-tree-openingangle}
\end{align}
The smaller the $\theta_c$ is, the more resolved the $j$-node becomes, and the longer and more precise the calculation becomes.

The critical opening angle $\theta_c$ can exceed unity. For example, let us consider a case where the $i$-bin is the bin next to the maximum of the grouped $j$-bins and the mass distribution within $j$-bins is leaning toward the $i$-bin. The distance $D$, calculated between the $m_i$ and the average value of $m_j$, can become very small. At the same time, the length of the cell $L$, which is calculated between the maximum and the minimum of $m_j$ can become very large. In this case, the calculated opening angle $\theta$ can exceed unity, and the condition for grouping the $j$-bin in this case can be handled. Even though such a case rarely happens, the critical opening angle $\theta_c$ over unity can be considered.

The second condition is the maximum width of the dust parameter distribution after a coagulation $k_c$. The tree algorithm uses the average value for the $j$-node, whereas the more precise direct method calculates all the bins within the $j$-node. This leads to only one value of $k$ in the tree algorithm, compared to multiple values (or a distribution) of $k$ in the direct algorithm. A large variance in the $k$-dust coming from $j$-node distribution can lead to unrealistic numerical diffusion in the tree algorithm. We limit the difference between the virtual maximum and minimum of $k$ to weaken this diffusion. Since calculating the actual maximum and minimum of $k$ is computationally demanding, we assumed a monotonic increase in the dust properties before and after a coagulation. Then, we can use the virtual maximum of $k$ calculated from the maximum of $j$-node values, and the virtual minimum likewise. The maximum and minimum are obtained for each element of the dust parameter vector. We defined it as resolved enough if
\begin{align}
 (\text{index of maximum virtual } k \text{-dust}) \nonumber\\ 
 - (\text{index of minimum virtual } k \text{-dust}) &< k_c, \label{eq-tree-kc}
\end{align}
where $k_c$ is the maximum width of the dust parameter distribution after a coagulation. The smaller the $k_c$ is, the more resolved the $j$-node becomes, and the longer and more precise the calculation becomes. Setting $k_c$ to be a very large number, such as $1000000$, is essentially the same as $k_c = \infty$, which means to disregard this second condition.

Note that this $k_c$ parameter turns out to be insignificant upon the parameter survey in later sections. This condition is included since we had an intuition that the outcome $k$, which was not in the original gravitational tree algorithm but is in this tree algorithm for coagulation, might affect the quality of the algorithm as explained above. To identify this question, we put in this parameter $k_c$ that can control the quality based on the distribution of the virtual $k$-bins, and addressed this parameter here.

The third condition is that the $i$-dust property is not contained within the $j$-nodes. If at least one of the dust properties of $i$ and $j$ matches, the $j$-bins cannot be grouped and must be calculated separately. This condition intends to separate the $i<j$ part, where it extends the assumption, and the $i>j$ part, where it does not need the extension of the assumption. The $i=j$ part is also calculated separately. This condition does not have a parameter that can be changed.

The pseudocode for the one-component tree method is shown in the Appendix \ref{sec-a-additionalkernels}.

\subsection{Strengths and limitations of the tree algorithm for the Smoluchowski coagulation equation} \label{sec-3-4-procon}

\begin{figure}
  \resizebox{\hsize}{!}{\includegraphics{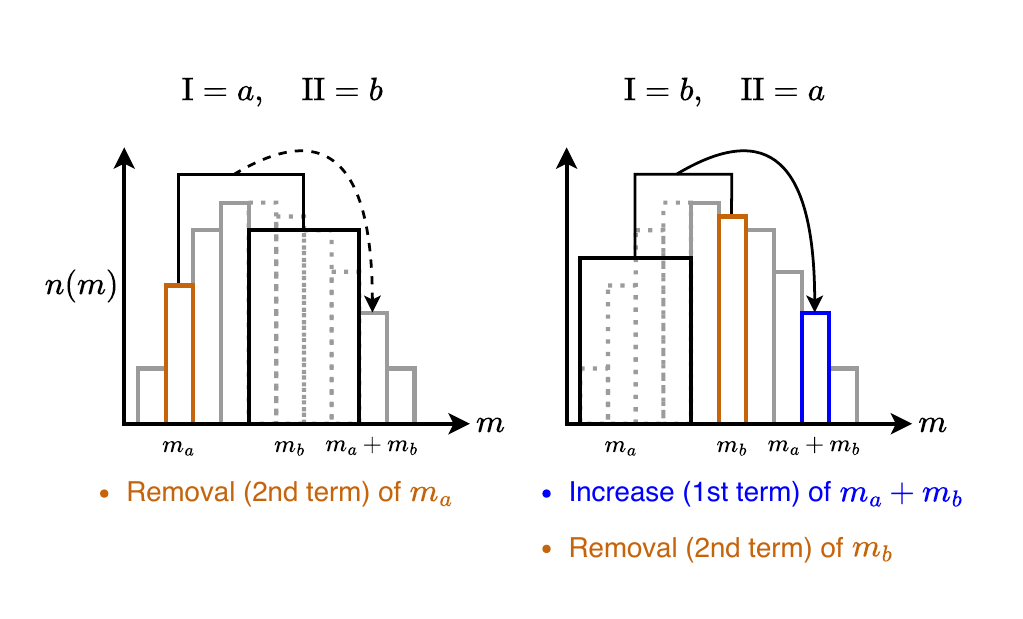}}
  \caption{Illustration of symmetry breaking in the tree algorithm for the SCE. The coagulation of mass $m=m_a$ and $m=m_b$ aggregates is calculated by one operation of the increase with the first term in the equation and two operations of the decrease with the second term. If $m_a < m_b$, the increase of $m_a + m_b$ and the decrease of $m_b$ is calculated with one grouping of bins (Right panel), and the decrease of $m_a$ is calculated with another grouping of bins (Left panel). Since these two groupings are different, the total mass exchange is generally non-zero, which means that the total mass does not conserve at the formalization. An operation cannot be performed on the grouped (II or $j$) bins since it increases the number of operations, diminishing the advantage of the low computational cost of the tree method.}
  \label{fig-3-symmetry}
\end{figure}

The main advantage of the tree algorithm is its low calculation cost. The overall temporal cost is $\mathcal{O} (d N^d \log N)$, where $d$ is the number of components and $N$ is the number of bins per component. For example, the tree algorithm for the one-component SCE costs $\mathcal{O}(N \log N)$ and for the two-component $\mathcal{O}(N^2 \log N)$. The breakdown of the time complexity is as follows: the $i$-dust loop costs $\mathcal{O} (N^d)$, and the $j$-dust scan using the tree costs as much as the height of the tree, which is $\mathcal{O} (\log N^d) = \mathcal{O} (d \log N)$ \citep{knuthArtComputerProgramming1997}. The tree construction only takes place at initialization, and thus does not affect the computational speed. Note that this construction also takes $\mathcal{O} (d N^d \log N)$. The update of the node data, such as total mass, for every time step also takes $\mathcal{O} (d N^d \log N)$.

A limitation of the tree algorithm is that the bin grouping introduces additional approximation errors. For example, for some of the parameters, the total mass is not conserved. The coagulation equation conserves mass, so its numerical scheme is also desirable to conserve mass. The mass conservation violation of the tree method arises from the breaking of symmetry in the mass transfer from smaller to larger mass bins by coagulation (Fig. \ref{fig-3-symmetry}). The two integrals in the SCE need to be matched during the mass transfer, but the tree algorithm evaluates them independently, leading to the symmetry violation. This is analogous to the breach of Newton's third law or total momentum conservation in the $N$-body tree algorithms. This is discussed further in Section \ref{sec-6-3-variants}.

\section{Methods of algorithm evaluation}\label{sec-4-methods}

This section explains the methods of algorithm evaluation. We measured the speed and errors of the conventional direct method and our tree method, changing the number of components $d$, kernel $R$, number of bins per component $N$, opening angle $\theta_c$, maximum dust distribution width after a coagulation $k_c$, and time integration method. The number of components $d=1$ corresponds to the mass-only case, and $d=2$ corresponds to cases such as mass-and-volume and mass-and-charge.

We show the test calculations for two kernels: the constant kernel and the additive kernel. The constant kernel is  $R(m_i, m_j)=1$. The additive kernel is $R(m_i, m_j)=(m_i + m_j)/2$ for one-component and $R(m_i, v_i, m_j, v_j) = (m_i + v_i + m_j + v_j)/4$ for two-component. The main text omits the multiplicative kernel since it describes gelation, which is beyond our interest, and the analytic solution for the multi-component multiplicative kernel equation has not been found. The one-component multiplicative kernel is also tested in Appendix \ref{sec-a-additionalkernels}. 

The initial condition is set such that analytic solutions have been derived. For the one-component, it is $\omega_i(0) = N_0 \exp(-m_i) \Delta m_i$, where the total number of aggregates is normalized to be 1 by $N_0^{-1} = \sum_i \exp(-m_i) \Delta m_i$. For the two-component, it is $\omega_{m_i, v_i}(0) = N_0 \exp(-m_i - v_i) \Delta m_i \Delta v_i$, where $N_0^{-1} = \sum_{m_i, v_i} \exp(-m_i - v_i) \Delta m_i \Delta v_i$. The analytic solutions for these kernels are shown in Section \ref{sec-2-onecomponent}  and \ref{sec-2-multicomponent}. There are also analytic solutions for the initial condition where the delta function is used. These cases are also tested in Appendix \ref{sec-a-additionalkernels}.

It is rather meaningless to compare results across different kernels; the important thing is the differences in the various algorithm parameters that affect the results for the same kernel. Testing with different kernels is mainly for algorithm validation, to determine whether the algorithms work for other kernels. 

We used the classic fourth-order Runge-Kutta method of coefficients 1/6, 1/3, 1/3, and 1/6 to time integrate Eq. (\ref{eq-1D-continuous-SCE-discretized}) and Eq. (\ref{eq-multiD-continuous-SCE-discretized}). We tested two different timesteps: (1) constant time step and (2) adaptive time step. For the constant time step, Table \ref{tab-1-mainkernels} shows the time integration parameters for each kernel. For the adaptive time step, we followed \citet{stammlerDustPyPythonPackage2022}; it is calculated such that the dust density $\omega$ does not become negative in a first-order Euler scheme considering negative elements of the $\Delta \omega$, with a coefficient of $0.05$:
\begin{align}
  \Delta t_\mathrm{adaptive} = 0.05 \times \min_i \left|\frac{\omega_i}{\min (\Delta \omega_{i}, 0)}\right|.
\end{align}
These values of the constant time step and the coefficient for the adaptive time step were set so that the calculation does not crash. The implications of using the explicit time integration methods are discussed further in Section \ref{sec-6-2-timeintegration}.

\begin{table}
\caption{Timesteps and maximum simulation times used in algorithm evaluation when using a constant time step. 
}
\label{tab-1-mainkernels}
\centering
\begin{tabular}{lcc}
\hline\hline
Kernel                               & Constant & Additive \\
\hline
Constant time step $\Delta t_\mathrm{const}$                  & $0.5$ & $0.005$\\
Max simulation time $t_\mathrm{max}$ & $10000$ & $24$ \\
\hline\hline
\end{tabular}
\tablefoot{These values are adopted to conform to \cite{leeValidityCoagulationEquation2000}.}
\end{table}

For each of these kernels, we calculated coagulations using the algorithms with parameters in Table \ref{tab-2-exploredparameters}. We use the number of bins per component $N = N_\mathrm{bd} (N_\mathrm{dec} + 1) + 1$ when showing and discussing results, where $N_\mathrm{bd}$ is the number of bins in the integer region per component and also the number of bins per decade, and $N_\mathrm{dec} = 14$ is the number of decades per component. The fiducial value for the number of bins per component is $N = 241$. The fiducial values are chosen to balance the speed and accuracy based on test calculations. As shown in the results, the fiducial $k_c=1000000$ performs better in terms of both speed and accuracy compared with other values of $k_c$. The fiducial $\theta_c$ and $N$ are chosen to balance speed and accuracy.

\begin{table*}
\caption{Algorithm parameters explored in the evaluation.}
\label{tab-2-exploredparameters}
\centering
\begin{tabular}{lccc}
\hline\hline
Parameter & Symbol & Tested Values & Fiducial Value \\
\hline
Time step & - & Constant, Adaptive & Adaptive\\
Number of bins per decade per component & $N_\mathrm{bd}$ & $4, 8, 12, 16, 20, 30, 40$ & $16$ \\
Critical opening angle & $\theta_c$ & $0.01, 0.02, 0.05, 0.1, 0.2, 0.5, 1, 2, 5, 10, 20, 50, 100$ & $1$\\
Max distribution width after coagulation & $k_c$ & $1, 2, 5, 1000000$ & $1000000$\\
\hline\hline
\end{tabular}
\end{table*}

We measured the following three indices in the algorithm evaluation: (1) Wall-clock time $T$, (2) $L_2$ error $\varepsilon_2$, and (3) Relative error of total mass $\Delta M$. The wall-clock time $T$ is the total execution time the algorithm took to compute. The $L_2$ error $\varepsilon_2$ is the sum of squared differences at the final state ($t=t_\mathrm{max}$) between the analytic distribution $m^2 n_{\mathrm{a}, {\boldsymbol{i}}}$ and the calculated distribution $m^2 n_{\mathrm{c}, {\boldsymbol{i}}}$:
\begin{align}
  n_{\mathrm{x}, {m_i}} &= \Delta m_i^{-1}\omega_{\mathrm{a}, {m_i}} (t_\mathrm{max})\\
  n_{\mathrm{x}, {m_i, v_i}} &= \Delta m_i^{-1} \Delta v_i^{-1}\omega_{\mathrm{a}, {m_i, v_i}} (t_\mathrm{max})\\
  \varepsilon_2 &= \begin{cases}
      \displaystyle \sum_{m_i} \Delta m_{i} m_i^4 \left[ n_{\mathrm{a}, {m_i}} - n_{\mathrm{c}, {m_i}} \right]^2\\
      \qquad (d=1),\\
      \displaystyle \sum_{m_i, v_i} \Delta m_{i} \Delta v_i m_i^2 v_i^2 \left[ n_{\mathrm{a}, {m_i, v_i}} - n_{\mathrm{c}, {m_i, v_i}} \right]^2\\
      \qquad (d=2).
  \end{cases}
\end{align}
We used $m^2 n$ for one-component and $m v n$ for two-component to match the figures of example calculations (ex., Fig. \ref{fig-res-1D-simpleex}). The relative error of total mass $\Delta M$ is the absolute difference of the total mass at the final state $M(t_\mathrm{max})$ and that at the initial state $M(0)$, where
\begin{align}
  M(t) &= \begin{cases}
      \displaystyle \sum_{m_i} \Delta m_{i} m_i n_\mathrm{c} (m_i),\quad (d=1),\\
      \displaystyle \sum_{m_i, v_i} \Delta m_{i} \Delta v_i m_i n_\mathrm{c} (m_i, v_i),\quad (d=2),
  \end{cases}\\
  \Delta M &= |M(t_\mathrm{max}) - M(0)|.
\end{align}

We used our Mac Studio (2022) for the algorithm evaluation. Its CPU is an Apple M1 Ultra with 20 cores and threads. Since the algorithms are not parallelized, we evaluated them on a single core and thread. There are two types of cores in the CPU: 3.20 GHz and 2.06 GHz, and a simple test using Activity Monitor showed that the evaluation was run on a 3.20 GHz core. The computer has 128 GB of memory. The OS is macOS Ventura 13.5.1.

\section{Results of algorithm evaluation}\label{sec-5-results}

This section shows the results for the algorithm evaluation. In Section \ref{sec-5-example}, we present examples of the coagulation computations using the direct and tree methods. In Sections \ref{sec-5-N} \ref{sec-5-adaptive}, \ref{sec-5-theta_c}, and \ref{sec-5-k_c}, we show the effects of simulation parameters $N$, time step, $\theta_c$, and $k_c$, respectively, on the computational time and accuracy. Finally, in Section \ref{sec-5-tradeoff}, we demonstrate the trade-off between time and accuracy. Additional tests, for the one-component multiplicative kernel case and the cases with the delta function as the initial condition, are shown in the Appendix \ref{sec-a-additionalkernels}.

\subsection{Examples of coagulation calculation} \label{sec-5-example}

\begin{figure*}
  \centering
  \includegraphics[width=17cm]{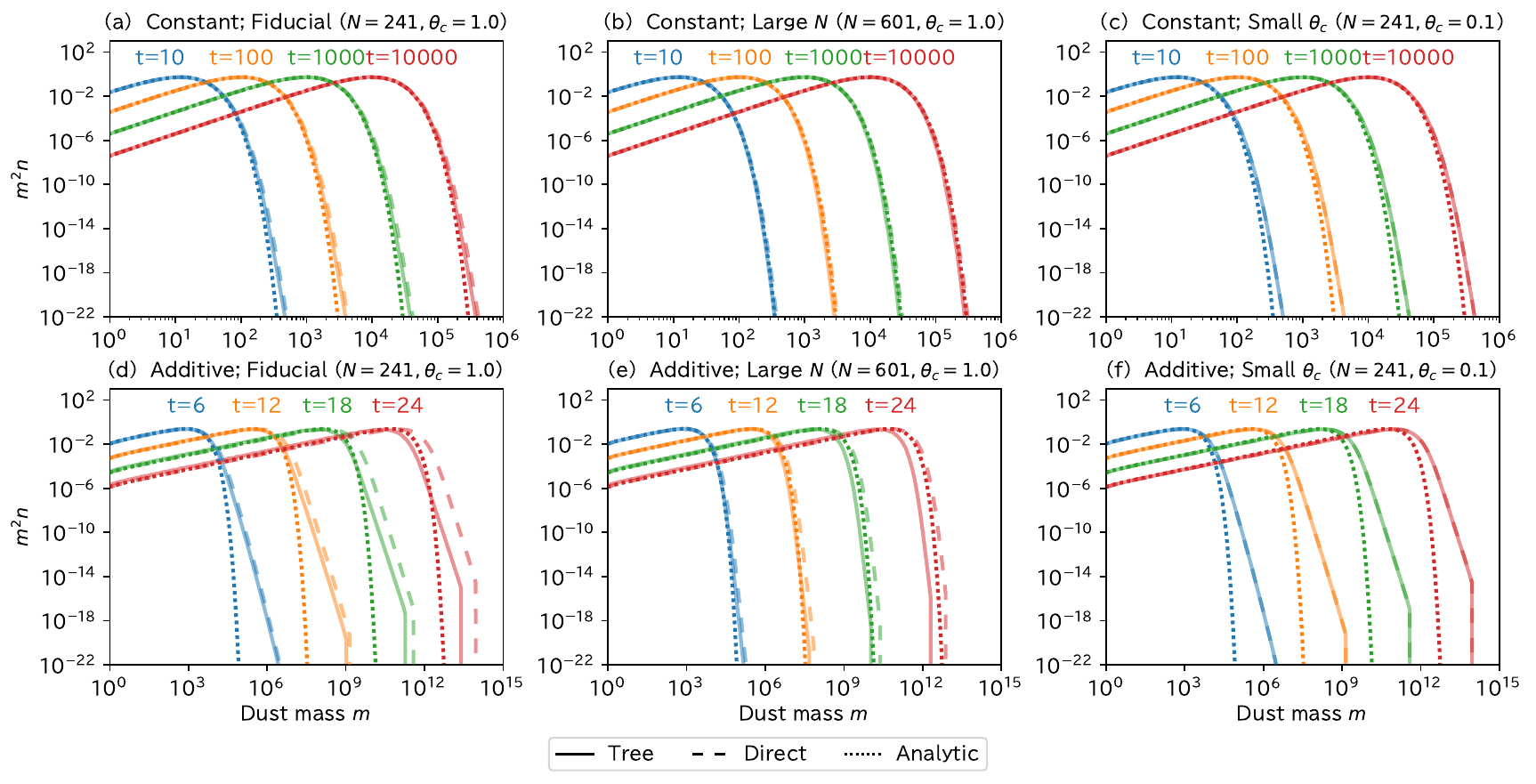}
  \caption{Analytic solutions and results of numerical calculation of one-component coagulation equation. The top panels (a), (b), and (c) are for the constant kernel, and the bottom panels (d), (e), and (f) are for the additive kernel. The left panels (a) and (d) are calculated with fiducial parameters $N_\mathrm{bd} = 16$ (i.e., $N=241$) and $\theta_c = 1$, the middle panels (b) and (e) are calculated with a finer grid $N_\mathrm{bd} = 40$ (i.e., $N = 601$) and $\theta_c = 1$, and the right panels (c) and (f) are calculated with a finer bin grouping $N_\mathrm{bd} = 16$ (i.e., $N=241$) and $\theta_c = 0.1$. In each plot, the sets of lines with four different colors show the snapshots at four different times. In each set, the lines show the tree method with the given $\theta_c$ and $k_c = 1000000$ (solid line), the direct method (dashed line), and the analytic solution (dotted line). All numerical results are calculated with adaptive time stepping.}
  \label{fig-res-1D-simpleex}
\end{figure*}

\begin{figure*}
  \centering
  \includegraphics[width=17cm]{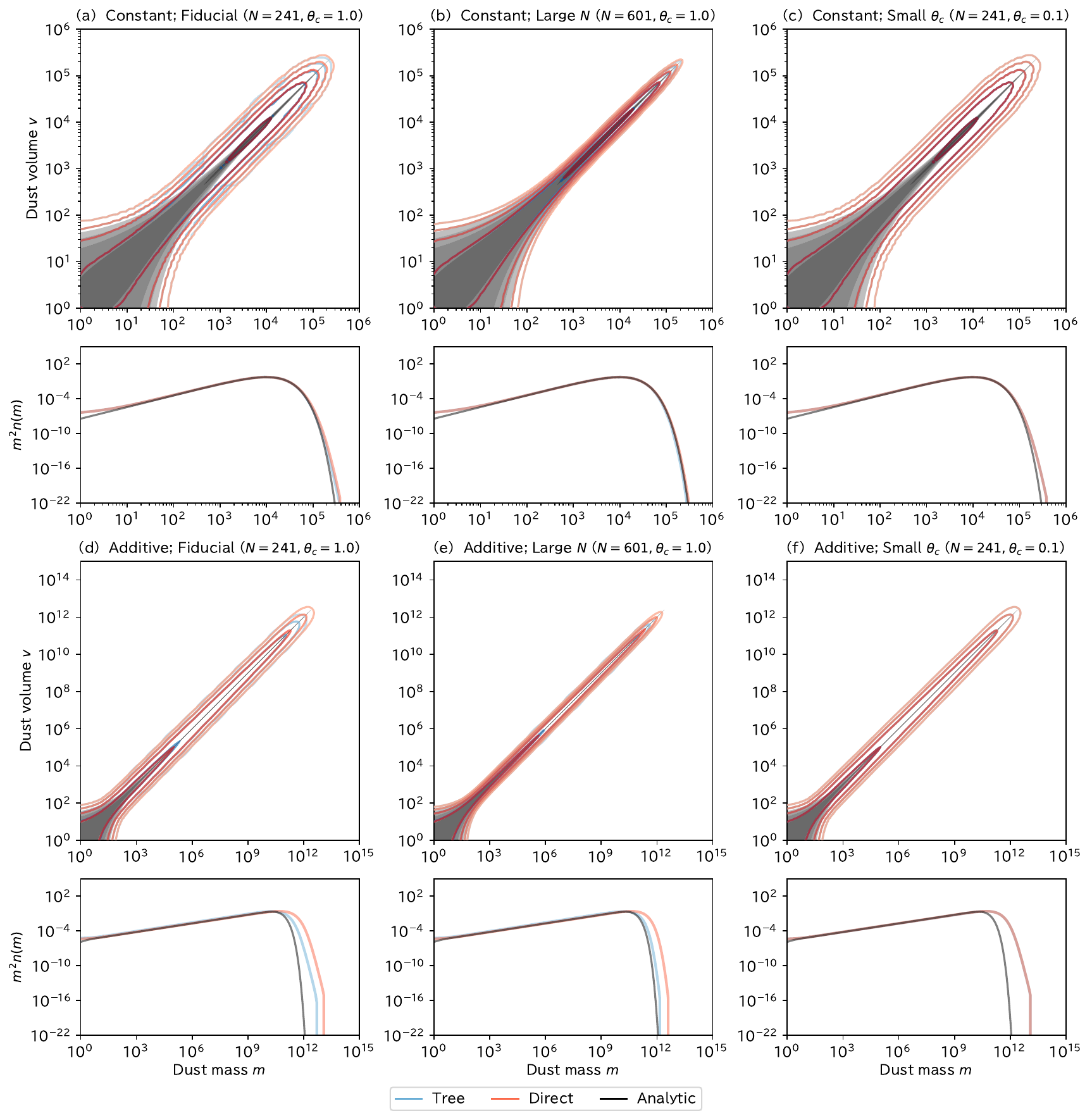}
  \caption{Analytic solutions and results of numerical calculation of the two-component coagulation equation. The panel layout is the same as Fig. \ref{fig-res-1D-simpleex}, with an additional integrated distribution panel for each case. The upper panel in each subplot (a) through (f) shows the two-dimensional distribution of $m v n(m, v)$, with contour line levels of $10^{-16}, 10^{-12}, 10^{-8}, 10^{-4}$, and $1$. The bottom panel in each subplot shows the one-dimensional distribution integrated over the secondary component (volume $v$) to get the distribution of $m^2 n(m)$, to match Fig. \ref{fig-res-1D-simpleex}. The panels show the snapshot at the end of the simulation $t=t_\mathrm{max}$. Three different colors show the three different schemes: the tree method with $\theta_c = 1$ (blue contour lines), the direct method (red contour lines), and the analytic solution (monochromatic filled contour or grey line). For most cases, the contour lines of the direct method and the tree method overlap in the 2D plot.}
  \label{fig-res-2D-simpleex}
\end{figure*}

Figure \ref{fig-res-1D-simpleex} shows the time-evolutions of the one-component ($d=1$) dust mass distributions. For all cases, the direct and tree methods are in good overall agreement with the analytic solutions. The main error occurs in the larger tail of the distribution, where the numerical methods produce a heavier tail, leading to a larger maximum dust size. The error is smaller in the constant kernel than in the additive kernel, and is also smaller in the large $N$ case than in the fiducial case. This error arises from the numerical diffusion and is common among coagulation calculations \citep[see][]{drazkowskaModelingDustGrowth2014}. The fiducial plots (a) and (d) show that the maximum dust size (i.e., $x$-intercept) of the tree method is closer to the analytic one than the direct method. This is counterintuitive. It does not mean that the tree method calculates precisely, but the discrepancy is smaller because of another error: the underestimation of maximum dust size due to bin grouping. The error is smaller in the small $\theta_c$ case than in the fiducial case. The underestimation error due to the bin grouping cancels out with the tail overestimation from the numerical diffusion, causing the maximum dust size of the tree method to be closer to the analytic solution than the direct method. These two errors: the overestimation due to the numerical diffusion and the underestimation due to the bin grouping, affect the following error analysis and are discussed further in Section \ref{sec-6-1-interpretation}. The numerical solutions have a small bump at $m=N_\mathrm{bd} m_0 = 16m_0\ \mathrm{or}\ 40m_0$, where the gridding changes from linear to logarithmic. The change in the slope at the larger tail of the numerical distribution, especially for the additive kernel, is caused by forcing $\omega$ to be zero when $\omega < 10^{-40}$.

Figure \ref{fig-res-2D-simpleex} shows the two-component ($d=2$) dust distributions at the end of the simulation. For all cases, the direct and tree methods are in good overall agreement with the analytic solutions. Since the initial condition is exponentially decreasing $n(0, m, v) = N_0/(m_0 v_0) \exp (-m/m_0 - v/v_0)$, infinitesimal dust aggregates with $m < 1$ (i.e., area left of the $m=10^0$-line in the figure), or $v < 1$ (i.e., area below the $v=10^0$-line in the figure) exist mathematically. These aggregates disperse the monomer $m=v=1$ aggregates to the right or vertically above, which goes against our physical intuition. The sharp "tip" of the two-dimensional distribution becomes wider and duller in the numerical solutions compared with the analytical solution, which comes from the numerical diffusion. This numerical diffusion is weaker in the large $N$ case than in the fiducial case. In the non-integrated 2D panels, the difference between the direct method and the tree method is small, but the difference becomes larger to a recognizable degree in the integrated panels. The integrated distribution is similar to the ones in the one-component case, and the two errors can be discussed similarly. The methods in larger $N$ cases have a smaller numerical diffusion error than in the fiducial cases, and the tree method in the smaller $\theta_c$ cases has a smaller underestimation error than in the fiducial cases.

\subsection{Effects of $N$ on speed and accuracy} \label{sec-5-N}

\begin{figure*}
  \centering
  \includegraphics[width=17cm]{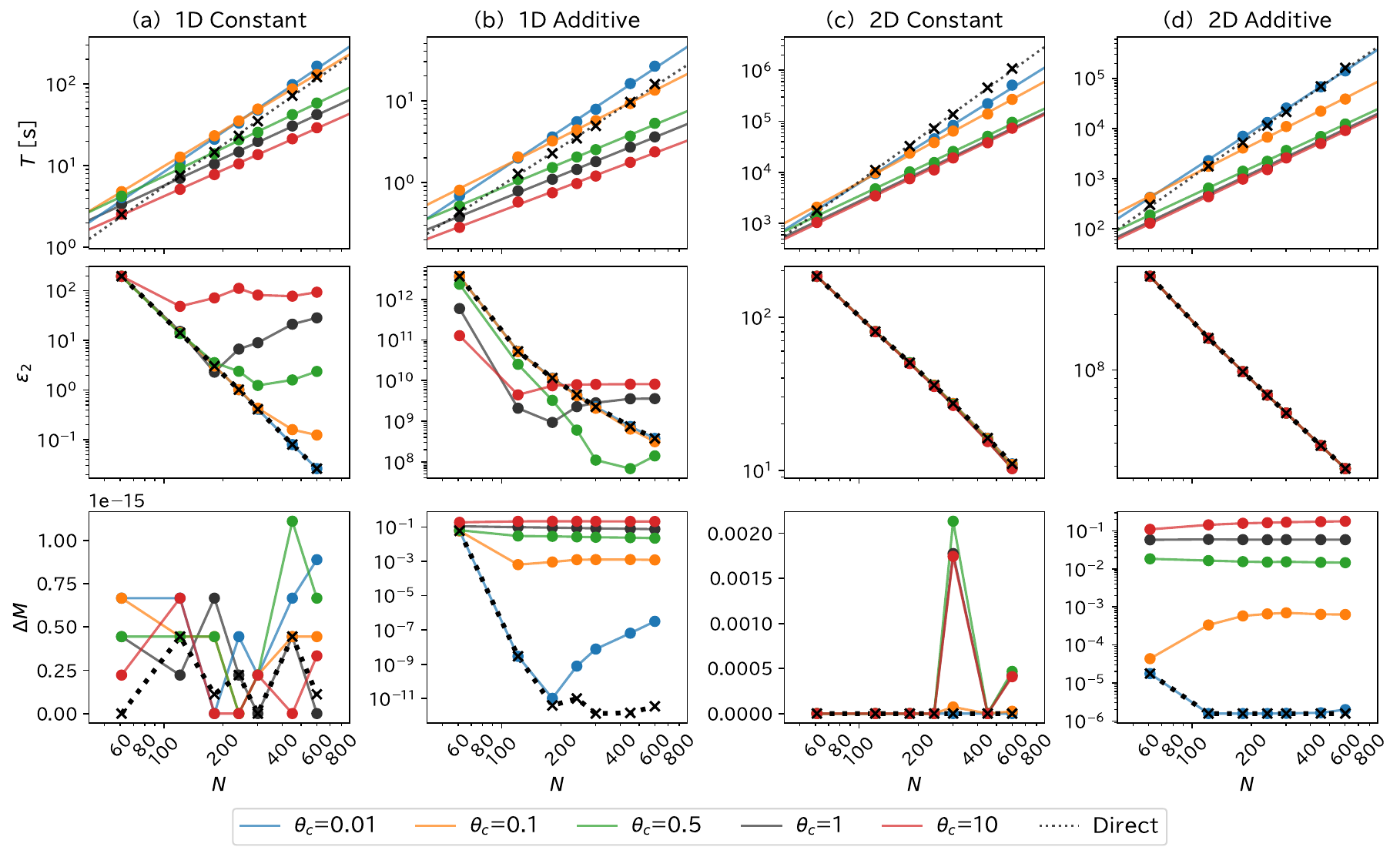}
  \caption{The effect of the number of bins per component $N$ on the wall-clock time $T$ (the top panels), the $L_2$ error $\varepsilon_2$ (the middle panels), and the relative error of the total mass $\Delta M$ (the bottom panels) for the constant time step. From the left, the panels show the results for the (a) one-component constant kernel, (b) one-component additive kernel, (c)  two-component constant kernel, and (d) two-component additive kernel. In each plot, different colors correspond to the direct method, or the different values of $\theta_c$ for the tree methods. All tree method cases are calculated with $k_c=1000000$. For the wall-clock time $T$, the lines were fitted by the least squares method, using the logarithmic values of $N$ and $T$.}
  \label{fig-5-N-consttimestep}
\end{figure*}

Figure \ref{fig-5-N-consttimestep} shows the effect of the number of bins $N$. The fiducial setting of the time stepping is adaptive, but the $N$ vs. $T$ graph for the adaptive time step is hard to understand, so here we show the results for the constant time stepping for $N$ vs. $T$ only. Other graphs are calculated using the adaptive time step. The following sub-sub-sections discuss the figure from top to bottom: $N$ vs. $T$, $N$ vs. $\varepsilon_2$, and $N$ vs. $\Delta M$. 

\subsubsection{$N$ vs. $T$}

\begin{table}
\caption{Scaling exponents for $N$ vs. $T$, for constant time step.}
\label{tab-5-scaling-N_2_time-consttimestep}
\centering
\begin{tabular}{l|cccc}
\hline\hline
& \multicolumn{2}{c}{1D} & \multicolumn{2}{c}{2D} \\
Kernel & Constant & Additive & Constant & Additive \\
\hline
$\theta_c = 0.01$ & $1.62$ & $1.58$ & $2.40$ & $2.55$ \\
$\theta_c = 0.1$  & $1.45$ & $1.21$ & $2.09$ & $1.97$ \\
$\theta_c = 0.5$  & $1.15$ & $0.99$ & $1.81$ & $1.82$ \\
$\theta_c = 1$    & $1.11$ & $0.97$ & $1.83$ & $1.85$ \\
$\theta_c = 10$   & $1.07$ & $0.91$ & $1.84$ & $1.86$ \\
Direct            & $1.69$ & $1.56$ & $2.80$ & $2.75$ \\
\hline\hline
\end{tabular}
\end{table}

The topmost panels of Figure \ref{fig-5-N-consttimestep} show the wall-clock time $T$ dependence on $N$. The scaling exponents for this are shown in Table \ref{tab-5-scaling-N_2_time-consttimestep}.

First, we detail the one-component constant kernel case. The direct method scales as $\mathcal{O} (N^{1.69})$, which is better than the theoretical estimate of $\mathcal{O} (N^2)$. The tree method scales as $\mathcal{O} (N^{1.62})$ to $\mathcal{O} (N^{1.07})$ for $\theta_c=0.01$ to $\theta_c=10$, respectively, which is consistent with the theoretical estimate of $\mathcal{O} (N \log N)$, considering adequately large $\theta_c$ case. The tree algorithm with $\theta_c = 10$ is faster than the direct algorithm at all $N$, and those with $\theta_c = 0.5, 1$ are faster than the direct algorithm at $N > 200$. The tree algorithms with $\theta_c = 0.01, 0.1$ are slower than the direct algorithm at all $N$. At the fiducial value $N=241$, the tree method with $\theta_c = 1$ takes 14.7 seconds while the direct method takes 23.1 seconds to compute; i.e., the tree method was 1.57 times faster.

Next, we continue to the two-component constant kernel case. The direct method scales as $\mathcal{O} (N^{2.80})$, which is better than the theoretical estimate of $\mathcal{O} (N^4)$. The tree method scales as $\mathcal{O} (N^{2.40})$ to $\mathcal{O} (N^{1.84})$ for $\theta_c=0.01$ to $\theta_c=10$, respectively, which is better than the theoretical estimate of $\mathcal{O} (N^2 \log N)$, considering adequately large $\theta_c$ case. The tree algorithm with parameters explored in this evaluation is faster than the direct algorithm for all $N$, except the $\theta_c = 0.1, N=61$ case. At the fiducial value $N=241$, the tree method with $\theta_c = 1$ takes 12146.7 seconds while the direct method takes 72253.5 seconds to compute; i.e, the tree method was $5.9$ times faster. For the large $N=601$ case, the tree method with $\theta_c = 1$ takes 78762.0 seconds while the direct method takes 1080260 seconds to compute; i.e., the tree method was 13.7 times faster.

For both the one-component and two-component cases, the tree algorithm with a smaller $\theta_c$, which treats bins more precisely, has worse scale than the ones with a bigger $\theta_c$, approaching the scaling of the direct algorithm. The scaling for both methods improves for the two-component than the one-component, compared with the theoretical estimates. The algorithmic scalings are better than the theoretical estimates of $\mathcal{O} (N^{2d})$ for the direct algorithm and $\mathcal{O} (d N^d \log N)$ for the tree algorithm. This can be attributed to the step where the algorithm skips empty bins, as explained in Section \ref{sec-3-3-treeSCEdetail}, which considerably reduces computational time.

For the additive kernel cases, the overall trend is the same as that of the constant kernel.

\subsubsection{$N$ vs. $\varepsilon_2$}

The middle-row panels of Figure \ref{fig-5-N-consttimestep} show the $L_2$ error $\varepsilon_2$ dependence on $N$.

First, we detail the one-component constant kernel case. The error of the direct method converges to zero as $N$ increases, with scaling of about $\varepsilon_2 \sim \mathcal{O} (N^{-4})$. However, for the tree method, the error hits its minimum at some point, depending on the critical opening angle $\theta_c$. The tree method with a smaller $\theta_c$, which treats bins more precisely, converges to the direct method: the minimum of the error decreases, and the $N$ at which it occurs increases.

Next, for the one-component additive kernel, the trends for the direct method and the tree method with a smaller $\theta_c$ are the same as the constant kernel, but the trends for the tree method with a larger $\theta_c$ are different. Here, the error for the tree method with a large $\theta_c (\geq 1)$ stays almost the same for all $N$, and it makes the error in a small $N$ lower than that of the direct method. We discuss the cause of these strange behaviors of the errors in Section \ref{sec-6-1-interpretation}.

Finally, for the two-component case, all errors behave similarly to those of the one-component direct method. This means that the effect of the critical opening angle $\theta_c$ on the error is limited. Instead, the effect of the numerical diffusion coming from the number of bins $N$ is dominant. This diffusion can be attributed to the sharp tip in the analytic two-dimensional distribution becoming wider and duller in the numerical solutions. The scaling of the error decrease is worse than the one-component, with scaling of about $\varepsilon_2 \sim \mathcal{O} (N^{-1})$. This indicates that the numerical diffusion should be considered more carefully in multi-component cases than in the one-component case. We note that if the two-component distribution is integrated and compared with the analytical solutions, the same trend as the one-component, of the error for the tree method not converging to zero, appears.

\subsubsection{$N$ vs. $\Delta M$} \label{sec-5-N_2_relerr}

The bottommost panels of Figure \ref{fig-5-N-consttimestep} show the relative error of the total mass $\Delta M$ dependence on $N$.

In all cases, the error is almost constant for all $N$'s, with several exceptions. Its value is machine-epsilon level in the one-component or two-component constant kernel cases with all schemes, and in all kernels with the direct method. In the additive kernel cases, the tree method has a larger error in the total mass, up to 30 percent for the fiducial parameters. As aforementioned in Section \ref{sec-3-4-procon}, the tree method breaks symmetry in the integral, causing this total mass error. Using a smaller value of $\theta_c$ decreases this error, converging to the result of the direct method. The exception of the direct method not conserving mass in the additive kernel at a small $N$ comes from the tail of the mass distribution running over the defined computation mass grid. In the two-component constant kernel case, the tree method with $N=20, 40$ behaves strangely.

\subsection{Effects of time step choice on speed and accuracy} \label{sec-5-adaptive}

\begin{figure*}
  \centering
  \includegraphics[width=17cm]{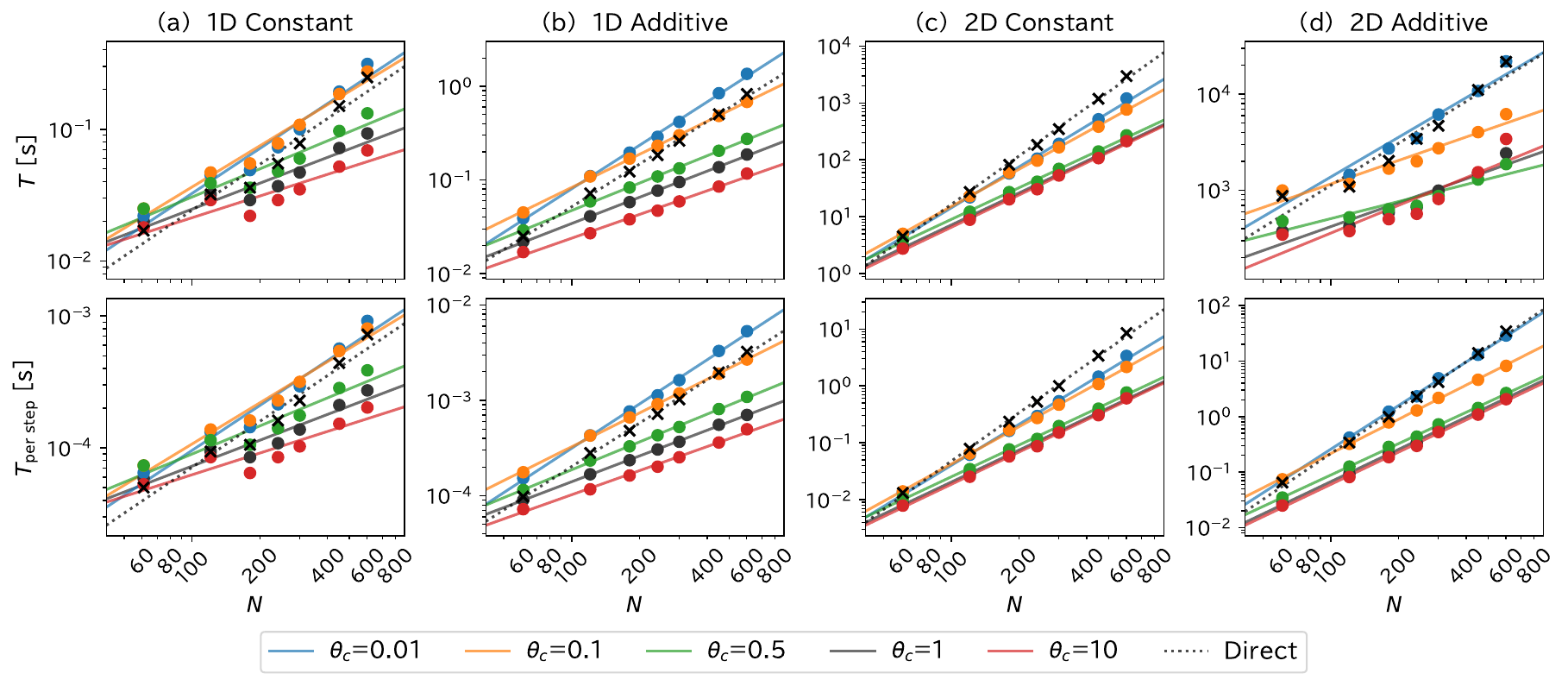}
  \caption{The effect of the number of bins per component $N$ on the wall-clock time $T$ (top panels) and the wall-clock time per step $T_\text{per step}$ (bottom panels) for the adaptive time step. From the left, the panels show the results for the (a) one-component constant kernel, (b) one-component additive kernel, (c)  two-component constant kernel, and (d) two-component additive kernel. In each plot, different colors correspond to the direct method, or the different values of $\theta_c$ for the tree methods. The tree method is calculated with $k_c=1000000$. The lines were fitted by the least squares method, using the logarithmic values of $N$ and $T$.}
  \label{fig-5-N-adaptivetimestep}
\end{figure*}

\begin{table}
\caption{Scaling exponents for $N$ vs. $T$, for adaptive time step.}
\label{tab-5-scaling-N_2_time-adaptimestep}
\centering
\begin{tabular}{l|cccc}
\hline\hline
& \multicolumn{2}{c}{1D} & \multicolumn{2}{c}{2D} \\
Kernel & Constant & Additive & Constant & Additive \\
\hline
$\theta_c = 0.01$ & $1.13$ & $1.54$ & $2.39$ & $1.37$ \\
$\theta_c = 0.1$  & $1.04$ & $1.17$ & $2.18$ & $0.81$ \\
$\theta_c = 0.5$  & $0.71$ & $0.97$ & $1.85$ & $0.59$ \\
$\theta_c = 1$    & $0.65$ & $0.93$ & $1.87$ & $0.83$ \\
$\theta_c = 10$   & $0.55$ & $0.84$ & $1.89$ & $0.96$ \\
Direct            & $1.16$ & $1.51$ & $2.83$ & $1.45$ \\
\hline\hline
\end{tabular}
\end{table}

Figure \ref{fig-5-N-adaptivetimestep} shows the $N$-scaling of the wall-clock time $T$ and the wall-clock time per step $T_\text{per step}$ for the adaptive time step. The scaling exponents for $T$ are shown in Table \ref{tab-5-scaling-N_2_time-adaptimestep}.

First, we detail the general trend compared with the figures of the constant time step (topmost panels of Fig. \ref{fig-5-N-consttimestep}). Compared with the constant time step, the adaptive time step accelerates the calculation by a factor of tens or even hundreds. The intuitive trend of $T$ (upper panels) increasing with $N$ is the same as with a constant time step, but the details are hard to interpret because the number of steps is adaptively set. The behavior of $T_\text{per step}$ (lower panels) is mostly natural, resembling the ones in the constant time step (topmost panels of Fig. \ref{fig-5-N-consttimestep}), except the one-component constant kernel case. In the one-component constant kernel case, the scatter plot of the adaptive time step does not lie on a line and is not monotonically increasing with $N$. In the two-component additive kernel case, the scatter plot for the total wall-clock time $T$ seems to behave like a quadratic curve. The other two kernels are more natural, with data points on top of the fitted lines in both scatter plots. 

The choice of the constant time step or the adaptive time step has almost no effect on the accuracy (i.e., the shape of the distribution, $\varepsilon_2$, and $\Delta M$), so these figures are omitted for space.

\subsection{Effects of $\theta_c$ on speed and accuracy} \label{sec-5-theta_c}

\begin{figure*}
  \centering
  \includegraphics[width=17cm]{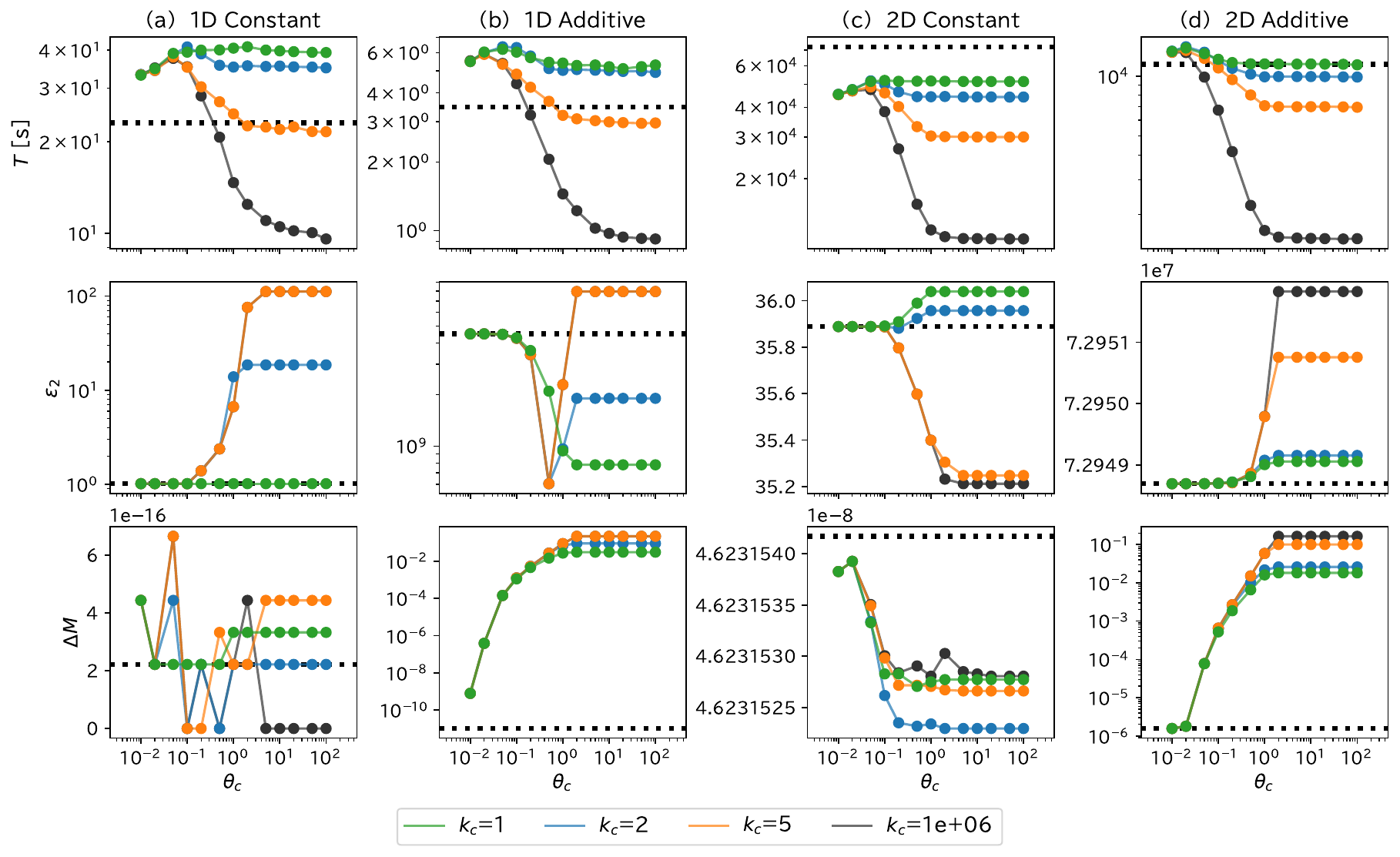}
  \caption{The effect of the critical opening angle $\theta_c$ on the wall-clock time $T$ (the top panels), the $L_2$ error $\varepsilon_2$ (the middle panels), and the relative error of the total mass $\Delta M$ (the bottom panels) for the constant time step. From the left, the panels show the results for the (a) one-component constant kernel, (b) one-component additive kernel, (c)  two-component constant kernel, and (d) two-component additive kernel. Since the critical opening angle $\theta_c$ only affects the tree method, the results for the direct method are shown using a black dotted horizontal line. In each plot, different colors correspond to the different values of $k_c$ for the tree methods. All methods are calculated with $N=241$.}
  \label{fig-5-theta_c}
\end{figure*}

Figure \ref{fig-5-theta_c} shows the effect of the critical opening angle $\theta_c$ for the constant time step. The following sub-sub-sections discuss the figure from top to bottom: $\theta_c$ vs. $T$, $\theta_c$ vs. $\varepsilon_2$, and $\theta_c$ vs. $\Delta M$. 

\subsubsection{$\theta_c$ vs. $T$}

The topmost panels of Figure \ref{fig-5-theta_c} show the wall-clock time $T$ dependence on the critical opening angle $\theta_c$.

Overall, from small $\theta_c$ to large $\theta_c$, the time for the tree algorithm increases, then at $\theta_c \approx 0.05$, it reaches a maximum (i.e., the computation speed hits a minimum), and finally it decreases and converges to a finite value. The wall-clock time at a small $\theta_c = 0.01$ does not depend on $k_c$. The time maximum at $\theta_c \approx 0.05$ is larger for the one-component than the two-component and also for a smaller $k_c$ than a larger $k_c$. Finally, the time at a large $\theta_c$ is almost a constant at $\theta_c \gtrsim 5$, and it decreases as $k_c$ increases. For the one-component, the tree method is faster than the direct method only when both criteria are sufficiently large, i.e., $\theta_c \gtrsim 0.5$ and $k_c = 1000000$. For the two-component, the tree method is faster than the direct method across most of the explored parameters.

\subsubsection{$\theta_c$ vs. $\varepsilon_2$}

The middle-row panels of Figure \ref{fig-5-theta_c} show the $L_2$ error $\varepsilon_2$ dependence on the critical opening angle $\theta_c$.

First, we explain the one-component constant kernel. There is a clear trend that a large $\theta_c$ makes the tree algorithm less accurate, which is consistent with the theoretical prediction. First, at $\theta_c \lesssim 0.5$, which groups bins finely, the tree method maintains an error equivalent to that of the direct method. Then at $\theta_c \approx 1$, the error increases sharply. Finally, at $\theta_c \gtrsim 5$, which groups bins coarsely, the error becomes almost constant. The increase of the error at $\theta_c \approx 1$ is larger for a large $k_c$, and a small $k_c=1$ removes the increase, making it constant for all $\theta_c$.

Next, for the one-component additive kernel, there is an additional error minimum at $\theta_c=0.5$. This minimum makes it more accurate than the direct method at the point. For a large $k_c$, the error sharply increases at $\theta_c \approx 1$, making it less accurate than the direct method, but for a small $k_c = 1$, it maintains its low error. As a result, the $L_2$ error becomes smaller as $\theta_c$ becomes larger for the additive kernel with $k_c = 1$, which is against the theoretical estimate.

Finally, for the two-component cases, the $L_2$ error stays nearly the same for any value of $\theta_c$, paying attention to the $y$-scale. This means that the critical opening angle $\theta_c$ has a small effect compared with the number of bins $N$. This is consistent with the results for the $N$ vs. $\varepsilon_2$ graph (The middle-row panels of Fig. \ref{fig-5-N-consttimestep}). The error of the tree method converges to the value of the direct method at a small $\theta_c$. In the two-component constant kernel case, the tree method with large $\theta_c$ and $k_c$ yields a smaller error than the direct method, which is not straightforward to explain.

\subsubsection{$\theta_c$ vs. $\Delta M$}

The bottommost panels of Figure \ref{fig-5-theta_c} show the relative error of the total mass $\Delta M$ dependence on the critical opening angle $\theta_c$.

As explained in Section \ref{sec-5-N_2_relerr}, all schemes maintain the machine epsilon level of the error for the constant kernel cases, and the direct method maintains it for all cases. For the additive kernel cases, similar to the $L_2$ error $\varepsilon_2$, the $\Delta M$ error increases as $\theta_c$ increases. The increase only happens at $\theta_c \lesssim 1$, and is constant at $\theta_c \gtrsim 1$. A large $k_c$ increases the error, but its effect is small.

For the two-component, the $\Delta M$ error is larger than in the one-component, with about several percent to 30 percent. 

\subsection{Effects of $k_c$ on speed and accuracy} \label{sec-5-k_c}

\subsubsection{$k_c$ vs. $T$}

From the topmost panels of Figure \ref{fig-5-theta_c}, a larger value of $k_c$, which coarsely treats bins, decreases the time for all cases. This is consistent with the theoretical prediction. The decrease only occurs for $\theta_c \gtrsim 0.1$, where the bin grouping takes effect.  

\subsubsection{$k_c$ vs. $\varepsilon_2$}

From the middle-row panels of Figure \ref{fig-5-theta_c}, a larger value of $k_c$ increases the $L_2$ error $\varepsilon_2$ for the one-component case at a large $\theta_c$. This is consistent with the theoretical prediction. For the two-component case, the effect of $k_c$ is subtle.

\subsubsection{$k_c$ vs. $\Delta M$}

For the constant kernel cases, since all schemes conserve the total mass, $k_c$ does not affect this error. For the additive kernel cases, a larger value of $k_c$ increases the $\Delta M$ error (bottommost panels of Fig. \ref{fig-5-theta_c}). This is consistent with the theoretical prediction.

\subsection{Trade-off between speed and accuracy} \label{sec-5-tradeoff}

\begin{figure*}
  \centering
  \includegraphics[width=17cm]{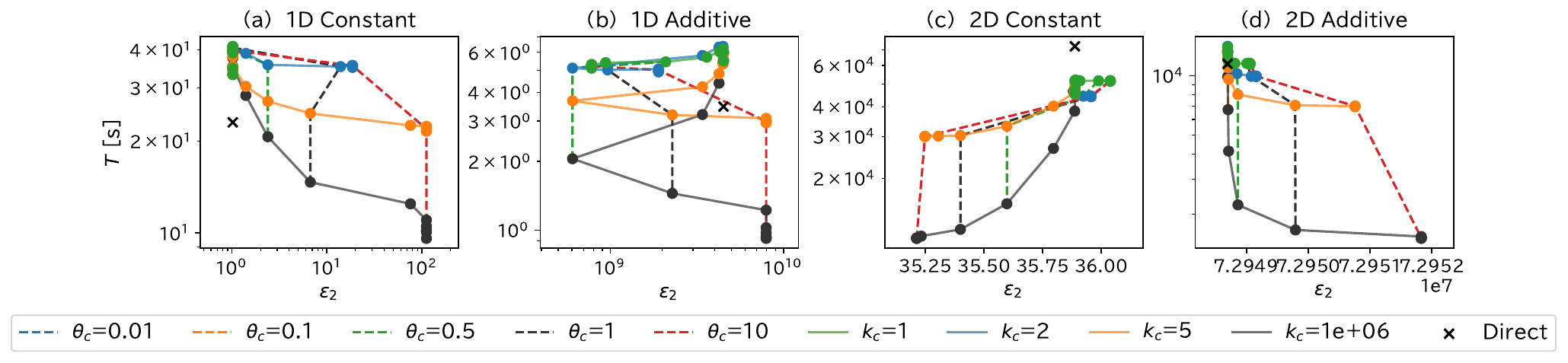}
  \caption{$L_2$ error $\varepsilon_2$ versus wall-clock time $T$ with $N=241$. Each point in a line corresponds to different values of $\theta_c$ and $k_c$. From the left, the panels show the results for the (a) one-component constant kernel, (b) one-component additive kernel, (c)  two-component constant kernel, and (d) two-component additive kernel. The parameter can be considered good if the point is located lower (faster) and further to the left (accurate).}
  \label{fig-5-ssd_2_time-theta_ck_c-N-241}
\end{figure*}

\begin{figure*}
  \centering
  \includegraphics[width=17cm]{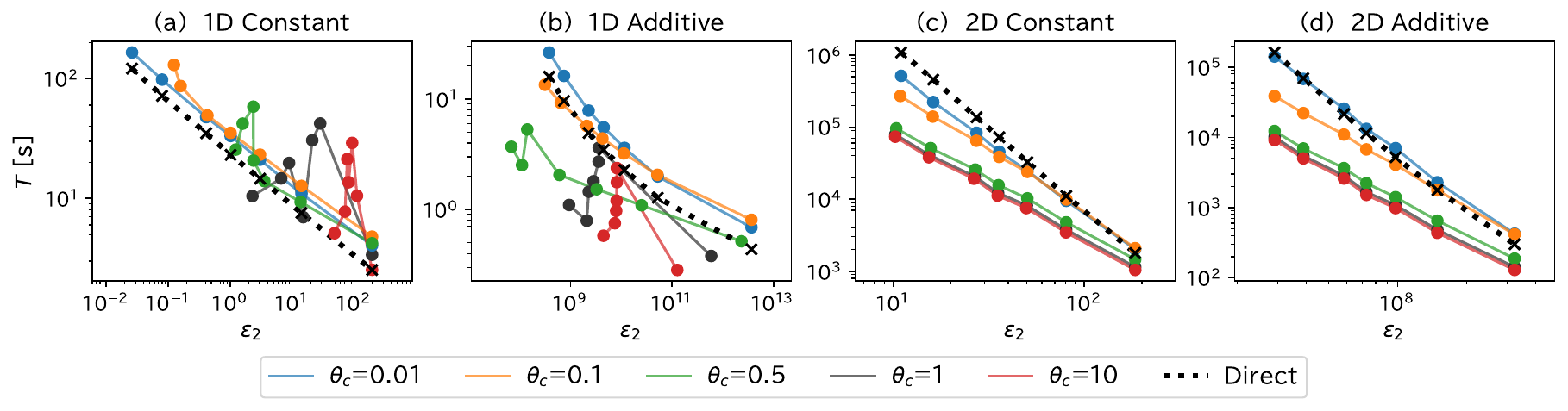}
  \caption{Same as Fig. \ref{fig-5-ssd_2_time-theta_ck_c-N-241}, but for a constant $k_c=1000000$. Each point in a line corresponds to different values of $N$, where the most time-consuming and most accurate one (top-left) is $N=40 \times (14 + 1) + 1=601$ and the fastest and least accurate one (bottom-right) is $N=4 \times (14 + 1) + 1 = 61$.}
  \label{fig-5-ssd_2_time-theta_c-k_c-1000000}
\end{figure*}

Figure \ref{fig-5-ssd_2_time-theta_ck_c-N-241} shows the trade-off between the wall-clock time $T$ and the $L_2$ error $\varepsilon_2$, with constant $N$. A point in the bottom-left part of the figure represents an algorithm parameter that is both time-efficient and accurate; a point in the top-right part opposite. The tree algorithm becomes slower and more accurate by making $\theta_c$ and $k_c$ smaller. We can see that for the one-component, the tree algorithm is faster than the direct method for large values of $\theta_c$ and $k_c$. For the two-component, the tree algorithm outperforms the direct algorithm at most of the parameter values. The tree algorithm with a larger $k_c$ can be faster and more accurate than those with a smaller $k_c$.

Figure \ref{fig-5-ssd_2_time-theta_c-k_c-1000000} shows the trade-off between the wall-clock time $T$ and the $L_2$ error $\varepsilon_2$, with constant $k_c$. Especially for the two-component cases, both algorithms have a scaling law where as $N$ increases, the time increases and the error decreases. To achieve a desirable error (i.e., to make the point go left of a vertical line in the figure), sufficiently large $N$ and $\theta_c$ are required. The implications for the physical applications are discussed in Section \ref{sec-6-1-interpretation}.

\section{Discussion}\label{sec-6-discussion}

\subsection{Interpretation of results} \label{sec-6-1-interpretation}

The results suggest that our tree method is suitable for the SCE with two or more components. Through parameter search of the algorithm, we find that in the one-component case, the tree algorithm is faster than the direct algorithm for a specific range of parameters. Furthermore, in the two-component case, the tree algorithm outperforms the direct algorithm in all parameter regions surveyed.

Out of these regions, a sufficiently large value of $k_c = 1000000$ performed better in both time and accuracy than those with smaller values of $k_c$ (Fig. \ref{fig-5-ssd_2_time-theta_ck_c-N-241}). This means that the condition (2) ($k_c$) for the bin grouping, explained in Section \ref{sec-3-3-treeSCEdetail}, is unnecessary and can be removed.

Similarly, a large value of $\theta_c$ performed better in both time and accuracy compared with those with a smaller value of $\theta_c$, especially in the two-component cases. As we discuss later, this is because in the two-component, the numerical diffusion coming from the insufficient number of bins $N$ is dominant, and the effect of the bin grouping is obscured. This means that the tree method is especially effective in the multi-component coagulation.

We can decide which values of parameters $N$ and $\theta_c$ to use by balancing speed and accuracy based on performance evaluations. A comprehensive study including comparison with the Monte Carlo method by \citet{drazkowskaModelingDustGrowth2014} suggested $N_\mathrm{bd} = 20$ to $N_\mathrm{bd} = 40$, which corresponds to $N = 301$ to $N = 601$, to achieve sufficient accuracy. Comparing this with Fig. \ref{fig-5-ssd_2_time-theta_c-k_c-1000000}, we can infer that the best tree algorithm parameters are $N=601$ and a small value of $\theta_c$, such as $\theta_c = 0.1$ for the one-component. Although the bin grouping in the tree algorithm increases the error, it has the advantage of enabling the SCE calculation with two or more components by speeding it up.

One major problem found in the performance evaluation is the high numerical diffusion in the two-component coagulation. This diffusion is common to both the direct and tree method, as seen in exemplary figures (Fig. \ref{fig-res-2D-simpleex}). This caused a slow convergence in the $N$ vs. $\varepsilon_2$ figure (middle-row panels in Fig. \ref{fig-5-N-consttimestep}). We have concluded that, to calculate the multi-component dust coagulation in protoplanetary disks accurately, higher-order schemes are necessary. A promising candidate for a higher-order scheme is to change the Podolak algorithm that distributes the mass transfer between the two bins, into the extended cell-averaging technique (ECAT) \citep{kostoglouExtendedCellAverage2007, chaudhuryExtendedCellaverageTechnique2013}. This is essentially a higher-order Podolak algorithm that additionally considers the second-order moment of a distribution. However, the current ECAT does not consider the mass-transfer exceeding 15 digits, a machine-epsilon of the double-precision float. Such a modification is necessary to make it applicable to planet formation.

To make the tree method for the multi-component coagulation valid to use for planet formation, further investigations, including comparisons with Monte Carlo methods and performance evaluations with physical kernels, are necessary.

The $L_2$ error $\varepsilon_2$ of the tree method shows behaviors that are hard to understand. One of them is the response to the number of bins $N$ in the one-component case (middle-row panels of Fig. \ref{fig-5-N-consttimestep}). First, for the constant kernel, the direct method and the tree method with a small $\theta_c$ converge to zero as $N$ increases. The tree method with a large $\theta_c$ hits an error minimum at some $N$, and stays constant from there. For the additive kernel, the direct method and the tree method with a small $\theta_c$ behave the same as the constant kernel, but the tree method with a large $\theta_c$ has an almost-constant error for all $N$'s. Another strange behavior is the error's response to the opening angle $\theta_c$ for the additive kernel (Fig. \ref{fig-5-theta_c}). There is a local minimum at $\theta_c \approx 0.5$, and here, the error of the tree method is lower than that of the direct method.

The strange behaviors of the $L_2$ error $\varepsilon_2$ can be understood through the balance of two error origins: (1) the overestimation of the maximum dust size caused by the numerical diffusion, and (2) the underestimation of the peak size and the maximum dust size by the tree method when $\theta_c$ is large. Let us compare the left panels (fiducial cases) and the middle-column panels (large $N$ cases) of Fig. \ref{fig-res-1D-simpleex}. The former plot exhibits greater numerical diffusion, or the overestimation of the maximum dust size, than the latter. Additionally, the tree method with a larger $\theta_c$ (left panels of Fig. \ref{fig-res-1D-simpleex}) underestimates the peak and the maximum dust size compared with the direct method or the tree method with a smaller $\theta_c$ (right panels of Fig. \ref{fig-res-1D-simpleex}). The underestimation is prominent in the additive kernel. These two numerical effects can cancel each other out, resulting in the tree method performing better than the direct method coincidentally. The point at which this occurs corresponds to the $\theta_c$, for which the tree method has a better accuracy than the direct method in Fig. \ref{fig-5-theta_c}. This might also be a cause for the tree method with a large $\theta_c$ having an almost-constant error for the additive kernel (Fig. \ref{fig-5-theta_c}). Next, due to the numerical diffusion, the tail of the final dust distribution for the additive kernel exceeds the maximum mass bin calculated $m=N_\mathrm{bd} \times 10^{14}$. This leads to a large $\Delta M$ error at small $N$ (Fig. \ref{fig-5-N-consttimestep}). This also makes the error of the tree method with a large $\theta_c$ smaller than that of the direct method in small $N$ for the additive kernel.

The dependence of the wall-clock time $T$ on the opening angle $\theta_c$ is also hard to understand. The local maximum of time at $\theta_c \approx 0.5$ (topmost panels in Fig. \ref{fig-5-theta_c}) does not appear for the tree method in $N$-body gravity simulations \citep[see][]{hernquistPerformanceCharacteristicsTree1987}. A smaller value of the $k_c$ leads to a larger value at the local maximum, suggesting that the interaction of the grouping criteria $k_c$ and $\theta_c$ is important for interpretation. The overhead of the tree method might have caused this.

Our tree method does not exactly conserve the total mass for a specific range of parameters and kernels (e.g., bottommost panels in Fig. \ref{fig-5-N-consttimestep}). This drawback is analogous to the non-conservation of total momentum in the $N$-body gravity tree algorithm. Both of the non-conservation come from the asymmetry in calculating the pair-wise interactions. In the $N$-body tree algorithm, the gravitational force from the $i_1$-particle to the $j$-particle is grouped with the $i_2$-particle to the $j$-particle if $i_1$ and $i_2$ are close together. However, the gravitational force from the $j$-particle to the $i_1$-particle is calculated independently from the force from the $j$-particle to the $i_2$-particle. This is the asymmetry in calculating the interaction, and it causes the total momentum to be non-conserved. The asymmetry also exists in our tree method for coagulation calculation (Fig. \ref{fig-3-symmetry}), and it analogously causes the total mass to be non-conserved. However, for $N$-body gravity simulations, a tree algorithm that conserves the total momentum has been developed \citep{dehnenVeryFastMomentumconserving2000}. This is achieved by using the Taylor expansion in Cartesian coordinates. Development of a similar variant in the SCE tree algorithm is a future issue.

For the one-component constant kernel, the tree method conserves mass. We suppose that the integration of the coagulation term over the distribution for the constant kernel coincidentally matches exactly the rectangle integration method (Fig. \ref{fig-3-symmetry}). This can probably be rephrased as saying that in the constant kernel, higher-order terms of the Taylor expansion of the coagulation term are negligible.

\subsection{Time integration of the Smoluchowski coagulation equation} \label{sec-6-2-timeintegration}

We used the explicit classical fourth-order Runge-Kutta method for the time integration. The coagulation-fragmentation equation in planet formation is known to be stiff, and implicit integration has been used widely \citep[see][]{brauerCoagulationFragmentationRadial2008, stammlerDustPyPythonPackage2022}. However, the extension from one-component to multi-component enlarges the Jacobian matrix size from $N^2$ to $N^{2d}$, where $N$ is the number of bins per component and $d$ is the number of components. Calculating such a large matrix is practically impossible, lowering the superiority of implicit time integrations for multi-component equations. This Jacobian is also time-varying, which disallows matrix decomposition methods. To overcome these problems, we had to use explicit time integration. However, it actually turned out to be better than the implicit ones. First noted by \citet{rafikovFastON2Fragmentation2020}, although the implicit time integrations speed up the calculation by allowing larger time steps, realistic simulations require a moderately small time step to achieve a sufficiently small error. This drops the widely known advantage of the implicit methods being fast and stable, making them comparable to explicit methods. Therefore, we suggest solving the coagulation-fragmentation equations with explicit time-integration methods, even for the one-component and especially for the multi-component.

We have tested using the constant time step and the adaptive time step in the performance evaluation. The results suggest that using the adaptive one is better, since it is faster while having the same accuracy. Alternative ways of calculating the adaptive time steps, such as Runge--Kutta--Fehlberg method (RKF45) and Cash--Karp method, might perform better, but these are future issues.

There should be a restriction on the maximum time step for the pair-wise method with explicit time integration, since a time step too large causes the calculation to crash. Related time step restrictions, such as the CFL condition for the hyperbolic, conservative form of the SCE with explicit time integration \citep{laibeCourantFriedrichsLewy2022}, have been investigated. However, the CFL condition cannot be directly applied to our formulation because our formulation is not hyperbolic. A derivation of the corresponding time step restriction is an important topic for future work.

\subsection{Variants of tree algorithm for the Smoluchowski coagulation equation} \label{sec-6-3-variants}

Several variations of our tree method can be hinted at. First, we only tested using the simplest numerical integration: the rectangle rule of Newton-Cotes methods. Other sophisticated methods, such as Simpson's rule or power-law interpolations within bins \citep{leeValidityCoagulationEquation2000}, should perform better. This is analogous to implementing the multipole expansion in the tree method for $N$-body gravity simulations \citep[e.g.][]{greengardFastAlgorithmParticle1987, capuzzo-dolcettaComparisonFastMultipole1998, chengFastAdaptiveMultipole1999, dehnenFastMultipoleMethod2014}. Different definitions of the “distance” in dust coagulation, such as using the $L_\infty$ norm instead of the $L_2$ norm, may be used. More subtle optimizations, such as using arrays instead of structs with pointers, or caching values, may perform better.

This work focused on coagulation, but the tree method can, in principle, also be applied to fragmentation. The difference between coagulation and fragmentation lies in the number of bins for the collision outcome. In coagulation, only the bin of mass $m_i + m_j$ increases, whereas in fragmentation, an assumed fragmentation mass distribution is used to increase the bins of outcomes. This additional operation on bins requires extra time, but assuming a self-similarity distribution can decrease it to a reasonable time \citep{rafikovFastON2Fragmentation2020}.

A fast algorithm for coagulation and fragmentation, such as our tree method, benefits many astrophysical studies on dust distribution evolution. Multiple components of dust particles, including porosity, electric charge, chemical composition, and temperature, have been studied recently. These studies often assume empirical and approximate models, such as the mono-disperse models \citep[e.g.][]{michoulierCompactionFragmentationBouncing2024}. These models need to be checked with more accurate coagulation-fragmentation models, fitting the necessary parameters \citep[e.g.][]{pfeilTriPoDTriPopulationSize2024}. Our tree algorithm enables the multi-component distribution-aware dust coagulation calculation with accurate physical models in a spatial-1D (radius) disk, such as DustPy \citep{stammlerDustPyPythonPackage2022}, for the first time. With this algorithm, how multi-component dust aggregates grow in protoplanetary disks can be studied accurately, providing insights into refined planet formation models.

\section{Summary and conclusion}\label{sec-7-summary}

Multi-component dust aggregates play an important role in planet formation. The multi-component coagulation (and fragmentation) calculation is necessary, but its high computational cost is an obstacle. We developed a novel tree algorithm for the multi-component coagulation, inspired by the tree algorithm used in $N$-body gravity simulations. It assumes that coagulation results in a similar outcome if the two pairs of colliding dust aggregates have similar properties, and that the dust property ratio gives the similarity as a “distance”. This assumption enables us to group bins afar accordingly with the opening angle $\theta_c$ and the maximum dust distribution width $k_c$. By grouping the bins, the algorithm can calculate coagulation faster than the conventional direct method. We tested this by comparing it with the analytic solutions and the direct method. We assessed the dependence of the wall-clock time $T$, the $L_2$ error in the distribution function $\varepsilon_2$, and the relative error of the total mass conservation $\Delta M$, on the number of components $d=1, 2$, kernels $R$, number of bins $N$, critical opening angle $\theta_c$, maximum dust distribution width after a coagulation $k_c$, and time integration method. Our conclusions are as follows:
\begin{itemize}
    \item For the one-component, our tree method is faster than the direct method when the critical opening angle $\theta_c$ is sufficiently large. For the two-component, our tree method is systematically faster than the direct method across all parameter regions surveyed. For example, the tree method was 5.9 times faster than the direct method for the fiducial parameters, and 13.7 times faster for the larger $N=601$ case.
    \item The primary numerical parameter of the tree method is the critical opening angle $\theta_c$, which controls the trade-off between computational cost and accuracy effectively. A larger $\theta_c$ yields faster computation but larger errors. $k_c$ is found to be not important, and removing it makes the calculation faster while maintaining the same accuracy.
    \item The tree method reproduces the overall behavior well, with the main deviations arising from (1) an overestimation of the maximum dust size caused by the numerical diffusion, which is also present in the direct method, and (2) a slight underestimation of the peak and the maximum dust size due to bin grouping. The first error can be reduced by making $N$ larger, and the second error can be reduced by making $\theta_c$ smaller.
    \item For a specific range of parameters and kernels, the tree method does not exactly conserve the total mass. However, this deviation can be reduced by adopting a smaller critical opening angle $\theta_c$.
\end{itemize}
Altogether, our tree method is a first step toward a complete coagulation calculation for multi-component dust aggregates. In future work, we plan to apply this to the time evolution of the dust porosity distribution in protoplanetary disks, incorporating various phenomena. Developing a fully mass-conserving, higher-order formulation of the tree method is an important subject for future work. We also noted a possible variant of the tree method for fragmentation schemes applicable to planet formation.

\section*{Code availability}

The code is available upon reasonable request.

\begin{acknowledgements}
    The authors would like to deeply appreciate the anonymous referee for comments, which greatly helped in improving the manuscript. The authors would also like to appreciate Shota Notsu, Kazunari Iwasaki, and Yuki Kambara for fruitful discussions. The authors would also like to appreciate Sota Arakawa, Shubham Bhardwaj, Ryotaro Chiba, Uma Gorti, Shu-ichiro Inutsuka, Yuichiro Ishida, Hiroshi Kobayashi, Eiichiro Kokubo, Naoto Maki, Hiroki Nagakura, Hideko Nomura, Keiji Ohtsuki, Satoshi Okuzumi, Chen Peng-Fei, Shota Sato, Sin-iti Sirono, Hidekazu Tanaka, Kei Tanaka, and Yuki Yoshida for valuable comments. T.K.W. acknowledges support through the NAOJ Junior Fellow Program. This work was supported by JSPS KAKENHI Grant Number JP22K03680. This work was partially supported by Overseas Travel Fund for Students (2025) of Astronomical Science Program, The Graduate University for Advanced Studies, SOKENDAI. The authors acknowledge Grammarly's help in improving only the writing.

    \textit{Software}: \texttt{GNU Parallel} \citep{tangeGNUParallel202602222026}, \texttt{NumPy} \citep{harris2020array}, \texttt{Matplotlib} \citep{Hunter:2007}, \texttt{Pandas} \citep{reback2020pandas}, and \texttt{SciPy} \citep{virtanenSciPy10Fundamental2020}.
\end{acknowledgements}

\bibliographystyle{aa}
\bibliography{2025-tree}

@article{stammlerDustPyPythonPackage2022,
  title = {{{DustPy}}: {{A Python Package}} for {{Dust Evolution}} in {{Protoplanetary Disks}}},
  shorttitle = {{{DustPy}}},
  author = {Stammler, Sebastian Markus and Birnstiel, Tilman},
  year = 2022,
  month = aug,
  journal = {ApJ},
  volume = {935},
  number = {1},
  eprint = {2207.00322},
  primaryclass = {astro-ph},
  pages = {35},
  issn = {0004-637X, 1538-4357},
  doi = {10.3847/1538-4357/ac7d58},
  url = {http://arxiv.org/abs/2207.00322},
  urldate = {2023-07-12},
  archiveprefix = {arXiv}
}

@article{okuzumiRapidCoagulationPorous2012,
  title = {Rapid Coagulation of Porous Dust Aggregates Outside the Snow Line: {{A}} Pathway to Successful Icy Planetesimal Formation},
  shorttitle = {Rapid Coagulation of Porous Dust Aggregates Outside the Snow Line},
  author = {Okuzumi, Satoshi and Tanaka, Hidekazu and Kobayashi, Hiroshi and Wada, Koji},
  year = 2012,
  journal = {ApJ},
  volume = {752},
  number = {2},
  pages = {106},
  publisher = {IOP Publishing},
  doi = {10.1088/0004-637X/752/2/106},
  url = {https://iopscience.iop.org/article/10.1088/0004-637X/752/2/106/meta}
}

@article{kataokaFluffyDustForms2013,
  title = {Fluffy Dust Forms Icy Planetesimals by Static Compression},
  author = {Kataoka, Akimasa and Tanaka, Hidekazu and Okuzumi, Satoshi and Wada, Koji},
  year = 2013,
  month = sep,
  journal = {A\&A},
  volume = {557},
  pages = {L4},
  publisher = {EDP Sciences},
  issn = {0004-6361, 1432-0746},
  doi = {10.1051/0004-6361/201322151},
  url = {https://www.aanda.org/articles/aa/abs/2013/09/aa22151-13/aa22151-13.html},
  urldate = {2023-09-08},
  copyright = {\copyright{} ESO, 2013},
  langid = {english}
}

@article{adachiGasDragEffect1976,
  title = {The {{Gas Drag Effect}} on the {{Elliptic Motion}} of a {{Solid Body}} in the {{Primordial Solar Nebula}}},
  author = {Adachi, Isao and Hayashi, Chushiro and Nakazawa, Kiyoshi},
  year = 1976,
  month = dec,
  journal = {Prog. Theor. Phys.},
  volume = {56},
  number = {6},
  pages = {1756--1771},
  issn = {0033-068X},
  doi = {10.1143/PTP.56.1756},
  url = {https://doi.org/10.1143/PTP.56.1756},
  urldate = {2023-09-08}
}

@article{weidenschillingAerodynamicsSolidBodies1977,
  title = {Aerodynamics of Solid Bodies in the Solar Nebula},
  author = {Weidenschilling, S. J.},
  year = 1977,
  month = sep,
  journal = {MNRAS},
  volume = {180},
  number = {2},
  pages = {57--70},
  issn = {0035-8711},
  doi = {10.1093/mnras/180.2.57},
  url = {https://doi.org/10.1093/mnras/180.2.57},
  urldate = {2023-09-08}
}

@article{brauerCoagulationFragmentationRadial2008,
  title = {Coagulation, Fragmentation and Radial Motion of Solid Particles in Protoplanetary Disks},
  author = {Brauer, F. and Dullemond, C. P. and Henning, T.},
  year = 2008,
  month = mar,
  journal = {A\&A},
  volume = {480},
  pages = {859--877},
  issn = {0004-6361},
  doi = {10.1051/0004-6361:20077759},
  url = {https://ui.adsabs.harvard.edu/abs/2008A&A...480..859B},
  urldate = {2023-09-08}
}

@article{leeValidityCoagulationEquation2000,
  title = {On the {{Validity}} of the {{Coagulation Equation}} and the {{Nature}} of {{Runaway Growth}}},
  author = {Lee, Man Hoi},
  year = 2000,
  month = jan,
  journal = {Icarus},
  volume = {143},
  pages = {74--86},
  issn = {0019-1035},
  doi = {10.1006/icar.1999.6239},
  url = {https://ui.adsabs.harvard.edu/abs/2000Icar..143...74L},
  urldate = {2023-09-08}
}

@article{birnstielGasDustEvolution2010,
  title = {Gas- and Dust Evolution in Protoplanetary Disks},
  author = {Birnstiel, T. and Dullemond, C. P. and Brauer, F.},
  year = 2010,
  month = apr,
  journal = {A\&A},
  volume = {513},
  pages = {A79},
  publisher = {EDP Sciences},
  issn = {0004-6361, 1432-0746},
  doi = {10.1051/0004-6361/200913731},
  url = {https://www.aanda.org/articles/aa/abs/2010/05/aa13731-09/aa13731-09.html},
  urldate = {2023-09-08},
  copyright = {\copyright{} ESO, 2010},
  langid = {english}
}

@article{okuzumiNumericalModelingCoagulation2009,
  title = {Numerical {{Modeling}} of the {{Coagulation}} and {{Porosity Evolution}} of {{Dust Aggregates}}},
  author = {Okuzumi, Satoshi and Tanaka, Hidekazu and Sakagami, Masa-aki},
  year = 2009,
  month = feb,
  journal = {ApJ},
  volume = {707},
  number = {2},
  pages = {1247},
  publisher = {The American Astronomical Society},
  issn = {0004-637X},
  doi = {10.1088/0004-637X/707/2/1247},
  url = {https://dx.doi.org/10.1088/0004-637X/707/2/1247},
  urldate = {2023-09-09},
  langid = {english}
}

@article{rafikovFastON2Fragmentation2020,
  title = {A {{Fast O}}({{N2}}) {{Fragmentation Algorithm}}},
  author = {Rafikov, Roman R. and Silsbee, Kedron and Booth, Richard A.},
  year = 2020,
  month = apr,
  journal = {ApJS},
  volume = {247},
  number = {2},
  pages = {65},
  publisher = {The American Astronomical Society},
  issn = {0067-0049},
  doi = {10.3847/1538-4365/ab7b71},
  url = {https://dx.doi.org/10.3847/1538-4365/ab7b71},
  urldate = {2023-09-09},
  langid = {english}
}

@article{ossenkopfDustCoagulationDense1993,
  title = {Dust Coagulation in Dense Molecular Clouds : The Formation of Fluffy Aggregates.},
  shorttitle = {Dust Coagulation in Dense Molecular Clouds},
  author = {Ossenkopf, V.},
  year = 1993,
  month = dec,
  journal = {A\&A},
  volume = {280},
  pages = {617--646},
  issn = {0004-6361},
  url = {https://ui.adsabs.harvard.edu/abs/1993A&A...280..617O},
  urldate = {2023-09-11}
}

@article{ormelDustCoagulationProtoplanetary2007,
  title = {Dust Coagulation in Protoplanetary Disks: Porosity Matters},
  shorttitle = {Dust Coagulation in Protoplanetary Disks},
  author = {Ormel, C. W. and Spaans, M. and Tielens, A. G. G. M.},
  year = 2007,
  month = jan,
  journal = {A\&A},
  volume = {461},
  pages = {215--232},
  issn = {0004-6361},
  doi = {10.1051/0004-6361:20065949},
  url = {https://ui.adsabs.harvard.edu/abs/2007A&A...461..215O},
  urldate = {2023-09-11}
}

@article{hayashiStructureSolarNebula1981,
  title = {Structure of the {{Solar Nebula}}, {{Growth}} and {{Decay}} of {{Magnetic Fields}} and {{Effects}} of {{Magnetic}} and {{Turbulent Viscosities}} on the {{Nebula}}},
  author = {Hayashi, Chushiro},
  year = 1981,
  month = jan,
  journal = {Prog. Theor. Phys. Suppl.},
  volume = {70},
  pages = {35--53},
  issn = {0375-9687},
  doi = {10.1143/PTPS.70.35},
  url = {https://doi.org/10.1143/PTPS.70.35},
  urldate = {2023-09-13}
}

@article{cieslaEvolutionWaterDistribution2006,
  title = {The Evolution of the Water Distribution in a Viscous Protoplanetary Disk},
  author = {Ciesla, Fred J. and Cuzzi, Jeffrey N.},
  year = 2006,
  month = mar,
  journal = {Icarus},
  volume = {181},
  number = {1},
  pages = {178--204},
  issn = {0019-1035},
  doi = {10.1016/j.icarus.2005.11.009},
  url = {https://www.sciencedirect.com/science/article/pii/S0019103505004641},
  urldate = {2024-04-30}
}

@book{safronovEvolutionProtoplanetaryCloud1972,
  title = {Evolution of the Protoplanetary Cloud and Formation of the Earth and Planets.},
  author = {Safronov, V. S.},
  year = 1972,
  month = jan,
  journal = {Evolution of the protoplanetary cloud and formation of the earth and planets},
  publisher = {Israel Program for Scientific Translations, Keter Publishing House},
  url = {https://ui.adsabs.harvard.edu/abs/1972epcf.book.....S},
  urldate = {2024-07-31}
}

@article{johansenRapidPlanetesimalFormation2007,
  title = {Rapid Planetesimal Formation in Turbulent Circumstellar Disks},
  author = {Johansen, Anders and Oishi, Jeffrey S. and Low, Mordecai-Mark Mac and Klahr, Hubert and Henning, Thomas and Youdin, Andrew},
  year = 2007,
  month = aug,
  journal = {Nat},
  volume = {448},
  number = {7157},
  pages = {1022--1025},
  publisher = {Nature Publishing Group},
  issn = {1476-4687},
  doi = {10.1038/nature06086},
  url = {https://www.nature.com/articles/nature06086},
  urldate = {2024-07-31},
  copyright = {2007 Springer Nature Limited},
  langid = {english}
}

@article{barnesHierarchicalLogForcecalculation1986,
  title = {A Hierarchical {{O}}({{N}} Log {{N}}) Force-Calculation Algorithm},
  author = {Barnes, Josh and Hut, Piet},
  year = 1986,
  month = dec,
  journal = {Nat},
  volume = {324},
  number = {6096},
  pages = {446--449},
  publisher = {Nature Publishing Group},
  issn = {1476-4687},
  doi = {10.1038/324446a0},
  url = {https://www.nature.com/articles/324446a0},
  urldate = {2024-08-03},
  copyright = {1986 Springer Nature Limited},
  langid = {english}
}

@article{smoluchowskiDreiVortrageUber1916,
  title = {Drei Vortrage Uber Diffusion, Brownsche Molekularbewegung Und Koagulation von Kolloidteilchen},
  author = {von Smoluchowski, M.},
  year = 1916,
  journal = {Z Phys.},
  volume = {17},
  pages = {585--599},
  url = {https://cir.nii.ac.jp/crid/1573387448927618688},
  urldate = {2024-11-12}
}

@article{kovetzEffectCoalescenceCondensation1969,
  title = {The {{Effect}} of {{Coalescence}} and {{Condensation}} on {{Rain Formation}} in a {{Cloud}} of {{Finite Vertical Extent}}},
  author = {Kovetz, A. and Olund, B.},
  year = 1969,
  month = sep,
  journal = {J. Atmos. Sci.},
  issn = {1520-0469},
  url = {https://journals.ametsoc.org/view/journals/atsc/26/5/1520-0469_1969_026_1060_teocac_2_0_co_2.xml},
  urldate = {2024-11-17},
  chapter = {Journal of the Atmospheric Sciences},
  langid = {english}
}

@article{zsomOutcomeProtoplanetaryDust2010,
  title = {The Outcome of Protoplanetary Dust Growth: Pebbles, Boulders, or Planetesimals? - {{II}}. {{Introducing}} the Bouncing Barrier},
  shorttitle = {The Outcome of Protoplanetary Dust Growth},
  author = {Zsom, A. and Ormel, C. W. and G{\"u}ttler, C. and Blum, J. and Dullemond, C. P.},
  year = 2010,
  month = apr,
  journal = {A\&A},
  volume = {513},
  pages = {A57},
  publisher = {EDP Sciences},
  issn = {0004-6361, 1432-0746},
  doi = {10.1051/0004-6361/200912976},
  url = {https://www.aanda.org/articles/aa/abs/2010/05/aa12976-09/aa12976-09.html},
  urldate = {2025-01-29},
  copyright = {\copyright{} ESO, 2010},
  langid = {english}
}

@article{blumGrowthMechanismsMacroscopic2008,
  title = {The {{Growth Mechanisms}} of {{Macroscopic Bodies}} in {{Protoplanetary Disks}}},
  author = {Blum, J{\"u}rgen and Wurm, Gerhard},
  year = 2008,
  month = sep,
  journal = {ARA\&A},
  volume = {46},
  number = {Volume 46, 2008},
  pages = {21--56},
  publisher = {Annual Reviews},
  issn = {0066-4146, 1545-4282},
  doi = {10.1146/annurev.astro.46.060407.145152},
  url = {https://www.annualreviews.org/content/journals/10.1146/annurev.astro.46.060407.145152},
  urldate = {2025-01-29},
  langid = {english}
}

@article{ziffKineticsPolymerization1980,
  title = {Kinetics of Polymerization},
  author = {Ziff, Robert M.},
  year = 1980,
  month = aug,
  journal = {J. Stat. Phys.},
  volume = {23},
  number = {2},
  pages = {241--263},
  issn = {1572-9613},
  doi = {10.1007/BF01012594},
  url = {https://doi.org/10.1007/BF01012594},
  urldate = {2025-01-29},
  langid = {english}
}

@article{xuFastMonteCarlo2014,
  title = {Fast {{Monte Carlo}} Simulation for Particle Coagulation in Population Balance},
  author = {Xu, Zuwei and Zhao, Haibo and Zheng, Chuguang},
  year = 2014,
  month = aug,
  journal = {J. Aerosol Sci.},
  volume = {74},
  pages = {11--25},
  issn = {0021-8502},
  doi = {10.1016/j.jaerosci.2014.03.006},
  url = {https://www.sciencedirect.com/science/article/pii/S0021850214000469},
  urldate = {2025-01-30}
}

@article{matveevTensorTrainMonte2016,
  title = {Tensor Train versus {{Monte Carlo}} for the Multicomponent {{Smoluchowski}} Coagulation Equation},
  author = {Matveev, Sergey A. and Zheltkov, Dmitry A. and Tyrtyshnikov, Eugene E. and Smirnov, Alexander P.},
  year = 2016,
  month = jul,
  journal = {J. Comput. Phys.},
  volume = {316},
  pages = {164--179},
  issn = {0021-9991},
  doi = {10.1016/j.jcp.2016.04.025},
  url = {https://www.sciencedirect.com/science/article/pii/S0021999116300675},
  urldate = {2025-01-30}
}

@article{smirnovFastAccurateFinitedifference2016,
  title = {Fast and {{Accurate Finite-difference Method Solving Multicomponent Smoluchowski Coagulation Equation}} with {{Source}} and {{Sink Terms}}},
  author = {Smirnov, Alexander P. and Matveev, Sergey A. and Zheltkov, Dmitry A. and Tyrtyshnikov, Euegene E.},
  year = 2016,
  month = jan,
  journal = {Procedia Computer Science},
  series = {International {{Conference}} on {{Computational Science}} 2016, {{ICCS}} 2016, 6-8 {{June}} 2016, {{San Diego}}, {{California}}, {{USA}}},
  volume = {80},
  pages = {2141--2146},
  issn = {1877-0509},
  doi = {10.1016/j.procs.2016.05.533},
  url = {https://www.sciencedirect.com/science/article/pii/S1877050916310249},
  urldate = {2025-01-30}
}

@article{estradaSolvingCoagulationEquation2008,
  title = {Solving the {{Coagulation Equation}} by the {{Moments Method}}},
  author = {Estrada, P. R. and Cuzzi, J. N.},
  year = 2008,
  month = jul,
  journal = {ApJ},
  volume = {682},
  number = {1},
  pages = {515},
  issn = {0004-637X},
  doi = {10.1086/589685},
  url = {https://dx.doi.org/10.1086/589685},
  urldate = {2025-06-30},
  langid = {english}
}

@article{okuzumiELECTROSTATICBARRIERDUST2011,
  title = {{{ELECTROSTATIC BARRIER AGAINST DUST GROWTH IN PROTOPLANETARY DISKS}}. {{I}}. {{CLASSIFYING THE EVOLUTION OF SIZE DISTRIBUTION}}},
  author = {Okuzumi, Satoshi and Tanaka, Hidekazu and Takeuchi, Taku and Sakagami, Masa-aki},
  year = 2011,
  month = mar,
  journal = {ApJ},
  volume = {731},
  number = {2},
  pages = {95},
  publisher = {The American Astronomical Society},
  issn = {0004-637X},
  doi = {10.1088/0004-637X/731/2/95},
  url = {https://dx.doi.org/10.1088/0004-637X/731/2/95},
  urldate = {2025-01-30},
  langid = {english}
}

@article{krijtTransportCOProtoplanetary2018,
  title = {Transport of {{CO}} in {{Protoplanetary Disks}}: {{Consequences}} of {{Pebble Formation}}, {{Settling}}, and {{Radial Drift}}},
  shorttitle = {Transport of {{CO}} in {{Protoplanetary Disks}}},
  author = {Krijt, Sebastiaan and Schwarz, Kamber R. and Bergin, Edwin A. and Ciesla, Fred J.},
  year = 2018,
  month = aug,
  journal = {ApJ},
  volume = {864},
  number = {1},
  pages = {78},
  publisher = {The American Astronomical Society},
  issn = {0004-637X},
  doi = {10.3847/1538-4357/aad69b},
  url = {https://dx.doi.org/10.3847/1538-4357/aad69b},
  urldate = {2025-01-30},
  langid = {english}
}

@article{kobayashiEvolutionaryDescriptionGiant2017,
  title = {Evolutionary Description of Giant Molecular Cloud Mass Functions on Galactic Disks},
  author = {Kobayashi, Masato IN and Inutsuka, Shu-ichiro and Kobayashi, Hiroshi and Hasegawa, Kenji},
  year = 2017,
  journal = {ApJ},
  volume = {836},
  number = {2},
  pages = {175},
  publisher = {IOP Publishing},
  doi = {10.3847/1538-4357/836/2/175},
  url = {https://iopscience.iop.org/article/10.3847/1538-4357/836/2/175/meta},
  urldate = {2025-01-30}
}

@article{satoWaterDeliveryTerrestrial2016,
  title = {On the Water Delivery to Terrestrial Embryos by Ice Pebble Accretion},
  author = {Sato, Takao and Okuzumi, Satoshi and Ida, Shigeru},
  year = 2016,
  journal = {A\&A},
  volume = {589},
  pages = {A15},
  publisher = {EDP Sciences},
  doi = {10.1051/0004-6361/201527069},
  url = {https://www.aanda.org/articles/aa/abs/2016/05/aa27069-15/aa27069-15.html},
  urldate = {2025-01-31}
}

@article{silkStatisticalModelInitial1979,
  title = {A Statistical Model for the Initial Stellar Mass Function},
  author = {Silk, Joseph and Takahashi, Takamasa},
  year = 1979,
  journal = {ApJ},
  volume = {229},
  number = {1},
  pages = {242--256},
  publisher = {IOP},
  issn = {0004-637X},
  doi = {10.1086/156949},
  url = {https://adsabs.harvard.edu/full/1979ApJ...229..242S},
  urldate = {2026-02-21}
}

@article{wetherillComparisonAnalyticalPhysical1990,
  title = {Comparison of Analytical and Physical Modeling of Planetesimal Accumulation},
  author = {Wetherill, George W.},
  year = 1990,
  month = dec,
  journal = {Icarus},
  volume = {88},
  number = {2},
  pages = {336--354},
  issn = {0019-1035},
  doi = {10.1016/0019-1035(90)90086-O},
  url = {https://www.sciencedirect.com/science/article/pii/001910359090086O},
  urldate = {2025-01-31}
}

@article{ohtsukiArtificialAccelerationAccumulation1990,
  title = {Artificial Acceleration in Accumulation Due to Coarse Mass-Coordinate Divisions in Numerical Simulation},
  author = {Ohtsuki, Keiji and Nakagawa, Yoshitsugu and Nakazawa, Kiyoshi},
  year = 1990,
  journal = {Icarus},
  volume = {83},
  number = {1},
  pages = {205--215},
  publisher = {Elsevier},
  doi = {10.1016/0019-1035(90)90015-2},
  url = {https://www.sciencedirect.com/science/article/pii/0019103590900152},
  urldate = {2025-01-31}
}

@article{okuzumiElectricChargingDust2009,
  title = {Electric {{Charging}} of {{Dust Aggregates}} and Its {{Effect}} on {{Dust Coagulation}} in {{Protoplanetary Disks}}},
  author = {Okuzumi, Satoshi},
  year = 2009,
  month = jun,
  journal = {ApJ},
  volume = {698},
  pages = {1122--1135},
  publisher = {IOP},
  issn = {0004-637X},
  doi = {10.1088/0004-637X/698/2/1122},
  url = {https://ui.adsabs.harvard.edu/abs/2009ApJ...698.1122O},
  urldate = {2025-01-31}
}

@article{harris2020array,
  title = {Array Programming with {{NumPy}}},
  author = {Harris, Charles R. and Millman, K. Jarrod and {van der Walt}, St'efan J. and Gommers, Ralf and Virtanen, Pauli and Cournapeau, David and Wieser, Eric and Taylor, Julian and Berg, Sebastian and Smith, Nathaniel J. and Kern, Robert and Picus, Matti and Hoyer, Stephan and {van Kerkwijk}, Marten H. and Brett, Matthew and Haldane, Allan and {del R'{\i}o}, Jaime Fern'andez and Wiebe, Mark and Peterson, Pearu and {G'erard-Marchant}, Pierre and Sheppard, Kevin and Reddy, Tyler and Weckesser, Warren and Abbasi, Hameer and Gohlke, Christoph and Oliphant, Travis E.},
  year = 2020,
  month = sep,
  journal = {Nat},
  volume = {585},
  number = {7825},
  pages = {357--362},
  publisher = {{Springer Science and Business Media LLC}},
  doi = {10.1038/s41586-020-2649-2},
  url = {https://doi.org/10.1038/s41586-020-2649-2}
}

@article{Hunter:2007,
  title = {Matplotlib: {{A 2D}} Graphics Environment},
  author = {Hunter, J. D.},
  year = 2007,
  journal = {CiSE},
  volume = {9},
  number = {3},
  pages = {90--95},
  publisher = {IEEE COMPUTER SOC},
  doi = {10.1109/MCSE.2007.55}
}

@article{drazkowskaModelingDustGrowth2014,
  title = {Modeling Dust Growth in Protoplanetary Disks: {{The}} Breakthrough Case},
  shorttitle = {Modeling Dust Growth in Protoplanetary Disks},
  author = {Dr{\k a}{\.z}kowska, J. and Windmark, F. and Dullemond, C. P.},
  year = 2014,
  month = jul,
  journal = {A\&A},
  volume = {567},
  pages = {A38},
  issn = {0004-6361},
  doi = {10.1051/0004-6361/201423708},
  url = {https://ui.adsabs.harvard.edu/abs/2014A&A...567A..38D},
  urldate = {2025-02-02}
}

@article{teiserGrowthSuperlargePreplanetary2025,
  title = {The Growth of Super-Large Pre-Planetary Pebbles to an Impact Erosion Limit},
  author = {Teiser, J. and Penner, J. and Joeris, K. and Onyeagusi, F. C. and Kollmer, J. E. and Daab, D. and Wurm, G.},
  year = 2025,
  month = jan,
  journal = {Nat Astron},
  pages = {1--6},
  publisher = {Nature Publishing Group},
  issn = {2397-3366},
  doi = {10.1038/s41550-024-02470-x},
  url = {https://www.nature.com/articles/s41550-024-02470-x},
  urldate = {2025-02-17},
  copyright = {2025 The Author(s)},
  langid = {english}
}

@article{steinpilzElectricalChargingOvercomes2020,
  title = {Electrical Charging Overcomes the Bouncing Barrier in Planet Formation},
  author = {Steinpilz, Tobias and Joeris, Kolja and Jungmann, Felix and Wolf, Dietrich and Brendel, Lothar and Teiser, Jens and Shinbrot, Troy and Wurm, Gerhard},
  year = 2020,
  month = feb,
  journal = {Nat. Phys.},
  volume = {16},
  number = {2},
  pages = {225--229},
  publisher = {Nature Publishing Group},
  issn = {1745-2481},
  doi = {10.1038/s41567-019-0728-9},
  url = {https://www.nature.com/articles/s41567-019-0728-9},
  urldate = {2025-02-17},
  copyright = {2019 The Author(s), under exclusive licence to Springer Nature Limited},
  langid = {english}
}

@article{dehnenVeryFastMomentumconserving2000,
  title = {A {{Very Fast}} and {{Momentum-conserving Tree Code}}},
  author = {Dehnen, Walter},
  year = 2000,
  month = jun,
  journal = {ApJ},
  volume = {536},
  pages = {L39-L42},
  publisher = {IOP},
  issn = {0004-637X},
  doi = {10.1086/312724},
  url = {https://ui.adsabs.harvard.edu/abs/2000ApJ...536L..39D},
  urldate = {2025-02-19}
}

@article{lombartGrainGrowthAstrophysics2021,
  title = {Grain Growth for Astrophysics with Discontinuous {{Galerkin}} Schemes},
  author = {Lombart, Maxime and Laibe, Guillaume},
  year = 2021,
  month = mar,
  journal = {MNRAS},
  volume = {501},
  number = {3},
  pages = {4298--4316},
  issn = {0035-8711},
  doi = {10.1093/mnras/staa3682},
  url = {https://doi.org/10.1093/mnras/staa3682},
  urldate = {2025-02-26}
}

@article{tanakaSteadyStateSizeDistribution1996,
  title = {Steady-{{State Size Distribution}} for the {{Self-Similar Collision Cascade}}},
  author = {Tanaka, Hidekazu and Inaba, Satoshi and Nakazawa, Kiyoshi},
  year = 1996,
  month = oct,
  journal = {Icarus},
  volume = {123},
  pages = {450--455},
  issn = {0019-1035},
  doi = {10.1006/icar.1996.0170},
  url = {https://ui.adsabs.harvard.edu/abs/1996Icar..123..450T},
  urldate = {2025-02-26}
}

@article{lombartFragmentationDiscontinuousGalerkin2022,
  title = {Fragmentation with Discontinuous {{Galerkin}} Schemes: Non-Linear Fragmentation},
  shorttitle = {Fragmentation with Discontinuous {{Galerkin}} Schemes},
  author = {Lombart, Maxime and Hutchison, Mark and Lee, Yueh-Ning},
  year = 2022,
  month = dec,
  journal = {MNRAS},
  volume = {517},
  number = {2},
  pages = {2012--2027},
  issn = {0035-8711},
  doi = {10.1093/mnras/stac2232},
  url = {https://doi.org/10.1093/mnras/stac2232},
  urldate = {2025-02-26}
}

@article{lombartGeneralNonlinearFragmentation2024,
  title = {General Non-Linear Fragmentation with Discontinuous {{Galerkin}} Methods},
  author = {Lombart, Maxime and Br{\'e}hier, Charles-Edouard and Hutchison, Mark and Lee, Yueh-Ning},
  year = 2024,
  month = oct,
  journal = {MNRAS},
  volume = {533},
  number = {4},
  pages = {4410--4434},
  issn = {0035-8711},
  doi = {10.1093/mnras/stae2039},
  url = {https://doi.org/10.1093/mnras/stae2039},
  urldate = {2025-02-26}
}

@article{laibeCourantFriedrichsLewy2022,
  title = {On the {{Courant}}--{{Friedrichs}}--{{Lewy}} Condition for Numerical Solvers of the Coagulation Equation},
  author = {Laibe, Guillaume and Lombart, Maxime},
  year = 2022,
  month = mar,
  journal = {MNRAS},
  volume = {510},
  number = {4},
  pages = {5220--5225},
  issn = {0035-8711},
  doi = {10.1093/mnras/stab3499},
  url = {https://doi.org/10.1093/mnras/stab3499},
  urldate = {2025-02-26}
}

@article{liuHighOrderPositivity2019,
  title = {A {{High Order Positivity Preserving DG Method}} for {{Coagulation-Fragmentation Equations}}},
  author = {Liu, Hailiang and Gr{\"o}pler, Robin and Warnecke, Gerald},
  year = 2019,
  month = jan,
  journal = {SIAM J. Sci. Comput.},
  volume = {41},
  number = {3},
  pages = {B448-B465},
  publisher = {{Society for Industrial and Applied Mathematics}},
  issn = {1064-8275},
  doi = {10.1137/17M1150360},
  url = {https://epubs.siam.org/doi/10.1137/17M1150360},
  urldate = {2025-02-28}
}

@article{weiGPUacceleratedMonteCarlo2013,
  title = {{{GPU-accelerated Monte Carlo}} Simulation of Particle Coagulation Based on the Inverse Method},
  author = {Wei, J. and Kruis, F. E.},
  year = 2013,
  month = sep,
  journal = {J. Comput. Phys.},
  volume = {249},
  pages = {67--79},
  issn = {0021-9991},
  doi = {10.1016/j.jcp.2013.04.030},
  url = {https://www.sciencedirect.com/science/article/pii/S0021999113002957},
  urldate = {2025-02-28}
}

@article{virtanenSciPy10Fundamental2020,
  title = {{{SciPy}} 1.0: Fundamental Algorithms for Scientific Computing in {{Python}}},
  shorttitle = {{{SciPy}} 1.0},
  author = {Virtanen, Pauli and Gommers, Ralf and Oliphant, Travis E. and Haberland, Matt and Reddy, Tyler and Cournapeau, David and Burovski, Evgeni and Peterson, Pearu and Weckesser, Warren and Bright, Jonathan and Van Der Walt, St{\'e}fan J. and Brett, Matthew and Wilson, Joshua and Millman, K. Jarrod and Mayorov, Nikolay and Nelson, Andrew R. J. and Jones, Eric and Kern, Robert and Larson, Eric and Carey, C J and Polat, {\.I}lhan and Feng, Yu and Moore, Eric W. and VanderPlas, Jake and Laxalde, Denis and Perktold, Josef and Cimrman, Robert and Henriksen, Ian and Quintero, E. A. and Harris, Charles R. and Archibald, Anne M. and Ribeiro, Ant{\^o}nio H. and Pedregosa, Fabian and Van Mulbregt, Paul and {SciPy 1.0 Contributors} and Vijaykumar, Aditya and Bardelli, Alessandro Pietro and Rothberg, Alex and Hilboll, Andreas and Kloeckner, Andreas and Scopatz, Anthony and Lee, Antony and Rokem, Ariel and Woods, C. Nathan and Fulton, Chad and Masson, Charles and H{\"a}ggstr{\"o}m, Christian and Fitzgerald, Clark and Nicholson, David A. and Hagen, David R. and Pasechnik, Dmitrii V. and Olivetti, Emanuele and Martin, Eric and Wieser, Eric and Silva, Fabrice and Lenders, Felix and Wilhelm, Florian and Young, G. and Price, Gavin A. and Ingold, Gert-Ludwig and Allen, Gregory E. and Lee, Gregory R. and Audren, Herv{\'e} and Probst, Irvin and Dietrich, J{\"o}rg P. and Silterra, Jacob and Webber, James T and Slavi{\v c}, Janko and Nothman, Joel and Buchner, Johannes and Kulick, Johannes and Sch{\"o}nberger, Johannes L. and De Miranda Cardoso, Jos{\'e} Vin{\'i}cius and Reimer, Joscha and Harrington, Joseph and Rodr{\'i}guez, Juan Luis Cano and {Nunez-Iglesias}, Juan and Kuczynski, Justin and Tritz, Kevin and Thoma, Martin and Newville, Matthew and K{\"u}mmerer, Matthias and Bolingbroke, Maximilian and Tartre, Michael and Pak, Mikhail and Smith, Nathaniel J. and Nowaczyk, Nikolai and Shebanov, Nikolay and Pavlyk, Oleksandr and Brodtkorb, Per A. and Lee, Perry and McGibbon, Robert T. and Feldbauer, Roman and Lewis, Sam and Tygier, Sam and Sievert, Scott and Vigna, Sebastiano and Peterson, Stefan and More, Surhud and Pudlik, Tadeusz and Oshima, Takuya and Pingel, Thomas J. and Robitaille, Thomas P. and Spura, Thomas and Jones, Thouis R. and Cera, Tim and Leslie, Tim and Zito, Tiziano and Krauss, Tom and Upadhyay, Utkarsh and Halchenko, Yaroslav O. and {V{\'a}zquez-Baeza}, Yoshiki},
  year = 2020,
  month = mar,
  journal = {Nat Methods},
  volume = {17},
  number = {3},
  pages = {261--272},
  issn = {1548-7091, 1548-7105},
  doi = {10.1038/s41592-019-0686-2},
  url = {https://www.nature.com/articles/s41592-019-0686-2},
  urldate = {2025-06-13},
  langid = {english}
}

@article{kobayashiRapidFormationGasgiant2021,
  title = {Rapid {{Formation}} of {{Gas-giant Planets}} via {{Collisional Coagulation}} from {{Dust Grains}} to {{Planetary Cores}}},
  author = {Kobayashi, Hiroshi and Tanaka, Hidekazu},
  year = 2021,
  month = jan,
  journal = {ApJ},
  volume = {922},
  number = {1},
  pages = {16},
  publisher = {The American Astronomical Society},
  issn = {0004-637X},
  doi = {10.3847/1538-4357/ac289c},
  url = {https://dx.doi.org/10.3847/1538-4357/ac289c},
  urldate = {2025-06-28},
  langid = {english}
}

@article{zhaoDifferentiallyWeightedMonte2010,
  title = {A Differentially Weighted {{Monte Carlo}} Method for Two-Component Coagulation},
  author = {Zhao, Haibo and Kruis, F. Einar and Zheng, Chuguang},
  year = 2010,
  month = sep,
  journal = {J. Comput. Phys.},
  volume = {229},
  number = {19},
  pages = {6931--6945},
  issn = {0021-9991},
  doi = {10.1016/j.jcp.2010.05.031},
  url = {https://www.sciencedirect.com/science/article/pii/S0021999110002925},
  urldate = {2025-06-30}
}

@article{seigneurSimulationAerosolDynamics1986,
  title = {Simulation of {{Aerosol Dynamics}}: {{A Comparative Review}} of {{Mathematical Models}}},
  shorttitle = {Simulation of {{Aerosol Dynamics}}},
  author = {Seigneur, Christian and Hudischewskyj, A. Belle and Seinfeld, John H. and Whitby, Kenneth T. and Whitby, Evan R. and Brock, James R. and Barnes, Harold M.},
  year = 1986,
  month = jan,
  journal = {Aerosol Sci. Technol.},
  volume = {5},
  number = {2},
  pages = {205--222},
  publisher = {Taylor \& Francis},
  issn = {0278-6826},
  doi = {10.1080/02786828608959088},
  url = {https://doi.org/10.1080/02786828608959088},
  urldate = {2025-06-30},
  langid = {english}
}

@article{kumarSolutionPopulationBalance1996,
  title = {On the Solution of Population Balance Equations by Discretization---{{I}}. {{A}} Fixed Pivot Technique},
  author = {Kumar, Sanjeev and Ramkrishna, D.},
  year = 1996,
  month = apr,
  journal = {Chem. Eng. Sci.},
  volume = {51},
  number = {8},
  pages = {1311--1332},
  issn = {0009-2509},
  doi = {10.1016/0009-2509(96)88489-2},
  url = {https://www.sciencedirect.com/science/article/pii/0009250996884892},
  urldate = {2025-06-30}
}

@article{gunawanHighResolutionAlgorithms2004,
  title = {High Resolution Algorithms for Multidimensional Population Balance Equations},
  author = {Gunawan, Rudiyanto and Fusman, Irene and Braatz, Richard D.},
  year = 2004,
  journal = {AIChE J},
  volume = {50},
  number = {11},
  pages = {2738--2749},
  issn = {1547-5905},
  doi = {10.1002/aic.10228},
  url = {https://onlinelibrary.wiley.com/doi/abs/10.1002/aic.10228},
  urldate = {2025-06-30},
  copyright = {Copyright \copyright{} 2004 American Institute of Chemical Engineers (AIChE)},
  langid = {english}
}

@article{oseledetsTensorTrainDecomposition2011,
  title = {Tensor-{{Train Decomposition}}},
  author = {Oseledets, I. V.},
  year = 2011,
  month = jan,
  journal = {SIAM J. Sci. Comput.},
  volume = {33},
  number = {5},
  pages = {2295--2317},
  publisher = {{Society for Industrial and Applied Mathematics}},
  issn = {1064-8275},
  doi = {10.1137/090752286},
  url = {https://epubs.siam.org/doi/abs/10.1137/090752286},
  urldate = {2025-06-30}
}

@article{xuAcceleratingPopulationBalanceMonte2015,
  title = {Accelerating Population Balance-{{Monte Carlo}} Simulation for Coagulation Dynamics from the {{Markov}} Jump Model, Stochastic Algorithm and {{GPU}} Parallel Computing},
  author = {Xu, Zuwei and Zhao, Haibo and Zheng, Chuguang},
  year = 2015,
  month = jan,
  journal = {J. Comput. Phys.},
  volume = {281},
  pages = {844--863},
  issn = {0021-9991},
  doi = {10.1016/j.jcp.2014.10.055},
  url = {https://www.sciencedirect.com/science/article/pii/S0021999114007438},
  urldate = {2025-06-30}
}

@inproceedings{zagidullinSupercomputerModellingSpatiallyheterogeneous2019,
  title = {Supercomputer {{Modelling}} of {{Spatially-heterogeneous Coagulation}} Using {{MPI}} and {{CUDA}}},
  booktitle = {Supercomputing},
  author = {Zagidullin, Rishat and Smirnov, Alexander and Matveev, Sergey and Tyrtyshnikov, Eugene},
  editor = {Voevodin, Vladimir and Sobolev, Sergey},
  year = 2019,
  pages = {403--414},
  publisher = {Springer International Publishing},
  address = {Cham},
  doi = {10.1007/978-3-030-36592-9_33},
  isbn = {978-3-030-36592-9},
  langid = {english}
}

@incollection{zhangNanoparticleAggregationPrinciples2014,
  title = {Nanoparticle {{Aggregation}}: {{Principles}} and {{Modeling}}},
  shorttitle = {Nanoparticle {{Aggregation}}},
  booktitle = {Nanomaterial: {{Impacts}} on {{Cell Biology}} and {{Medicine}}},
  author = {Zhang, Wen},
  editor = {Capco, David G. and Chen, Yongsheng},
  year = 2014,
  pages = {19--43},
  publisher = {Springer Netherlands},
  address = {Dordrecht},
  doi = {10.1007/978-94-017-8739-0_2},
  url = {https://doi.org/10.1007/978-94-017-8739-0_2},
  urldate = {2025-06-30},
  isbn = {978-94-017-8739-0},
  langid = {english}
}

@article{solsvikFoundationPopulationBalance2015,
  title = {The {{Foundation}} of the {{Population Balance Equation}}: {{A Review}}},
  shorttitle = {The {{Foundation}} of the {{Population Balance Equation}}},
  author = {Solsvik, Jannike and {and Jakobsen}, Hugo A.},
  year = 2015,
  month = apr,
  journal = {J. Dispers. Sci. Technol.},
  volume = {36},
  number = {4},
  pages = {510--520},
  publisher = {Taylor \& Francis},
  issn = {0193-2691},
  doi = {10.1080/01932691.2014.909318},
  url = {https://doi.org/10.1080/01932691.2014.909318},
  urldate = {2025-06-30}
}

@article{maHighResolutionSimulationMultidimensional2002,
  title = {High-{{Resolution Simulation}} of {{Multidimensional Crystal Growth}}},
  author = {Ma, David L. and Tafti, Danesh K. and Braatz, Richard D.},
  year = 2002,
  month = dec,
  journal = {Ind. Eng. Chem. Res.},
  volume = {41},
  number = {25},
  pages = {6217--6223},
  publisher = {American Chemical Society},
  issn = {0888-5885},
  doi = {10.1021/ie010680u},
  url = {https://doi.org/10.1021/ie010680u},
  urldate = {2025-07-03}
}

@article{hulburtProblemsParticleTechnology1964,
  title = {Some Problems in Particle Technology: {{A}} Statistical Mechanical Formulation},
  shorttitle = {Some Problems in Particle Technology},
  author = {Hulburt, H. M. and Katz, S.},
  year = 1964,
  month = aug,
  journal = {Chem. Eng. Sci.},
  volume = {19},
  number = {8},
  pages = {555--574},
  issn = {0009-2509},
  doi = {10.1016/0009-2509(64)85047-8},
  url = {https://www.sciencedirect.com/science/article/pii/0009250964850478},
  urldate = {2025-07-03}
}

@article{ottoPopulationBalanceModeling2024,
  title = {Population {{Balance Modeling}} of {{Particle Size}} and {{Porosity}} in {{Fluidized Bed Spray Agglomeration}}},
  author = {Otto, Eric and Ajalova, Aisel and B{\"u}ck, Andreas and Tsotsas, Evangelos and Kienle, Achim},
  year = 2024,
  month = oct,
  journal = {Ind. Eng. Chem. Res.},
  volume = {63},
  number = {41},
  pages = {17545--17556},
  publisher = {American Chemical Society},
  issn = {0888-5885},
  doi = {10.1021/acs.iecr.4c01660},
  url = {https://doi.org/10.1021/acs.iecr.4c01660},
  urldate = {2025-07-03}
}

@article{laurenziGeneralAlgorithmExact2002,
  title = {A {{General Algorithm}} for {{Exact Simulation}} of {{Multicomponent Aggregation Processes}}},
  author = {Laurenzi, Ian J. and Bartels, John D. and Diamond, Scott L.},
  year = 2002,
  month = apr,
  journal = {J. Comput. Phys.},
  volume = {177},
  number = {2},
  pages = {418--449},
  issn = {0021-9991},
  doi = {10.1006/jcph.2002.7017},
  url = {https://www.sciencedirect.com/science/article/pii/S0021999102970178},
  urldate = {2025-07-03}
}

@article{nakagawaGrowthSedimentationDust1981,
  title = {Growth and Sedimentation of Dust Grains in the Primordial Solar Nebula},
  author = {Nakagawa, Yoshitsugu and Nakazawa, Kiyoshi and Hayashi, Chushiro},
  year = 1981,
  month = mar,
  journal = {Icarus},
  volume = {45},
  number = {3},
  pages = {517--528},
  issn = {0019-1035},
  doi = {10.1016/0019-1035(81)90018-X},
  url = {https://www.sciencedirect.com/science/article/pii/001910358190018X},
  urldate = {2025-07-07}
}

@article{dullemondDustCoagulationProtoplanetary2005,
  title = {Dust Coagulation in Protoplanetary Disks: {{A}} Rapid Depletion of Small Grains},
  shorttitle = {Dust Coagulation in Protoplanetary Disks},
  author = {Dullemond, C. P. and Dominik, C.},
  year = 2005,
  month = may,
  journal = {A\&A},
  volume = {434},
  number = {3},
  pages = {971--986},
  publisher = {EDP Sciences},
  issn = {0004-6361, 1432-0746},
  doi = {10.1051/0004-6361:20042080},
  url = {https://www.aanda.org/articles/aa/abs/2005/18/aa2080/aa2080.html},
  urldate = {2025-07-07},
  copyright = {\copyright{} ESO, 2005},
  langid = {english}
}

@article{nomuraDustSizeGrowth2006,
  title = {Dust {{Size Growth}} and {{Settling}} in a {{Protoplanetary Disk}}},
  author = {Nomura, Hideko and Nakagawa, Yoshitsugu},
  year = 2006,
  month = apr,
  journal = {ApJ},
  volume = {640},
  number = {2},
  pages = {1099},
  issn = {0004-637X},
  doi = {10.1086/500251},
  url = {https://dx.doi.org/10.1086/500251},
  urldate = {2025-07-07},
  langid = {english}
}

@article{alfonsoValidityKineticCollection2008,
  title = {The Validity of the Kinetic Collection Equation Revisited},
  author = {Alfonso, L. and Raga, G. B. and Baumgardner, D.},
  year = 2008,
  month = feb,
  journal = {Atmos. Chem. Phys.},
  volume = {8},
  number = {4},
  pages = {969--982},
  publisher = {Copernicus GmbH},
  issn = {1680-7316},
  doi = {10.5194/acp-8-969-2008},
  url = {https://acp.copernicus.org/articles/8/969/2008/},
  urldate = {2025-07-07},
  langid = {english}
}

@article{spougeAnalyticSolutionsSmoluchowskis1985,
  title = {Analytic Solutions to {{Smoluchowski}}'s Coagulation Equation: A Combinatorial Interpretation},
  shorttitle = {Analytic Solutions to {{Smoluchowski}}'s Coagulation Equation},
  author = {Spouge, J. L.},
  year = 1985,
  journal = {J. Phys. A, Math. Gen.},
  volume = {18},
  number = {15},
  pages = {3063},
  publisher = {IOP Publishing},
  url = {https://iopscience.iop.org/article/10.1088/0305-4470/18/15/028/meta},
  urldate = {2025-07-07}
}

@article{zsomRepresentativeParticleApproach2008,
  title = {A Representative Particle Approach to Coagulation and Fragmentation of Dust Aggregates and Fluid Droplets},
  author = {Zsom, A. and Dullemond, C. P.},
  year = 2008,
  month = oct,
  journal = {A\&A},
  volume = {489},
  number = {2},
  pages = {931--941},
  publisher = {EDP Sciences},
  issn = {0004-6361, 1432-0746},
  doi = {10.1051/0004-6361:200809921},
  url = {https://www.aanda.org/articles/aa/abs/2008/38/aa09921-08/aa09921-08.html},
  urldate = {2025-07-07},
  copyright = {\copyright{} ESO, 2008},
  langid = {english}
}

@article{pfeilTriPoDTriPopulationSize2024,
  title = {{{TriPoD}}: {{Tri-Population}} Size Distributions for {{Dust}} Evolution - {{Coagulation}} in Vertically Integrated Hydrodynamic Simulations of Protoplanetary Disks},
  shorttitle = {{{TriPoD}}},
  author = {Pfeil, Thomas and Birnstiel, Til and Klahr, Hubert},
  year = 2024,
  month = nov,
  journal = {A\&A},
  volume = {691},
  pages = {A45},
  publisher = {EDP Sciences},
  issn = {0004-6361, 1432-0746},
  doi = {10.1051/0004-6361/202449337},
  url = {https://www.aanda.org/articles/aa/abs/2024/11/aa49337-24/aa49337-24.html},
  urldate = {2025-07-07},
  copyright = {\copyright{} The Authors 2024},
  langid = {english}
}

@article{gruyPopulationBalanceAggregation2011,
  title = {Population Balance for Aggregation Coupled with Morphology Changes},
  author = {Gruy, Fr{\'e}d{\'e}ric},
  year = 2011,
  month = jan,
  journal = {Colloids and Surfaces A: Physicochemical and Engineering Aspects},
  volume = {374},
  number = {1},
  pages = {69--76},
  issn = {0927-7757},
  doi = {10.1016/j.colsurfa.2010.11.010},
  url = {https://www.sciencedirect.com/science/article/pii/S092777571000631X},
  urldate = {2025-07-11}
}

@article{osullivanConservativeFiniteVolume2022,
  title = {A Conservative Finite Volume Method for the Population Balance Equation with Aggregation, Fragmentation, Nucleation and Growth},
  author = {O'Sullivan, Daniel and Rigopoulos, Stelios},
  year = 2022,
  month = dec,
  journal = {Chem. Eng. Sci.},
  volume = {263},
  pages = {117925},
  issn = {0009-2509},
  doi = {10.1016/j.ces.2022.117925},
  url = {https://www.sciencedirect.com/science/article/pii/S0009250922005097},
  urldate = {2025-07-11}
}

@article{kostoglouEvolutionAggregateSize2001,
  title = {Evolution of Aggregate Size and Fractal Dimension during {{Brownian}} Coagulation},
  author = {Kostoglou, Margaritis and Konstandopoulos, Athanasios G},
  year = 2001,
  month = dec,
  journal = {J. Aerosol Sci.},
  volume = {32},
  number = {12},
  pages = {1399--1420},
  issn = {0021-8502},
  doi = {10.1016/S0021-8502(01)00056-8},
  url = {https://www.sciencedirect.com/science/article/pii/S0021850201000568},
  urldate = {2025-07-11}
}

@article{kostoglouBivariatePopulationDynamics2006,
  title = {Bivariate Population Dynamics Simulation of Fractal Aerosol Aggregate Coagulation and Restructuring},
  author = {Kostoglou, M. and Konstandopoulos, A. G. and Friedlander, S. K.},
  year = 2006,
  month = sep,
  journal = {J. Aerosol Sci.},
  volume = {37},
  number = {9},
  pages = {1102--1115},
  issn = {0021-8502},
  doi = {10.1016/j.jaerosci.2005.11.009},
  url = {https://www.sciencedirect.com/science/article/pii/S0021850205002211},
  urldate = {2025-07-11}
}

@article{zsomOutcomeProtoplanetaryDust2011,
  title = {The Outcome of Protoplanetary Dust Growth: Pebbles, Boulders, or Planetesimals? - {{III}}. {{Sedimentation}} Driven Coagulation inside the Snowline},
  shorttitle = {The Outcome of Protoplanetary Dust Growth},
  author = {Zsom, A. and Ormel, C. W. and Dullemond, C. P. and Henning, T.},
  year = 2011,
  month = oct,
  journal = {A\&A},
  volume = {534},
  pages = {A73},
  publisher = {EDP Sciences},
  issn = {0004-6361, 1432-0746},
  doi = {10.1051/0004-6361/201116515},
  url = {https://www.aanda.org/articles/aa/abs/2011/10/aa16515-11/aa16515-11.html},
  urldate = {2025-07-11},
  copyright = {\copyright{} ESO, 2011},
  langid = {english}
}

@article{kornetDiversityPlanetarySystems2001,
  title = {Diversity of Planetary Systems from Evolution of Solids in Protoplanetary Disks},
  author = {Kornet, K. and Stepinski, T. F. and R{\'o}{\.z}yczka, M.},
  year = 2001,
  month = oct,
  journal = {A\&A},
  volume = {378},
  number = {1},
  pages = {180--191},
  publisher = {EDP Sciences},
  issn = {0004-6361, 1432-0746},
  doi = {10.1051/0004-6361:20011183},
  url = {https://www.aanda.org/articles/aa/abs/2001/40/aa1537/aa1537.html},
  urldate = {2025-07-11},
  copyright = {\copyright{} ESO, 2001},
  langid = {english}
}

@article{stepinskiGlobalEvolutionSolid1996,
  title = {Global Evolution of Solid Matter in Turbulent Protoplanetary Disks. {{I}}. {{Aerodynamics}} of Solid Particles.},
  author = {Stepinski, T. F. and Valageas, P.},
  year = 1996,
  month = may,
  journal = {A\&A},
  volume = {309},
  pages = {301--312},
  issn = {0004-6361},
  url = {https://ui.adsabs.harvard.edu/abs/1996A&A...309..301S},
  urldate = {2025-07-11}
}

@article{stepinskiGlobalEvolutionSolid1997,
  title = {Global Evolution of Solid Matter in Turbulent Protoplanetary Disks. {{II}}. {{Development}} of Icy Planetesimals.},
  author = {Stepinski, T. F. and Valageas, P.},
  year = 1997,
  month = mar,
  journal = {A\&A},
  volume = {319},
  pages = {1007--1019},
  issn = {0004-6361},
  url = {https://ui.adsabs.harvard.edu/abs/1997A&A...319.1007S},
  urldate = {2025-07-11}
}

@article{garaudGrowthMigrationSolids2007,
  title = {Growth and {{Migration}} of {{Solids}} in {{Evolving Protostellar Disks}}. {{I}}. {{Methods}} and {{Analytical Tests}}},
  author = {Garaud, P.},
  year = 2007,
  month = feb,
  journal = {ApJ},
  volume = {671},
  number = {2},
  pages = {2091},
  issn = {0004-637X},
  doi = {10.1086/523090},
  url = {https://dx.doi.org/10.1086/523090},
  urldate = {2025-07-11},
  langid = {english}
}

@article{birnstielSimpleModelEvolution2012,
  title = {A Simple Model for the Evolution of the Dust Population in Protoplanetary Disks},
  author = {Birnstiel, T. and Klahr, H. and Ercolano, B.},
  year = 2012,
  month = mar,
  journal = {A\&A},
  volume = {539},
  pages = {A148},
  publisher = {EDP Sciences},
  issn = {0004-6361, 1432-0746},
  doi = {10.1051/0004-6361/201118136},
  url = {https://www.aanda.org/articles/aa/abs/2012/03/aa18136-11/aa18136-11.html},
  urldate = {2025-07-11},
  copyright = {\copyright{} ESO, 2012},
  langid = {english}
}

@article{krijtPanopticModelPlanetesimal2016,
  title = {A Panoptic Model for Planetesimal Formation and Pebble Delivery},
  author = {Krijt, S. and Ormel, C. W. and Dominik, C. and Tielens, A. G. G. M.},
  year = 2016,
  month = feb,
  journal = {A\&A},
  volume = {586},
  pages = {A20},
  publisher = {EDP Sciences},
  issn = {0004-6361, 1432-0746},
  doi = {10.1051/0004-6361/201527533},
  url = {https://www.aanda.org/articles/aa/abs/2016/02/aa27533-15/aa27533-15.html},
  urldate = {2025-07-11},
  copyright = {\copyright{} ESO, 2016},
  langid = {english}
}

@article{laibeSPHSimulationsGrain2008,
  title = {{{SPH}} Simulations of Grain Growth in Protoplanetary Disks},
  author = {Laibe, G. and Gonzalez, J.-F. and Fouchet, L. and Maddison, S. T.},
  year = 2008,
  month = aug,
  journal = {A\&A},
  volume = {487},
  number = {1},
  pages = {265--270},
  publisher = {EDP Sciences},
  issn = {0004-6361, 1432-0746},
  doi = {10.1051/0004-6361:200809522},
  url = {https://www.aanda.org/articles/aa/abs/2008/31/aa09522-08/aa09522-08.html},
  urldate = {2025-07-11},
  copyright = {\copyright{} ESO, 2008},
  langid = {english}
}

@article{windmarkPlanetesimalFormationSweepup2012,
  title = {Planetesimal Formation by Sweep-up: How the Bouncing Barrier Can Be Beneficial to Growth},
  shorttitle = {Planetesimal Formation by Sweep-Up},
  author = {Windmark, F. and Birnstiel, T. and G{\"u}ttler, C. and Blum, J. and Dullemond, C. P. and Henning, Th},
  year = 2012,
  month = apr,
  journal = {A\&A},
  volume = {540},
  pages = {A73},
  publisher = {EDP Sciences},
  issn = {0004-6361, 1432-0746},
  doi = {10.1051/0004-6361/201118475},
  url = {https://www.aanda.org/articles/aa/abs/2012/04/aa18475-11/aa18475-11.html},
  urldate = {2025-07-11},
  copyright = {\copyright{} ESO, 2012},
  langid = {english}
}

@article{garciaEvolutionPorousDust2020,
  title = {Evolution of Porous Dust Grains in Protoplanetary Discs--{{I}}. {{Growing}} Grains},
  author = {Garcia, Anthony JL and Gonzalez, Jean-Fran{\c c}ois},
  year = 2020,
  journal = {MNRAS},
  volume = {493},
  number = {2},
  pages = {1788--1800},
  publisher = {Oxford University Press},
  doi = {10.1093/mnras/staa382},
  url = {https://academic.oup.com/mnras/article-abstract/493/2/1788/5731863},
  urldate = {2025-07-11}
}

@article{michoulierCompactionFragmentationBouncing2024,
  title = {Compaction during Fragmentation and Bouncing Produces Realistic Dust Grain Porosities in Protoplanetary Discs},
  author = {Michoulier, St{\'e}phane and Gonzalez, Jean-Fran{\c c}ois and Price, Daniel J.},
  year = 2024,
  month = aug,
  journal = {A\&A},
  volume = {688},
  pages = {A31},
  publisher = {EDP Sciences},
  issn = {0004-6361, 1432-0746},
  doi = {10.1051/0004-6361/202449719},
  url = {https://www.aanda.org/articles/aa/abs/2024/08/aa49719-24/aa49719-24.html},
  urldate = {2025-07-11},
  copyright = {\copyright{} The Authors 2024},
  langid = {english}
}

@article{michoulierDustGrainShattering2022,
  title = {Dust Grain Shattering in Protoplanetary Discs: Collisional Fragmentation or Rotational Disruption?},
  shorttitle = {Dust Grain Shattering in Protoplanetary Discs},
  author = {Michoulier, St{\'e}phane and Gonzalez, Jean-Fran{\c c}ois},
  year = 2022,
  month = dec,
  journal = {MNRAS},
  volume = {517},
  number = {2},
  pages = {3064--3077},
  issn = {0035-8711},
  doi = {10.1093/mnras/stac2842},
  url = {https://doi.org/10.1093/mnras/stac2842},
  urldate = {2025-07-11}
}

@article{johansenCoagulationfragmentationModelTurbulent2008,
  title = {A Coagulation-Fragmentation Model for the Turbulent Growth and Destruction of Preplanetesimals},
  author = {Johansen, A. and Brauer, F. and Dullemond, C. and Klahr, H. and Henning, T.},
  year = 2008,
  month = aug,
  journal = {A\&A},
  volume = {486},
  number = {2},
  pages = {597--611},
  publisher = {EDP Sciences},
  issn = {0004-6361, 1432-0746},
  doi = {10.1051/0004-6361:20079232},
  url = {https://www.aanda.org/articles/aa/abs/2008/29/aa9232-07/aa9232-07.html},
  urldate = {2025-07-11},
  copyright = {\copyright{} ESO, 2008},
  langid = {english}
}

@article{dehnenFastMultipoleMethod2014,
  title = {A Fast Multipole Method for Stellar Dynamics},
  author = {Dehnen, Walter},
  year = 2014,
  month = sep,
  journal = {ComAC},
  volume = {1},
  number = {1},
  pages = {1},
  issn = {2197-7909},
  doi = {10.1186/s40668-014-0001-7},
  url = {https://doi.org/10.1186/s40668-014-0001-7},
  urldate = {2025-07-12}
}

@article{capuzzo-dolcettaComparisonFastMultipole1998,
  title = {A {{Comparison}} between the {{Fast Multipole Algorithm}} and the {{Tree-Code}} to {{Evaluate Gravitational Forces}} in 3-{{D}}},
  author = {{Capuzzo-Dolcetta}, R. and Miocchi, P.},
  year = 1998,
  month = jun,
  journal = {J. Comput. Phys.},
  volume = {143},
  number = {1},
  pages = {29--48},
  issn = {0021-9991},
  doi = {10.1006/jcph.1998.5949},
  url = {https://www.sciencedirect.com/science/article/pii/S0021999198959496},
  urldate = {2025-07-12}
}

@article{greengardFastAlgorithmParticle1987,
  title = {A Fast Algorithm for Particle Simulations},
  author = {Greengard, L and Rokhlin, V},
  year = 1987,
  month = dec,
  journal = {J. Comput. Phys.},
  volume = {73},
  number = {2},
  pages = {325--348},
  issn = {0021-9991},
  doi = {10.1016/0021-9991(87)90140-9},
  url = {https://www.sciencedirect.com/science/article/pii/0021999187901409},
  urldate = {2025-07-12}
}

@article{chengFastAdaptiveMultipole1999,
  title = {A {{Fast Adaptive Multipole Algorithm}} in {{Three Dimensions}}},
  author = {Cheng, H. and Greengard, L. and Rokhlin, V.},
  year = 1999,
  month = nov,
  journal = {J. Comput. Phys.},
  volume = {155},
  number = {2},
  pages = {468--498},
  issn = {0021-9991},
  doi = {10.1006/jcph.1999.6355},
  url = {https://www.sciencedirect.com/science/article/pii/S0021999199963556},
  urldate = {2025-07-12}
}

@article{chakrabortyNewFrameworkSolution2007,
  title = {A New Framework for Solution of Multidimensional Population Balance Equations},
  author = {Chakraborty, Jayanta and Kumar, Sanjeev},
  year = 2007,
  month = aug,
  journal = {Chem. Eng. Sci.},
  volume = {62},
  number = {15},
  pages = {4112--4125},
  issn = {0009-2509},
  doi = {10.1016/j.ces.2007.04.049},
  url = {https://www.sciencedirect.com/science/article/pii/S0009250907003806},
  urldate = {2025-07-12}
}

@article{nandanwarNewDiscretizationSpace2008,
  title = {A New Discretization of Space for the Solution of Multi-Dimensional Population Balance Equations},
  author = {Nandanwar, Mahendra N. and Kumar, Sanjeev},
  year = 2008,
  month = apr,
  journal = {Chem. Eng. Sci.},
  volume = {63},
  number = {8},
  pages = {2198--2210},
  issn = {0009-2509},
  doi = {10.1016/j.ces.2008.01.015},
  url = {https://www.sciencedirect.com/science/article/pii/S0009250908000316},
  urldate = {2025-07-12}
}

@article{singhSolutionBivariateAggregation2018,
  title = {Solution of Bivariate Aggregation Population Balance Equation: A Comparative Study},
  shorttitle = {Solution of Bivariate Aggregation Population Balance Equation},
  author = {Singh, Mehakpreet and Kaur, Gurmeet and De Beer, Thomas and Nopens, Ingmar},
  year = 2018,
  month = apr,
  journal = {Reac Kinet Mech Cat},
  volume = {123},
  number = {2},
  pages = {385--401},
  issn = {1878-5204},
  doi = {10.1007/s11144-018-1345-9},
  url = {https://doi.org/10.1007/s11144-018-1345-9},
  urldate = {2025-07-12},
  langid = {english}
}

@article{chauhanSolutionBivariatePopulation2012,
  title = {On the Solution of Bivariate Population Balance Equations for Aggregation: {{X}}--Discretization of Space for Expansion and Contraction of Computational Domain},
  shorttitle = {On the Solution of Bivariate Population Balance Equations for Aggregation},
  author = {Chauhan, Shivendra Singh and Chiney, Abhinandan and Kumar, Sanjeev},
  year = 2012,
  month = mar,
  journal = {Chem. Eng. Sci.},
  series = {4th {{International Conference}} on {{Population Balance Modeling}}},
  volume = {70},
  pages = {135--145},
  issn = {0009-2509},
  doi = {10.1016/j.ces.2011.10.005},
  url = {https://www.sciencedirect.com/science/article/pii/S0009250911007093},
  urldate = {2025-07-12}
}

@article{akimkinInhibitedCoagulationMicronsize2020,
  title = {Inhibited {{Coagulation}} of {{Micron-size Dust Due}} to the {{Electrostatic Barrier}}},
  author = {Akimkin, V. V. and Ivlev, A. V. and Caselli, P.},
  year = 2020,
  month = jan,
  journal = {ApJ},
  volume = {889},
  number = {1},
  pages = {64},
  publisher = {The American Astronomical Society},
  issn = {0004-637X},
  doi = {10.3847/1538-4357/ab6299},
  url = {https://dx.doi.org/10.3847/1538-4357/ab6299},
  urldate = {2025-07-12},
  langid = {english}
}

@article{krijtErosionLimitsPlanetesimal2015,
  title = {Erosion and the Limits to Planetesimal Growth},
  author = {Krijt, S. and Ormel, C. W. and Dominik, C. and Tielens, A. G. G. M.},
  year = 2015,
  month = feb,
  journal = {A\&A},
  volume = {574},
  pages = {A83},
  publisher = {EDP Sciences},
  issn = {0004-6361, 1432-0746},
  doi = {10.1051/0004-6361/201425222},
  url = {https://www.aanda.org/articles/aa/abs/2015/02/aa25222-14/aa25222-14.html},
  urldate = {2025-07-12},
  copyright = {\copyright{} ESO, 2015},
  langid = {english}
}

@misc{dyachenkoMosaicskeletonApproximationAll2025,
  title = {Mosaic-Skeleton Approximation Is All You Need for {{Smoluchowski}} Equations},
  author = {Dyachenko, Roman R. and Matveev, Sergey A. and Valiakhmetov, Bulat I.},
  year = 2025,
  month = jan,
  number = {arXiv:2501.10206},
  eprint = {2501.10206},
  primaryclass = {math},
  publisher = {arXiv},
  doi = {10.48550/arXiv.2501.10206},
  url = {http://arxiv.org/abs/2501.10206},
  urldate = {2025-07-12},
  archiveprefix = {arXiv}
}

@article{oshiroInvestigatingBouncingBarrier2025,
  title = {Investigating the {{Bouncing Barrier}} with {{Collision Simulations}} of {{Compressed Dust Aggregates}}},
  author = {Oshiro, Haruto and Tatsuuma, Misako and Okuzumi, Satoshi and Tanaka, Hidekazu},
  year = 2025,
  month = apr,
  journal = {ApJ},
  volume = {983},
  number = {1},
  pages = {75},
  publisher = {The American Astronomical Society},
  issn = {0004-637X},
  doi = {10.3847/1538-4357/adbf04},
  url = {https://dx.doi.org/10.3847/1538-4357/adbf04},
  urldate = {2025-07-12},
  langid = {english}
}

@article{tanakaDustGrowthSettling2005,
  title = {Dust {{Growth}} and {{Settling}} in {{Protoplanetary Disks}} and {{Disk Spectral Energy Distributions}}. {{I}}. {{Laminar Disks}}},
  author = {Tanaka, Hidekazu and Himeno, Youhei and Ida, Shigeru},
  year = 2005,
  month = may,
  journal = {ApJ},
  volume = {625},
  number = {1},
  pages = {414},
  issn = {0004-637X},
  doi = {10.1086/429658},
  url = {https://dx.doi.org/10.1086/429658},
  urldate = {2025-07-12},
  langid = {english}
}

@article{drazkowskaPlanetesimalFormationSweepup2013,
  title = {Planetesimal Formation via Sweep-up Growth at the Inner Edge of Dead Zones},
  author = {Dr{\k a}{\.z}kowska, Joanna and Windmark, F. and Dullemond, C. P.},
  year = 2013,
  journal = {A\&A},
  volume = {556},
  pages = {A37},
  publisher = {EDP Sciences},
  doi = {10.1051/0004-6361/201321566},
  url = {https://www.aanda.org/articles/aa/abs/2013/08/aa21566-13/aa21566-13.html},
  urldate = {2025-07-13}
}

@article{kataokaStaticCompressionPorous2013,
  title = {Static Compression of Porous Dust Aggregates},
  author = {Kataoka, A. and Tanaka, H. and Okuzumi, S. and Wada, K.},
  year = 2013,
  month = jun,
  journal = {A\&A},
  volume = {554},
  pages = {A4},
  publisher = {EDP Sciences},
  issn = {0004-6361, 1432-0746},
  doi = {10.1051/0004-6361/201321325},
  url = {https://www.aanda.org/articles/aa/abs/2013/06/aa21325-13/aa21325-13.html},
  urldate = {2025-07-17},
  copyright = {\copyright{} ESO, 2013},
  langid = {english}
}

@article{hernquistPerformanceCharacteristicsTree1987,
  title = {Performance {{Characteristics}} of {{Tree Codes}}},
  author = {Hernquist, Lars},
  year = 1987,
  month = aug,
  journal = {ApJS},
  volume = {64},
  pages = {715},
  publisher = {IOP},
  issn = {0067-0049},
  doi = {10.1086/191215},
  url = {https://ui.adsabs.harvard.edu/abs/1987ApJS...64..715H},
  urldate = {2025-08-08}
}

@article{winnOccurrenceArchitectureExoplanetary2015,
  title = {The {{Occurrence}} and {{Architecture}} of {{Exoplanetary Systems}}},
  author = {Winn, Joshua N. and Fabrycky, Daniel C.},
  year = 2015,
  month = aug,
  journal = {ARA\&A},
  volume = {53},
  pages = {409--447},
  issn = {0066-4146},
  doi = {10.1146/annurev-astro-082214-122246},
  url = {https://ui.adsabs.harvard.edu/abs/2015ARA&A..53..409W},
  urldate = {2025-08-12}
}

@article{ahearnCometsBuildingBlocks2011,
  title = {Comets as {{Building Blocks}}},
  author = {A'Hearn, Michael F.},
  year = 2011,
  month = sep,
  journal = {ARA\&A},
  volume = {49},
  pages = {281--299},
  issn = {0066-4146},
  doi = {10.1146/annurev-astro-081710-102506},
  url = {https://ui.adsabs.harvard.edu/abs/2011ARA&A..49..281A},
  urldate = {2025-08-12}
}

@article{obergProtoplanetaryDiskChemistry2023,
  title = {Protoplanetary {{Disk Chemistry}}},
  author = {{\"O}berg, Karin I. and Facchini, Stefano and Anderson, Dana E.},
  year = 2023,
  month = aug,
  journal = {ARA\&A},
  volume = {61},
  number = {1},
  pages = {287--328},
  issn = {0066-4146, 1545-4282},
  doi = {10.1146/annurev-astro-022823-040820},
  url = {https://www.annualreviews.org/content/journals/10.1146/annurev-astro-022823-040820},
  urldate = {2025-08-13},
  copyright = {http://creativecommons.org/licenses/by/4.0/},
  langid = {english}
}

@article{kazeevMultilevelToeplitzMatrices2013,
  title = {Multilevel {{Toeplitz Matrices Generated}} by {{Tensor-Structured Vectors}} and {{Convolution}} with {{Logarithmic Complexity}}},
  author = {Kazeev, Vladimir A. and Khoromskij, Boris N. and Tyrtyshnikov, Eugene E.},
  year = 2013,
  month = jan,
  journal = {SIAM J. Sci. Comput.},
  volume = {35},
  number = {3},
  pages = {A1511-A1536},
  issn = {1064-8275, 1095-7197},
  doi = {10.1137/110844830},
  url = {http://epubs.siam.org/doi/10.1137/110844830},
  urldate = {2025-09-08},
  langid = {english}
}

@article{planesCollisionsMicrosizedAggregates2021,
  title = {Collisions between Micro-Sized Aggregates: Role of Porosity, Mass Ratio, and Impact Velocity},
  shorttitle = {Collisions between Micro-Sized Aggregates},
  author = {Planes, Mar{\'i}a Bel{\'e}n and Mill{\'a}n, Emmanuel N and Urbassek, Herbert M and Bringa, Eduardo M},
  year = 2021,
  month = may,
  journal = {MNRAS},
  volume = {503},
  number = {2},
  pages = {1717--1733},
  issn = {0035-8711},
  doi = {10.1093/mnras/stab610},
  url = {https://doi.org/10.1093/mnras/stab610},
  urldate = {2025-09-08}
}

@article{shimakiLowvelocityCollisionsCentimetersized2012,
  title = {Low-Velocity Collisions between Centimeter-Sized Snowballs: {{Porosity}} Dependence of Coefficient of Restitution for Ice Aggregates Analogues in the {{Solar System}}},
  shorttitle = {Low-Velocity Collisions between Centimeter-Sized Snowballs},
  author = {Shimaki, Yuri and Arakawa, Masahiko},
  year = 2012,
  month = sep,
  journal = {Icarus},
  volume = {221},
  number = {1},
  pages = {310--319},
  issn = {0019-1035},
  doi = {10.1016/j.icarus.2012.08.005},
  url = {https://www.sciencedirect.com/science/article/pii/S0019103512003247},
  urldate = {2025-09-08}
}

@article{shimakiTensileStrengthElastic2021,
  title = {Tensile Strength and Elastic Properties of Fine-Grained Ice Aggregates: {{Implications}} for Crater Formation on Small Icy Bodies},
  shorttitle = {Tensile Strength and Elastic Properties of Fine-Grained Ice Aggregates},
  author = {Shimaki, Yuri and Arakawa, Masahiko},
  year = 2021,
  month = nov,
  journal = {Icarus},
  volume = {369},
  pages = {114646},
  issn = {0019-1035},
  doi = {10.1016/j.icarus.2021.114646},
  url = {https://www.sciencedirect.com/science/article/pii/S0019103521003067},
  urldate = {2025-09-08}
}

@article{blumExperimentsStickingRestructuring2000,
  title = {Experiments on {{Sticking}}, {{Restructuring}}, and {{Fragmentation}} of {{Preplanetary Dust Aggregates}}},
  author = {Blum, J{\"u}rgen and Wurm, Gerhard},
  year = 2000,
  month = jan,
  journal = {Icarus},
  volume = {143},
  number = {1},
  pages = {138--146},
  issn = {0019-1035},
  doi = {10.1006/icar.1999.6234},
  url = {https://www.sciencedirect.com/science/article/pii/S0019103599962346},
  urldate = {2025-09-08}
}

@article{estradaGlobalModelingNebulae2022,
  title = {Global {{Modeling}} of {{Nebulae}} with {{Particle Growth}}, {{Drift}}, and {{Evaporation Fronts}}. {{II}}. {{The Influence}} of {{Porosity}} on {{Solids Evolution}}},
  author = {Estrada, Paul R. and Cuzzi, Jeffrey N. and Umurhan, Orkan M.},
  year = 2022,
  month = aug,
  journal = {ApJ},
  volume = {936},
  number = {1},
  pages = {42},
  publisher = {The American Astronomical Society},
  issn = {0004-637X},
  doi = {10.3847/1538-4357/ac7ffd},
  url = {https://dx.doi.org/10.3847/1538-4357/ac7ffd},
  urldate = {2025-09-18},
  langid = {english}
}

@misc{reback2020pandas,
  title = {Pandas-Dev/Pandas: {{Pandas}}},
  author = {{The pandas development team}},
  year = 2020,
  month = feb,
  doi = {10.5281/zenodo.3509134},
  url = {https://doi.org/10.5281/zenodo.3509134},
  howpublished = {Zenodo}
}

@article{guttlerOutcomeProtoplanetaryDust2010,
  title = {The Outcome of Protoplanetary Dust Growth: Pebbles, Boulders, or Planetesimals?. {{I}}. {{Mapping}} the Zoo of Laboratory Collision Experiments},
  shorttitle = {The Outcome of Protoplanetary Dust Growth},
  author = {G{\"u}ttler, C. and Blum, J. and Zsom, A. and Ormel, C. W. and Dullemond, C. P.},
  year = 2010,
  month = apr,
  journal = {A\&A},
  volume = {513},
  pages = {A56},
  publisher = {EDP},
  issn = {0004-6361},
  doi = {10.1051/0004-6361/200912852},
  url = {https://ui.adsabs.harvard.edu/abs/2010A&A...513A..56G},
  urldate = {2025-11-17}
}

@misc{vaikundaramanMcdust2DMonte2025,
  title = {Mcdust: {{A 2D Monte Carlo}} Code for Dust Coagulation in Protoplanetary Disks},
  shorttitle = {Mcdust},
  author = {Vaikundaraman, Vignesh and Gurrutxaga, Nerea and Dr{\k a}{\.z}kowska, Joanna},
  year = 2025,
  month = jul,
  number = {arXiv:2507.21239},
  eprint = {2507.21239},
  primaryclass = {astro-ph},
  publisher = {arXiv},
  doi = {10.48550/arXiv.2507.21239},
  url = {http://arxiv.org/abs/2507.21239},
  urldate = {2025-12-15},
  archiveprefix = {arXiv}
}

@article{beutelEfficientSimulationStochastic2024,
  title = {Efficient Simulation of Stochastic Interactions among Representative {{Monte Carlo}} Particles},
  author = {Beutel, M. and Dullemond, C. P. and Strzodka, R.},
  year = 2024,
  month = apr,
  journal = {A\&A},
  volume = {684},
  pages = {A2},
  publisher = {EDP Sciences},
  issn = {0004-6361, 1432-0746},
  doi = {10.1051/0004-6361/202347304},
  url = {https://www.aanda.org/articles/aa/abs/2024/04/aa47304-23/aa47304-23.html},
  urldate = {2026-01-25},
  copyright = {\copyright{} The Authors 2024},
  langid = {english}
}

@article{hirashitaEvolutionDustPorosity2021,
  title = {Evolution of Dust Porosity through Coagulation and Shattering in the Interstellar Medium},
  author = {Hirashita, Hiroyuki and Il'in, Vladimir B and Pagani, Laurent and Lef{\`e}vre, Charl{\`e}ne},
  year = 2021,
  month = mar,
  journal = {MNRAS},
  volume = {502},
  number = {1},
  pages = {15--31},
  issn = {0035-8711},
  doi = {10.1093/mnras/staa4018},
  url = {https://doi.org/10.1093/mnras/staa4018},
  urldate = {2026-01-28}
}

@article{lebreuillyProtostellarCollapseSimulations2023,
  title = {Protostellar Collapse Simulations in Spherical Geometry with Dust Coagulation and Fragmentation},
  author = {Lebreuilly, Ugo and {Vallucci-Goy}, Valentin and Guillet, Vincent and Lombart, Maxime and Marchand, Pierre},
  year = 2023,
  journal = {MNRAS},
  volume = {518},
  number = {3},
  pages = {3326--3343},
  publisher = {Oxford University Press},
  doi = {10.1093/mnras/stac3220},
  url = {https://academic.oup.com/mnras/article-abstract/518/3/3326/6815728},
  urldate = {2026-01-28}
}

@article{alexopoulosPartDynamicEvolution2009,
  title = {Part {{V}}: {{Dynamic}} Evolution of the Multivariate Particle Size Distribution Undergoing Combined Particle Growth and Aggregation},
  shorttitle = {Part {{V}}},
  author = {Alexopoulos, A. H. and Roussos, A. and Kiparissides, C.},
  year = 2009,
  journal = {Chem. Eng. Sci.},
  volume = {64},
  number = {14},
  pages = {3260--3269},
  publisher = {Elsevier},
  url = {https://www.sciencedirect.com/science/article/pii/S0009250909002462},
  urldate = {2026-02-05}
}

@article{singhNewFiniteVolume2021,
  title = {New Finite Volume Approach for Multidimensional {{Smoluchowski}} Equation on Nonuniform Grids},
  author = {Singh, Mehakpreet},
  year = 2021,
  month = oct,
  journal = {Stud Appl Math},
  volume = {147},
  number = {3},
  pages = {955--977},
  issn = {0022-2526, 1467-9590},
  doi = {10.1111/sapm.12415},
  url = {https://onlinelibrary.wiley.com/doi/10.1111/sapm.12415},
  urldate = {2026-02-20},
  langid = {english}
}

@article{fernandez-diazExactSolutionSmoluchowskis2007,
  title = {Exact Solution of {{Smoluchowski}}'s Continuous Multi-Component Equation with an Additive Kernel},
  author = {{Fern{\'a}ndez-D{\'i}az}, J. M. and {G{\'o}mez-Garc{\'i}a}, G. J.},
  year = 2007,
  month = may,
  journal = {EPL},
  volume = {78},
  number = {5},
  pages = {56002},
  issn = {0295-5075},
  doi = {10.1209/0295-5075/78/56002},
  url = {https://doi.org/10.1209/0295-5075/78/56002},
  urldate = {2026-02-21},
  langid = {english}
}

@article{scottAnalyticStudiesCloud1968,
  title = {Analytic {{Studies}} of {{Cloud Droplet Coalescence I}}.},
  author = {Scott, William T.},
  year = 1968,
  month = jan,
  journal = {J. Atmospheric Sci.},
  volume = {25},
  pages = {54--65},
  publisher = {AMS},
  issn = {0022-4928},
  doi = {10.1175/1520-0469(1968)025<0054:ASOCDC>2.0.CO;2},
  url = {https://ui.adsabs.harvard.edu/abs/1968JAtS...25...54S},
  urldate = {2026-02-22}
}

@article{safronovParticularCaseSolution1962,
  title = {Particular Case of Solution to Coagulation Equation},
  author = {Safronov, V. S.},
  year = 1962,
  journal = {Dokl. Akad. Nauk SSSR},
  volume = {147},
  number = {1},
  pages = {64},
  publisher = {MEZHDUNARODNAYA KNIGA 39 DIMITROVA UL., 113095 MOSCOW, RUSSIA}
}

@article{ernstCoagulationProcessesPhase1984,
  title = {Coagulation Processes with a Phase Transition},
  author = {Ernst, M. H and Ziff, R. M and Hendriks, E. M},
  year = 1984,
  month = jan,
  journal = {Journal of Colloid and Interface Science},
  volume = {97},
  number = {1},
  pages = {266--277},
  issn = {0021-9797},
  doi = {10.1016/0021-9797(84)90292-3},
  url = {https://www.sciencedirect.com/science/article/pii/0021979784902923},
  urldate = {2026-02-22}
}

@misc{tangeGNUParallel202602222026,
  title = {{{GNU Parallel}} 20260222 ('{{Epstein}} Files')},
  author = {Tange, Ole},
  year = 2026,
  month = feb,
  doi = {10.5281/zenodo.18735643},
  url = {https://zenodo.org/records/18735643},
  urldate = {2026-03-08},
  howpublished = {Zenodo}
}

@article{muellerZurAllgemeinenTheorie1928,
  title = {{Zur allgemeinen Theorie ser raschen Koagulation: Die Koagulation von St\"abchen- und Bl\"attchenkolloiden; die Theorie beliebig polydisperser Systeme und der Str\"omungskoagulation}},
  shorttitle = {{Zur allgemeinen Theorie ser raschen Koagulation}},
  author = {M{\"u}ller, Hans},
  year = 1928,
  month = nov,
  journal = {Kolloidchem Beih},
  volume = {27},
  number = {6-12},
  pages = {223--250},
  issn = {0071-8017},
  doi = {10.1007/BF02558510},
  url = {http://link.springer.com/10.1007/BF02558510},
  urldate = {2026-04-07},
  copyright = {http://www.springer.com/tdm},
  langid = {ngerman}
}

@article{schumannTheoreticalAspectsSize1940,
  title = {Theoretical Aspects of the Size Distribution of Fog Particles},
  author = {Schumann, T. E. W.},
  year = 1940,
  month = apr,
  journal = {Quart J Royal Meteoro Soc},
  volume = {66},
  number = {285},
  pages = {195--208},
  issn = {0035-9009, 1477-870X},
  doi = {10.1002/qj.49706628508},
  url = {https://rmets.onlinelibrary.wiley.com/doi/10.1002/qj.49706628508},
  urldate = {2026-04-07},
  langid = {english}
}

@article{melzakScalarTransportEquation1957,
  title = {A Scalar Transport Equation},
  author = {Melzak, ZA87880},
  year = 1957,
  journal = {Trans. Am. Math. Soc.},
  volume = {85},
  number = {2},
  eprint = {1992943},
  eprinttype = {jstor},
  pages = {547--560},
  publisher = {JSTOR},
  url = {https://www.jstor.org/stable/1992943?casa_token=4eRXI2c2dQYAAAAA:zsAsge0mQdnJeujgvHjnlJojR5Klfd8Ynyry0kKMN5hj87zzlRqZFxmQUOKNPbeP6olrbxQEzw-PC1tWVWxmE3yP1JgF5576lINwvpjrzLoCf6LTDwQ},
  urldate = {2026-04-08}
}

@article{golovinSolutionCoagulationEquation1963,
  title = {The Solution of the Coagulation Equation for Cloud Droplets in a Rising Air Current},
  author = {Golovin, A. M.},
  year = 1963,
  journal = {No Title},
  volume = {5},
  pages = {783},
  url = {https://cir.nii.ac.jp/crid/1370306904406010526},
  urldate = {2026-04-08}
}

@article{mcleodScalarTransportEquation1964,
  title = {On the Scalar Transport Equation},
  author = {McLeod, John Bryce},
  year = 1964,
  journal = {Proc. Lond. Math. Soc.},
  volume = {3},
  number = {3},
  pages = {445--458},
  publisher = {Oxford University Press},
  url = {https://academic.oup.com/plms/article-abstract/s3-14/3/445/1565455},
  urldate = {2026-04-08}
}

@article{gelbardCoagulationGrowthMulticomponent1978,
  title = {Coagulation and Growth of a Multicomponent Aerosol},
  author = {Gelbard, Fred M. and Seinfeld, John H.},
  year = 1978,
  journal = {J. Colloid Interface Sci.},
  volume = {63},
  number = {3},
  pages = {472--479},
  publisher = {Academic Press},
  url = {https://elibrary.ru/item.asp?id=31205537},
  urldate = {2026-04-09}
}

@article{gelbardSimulationMulticomponentAerosol1980,
  title = {Simulation of Multicomponent Aerosol Dynamics},
  author = {Gelbard, Fred and Seinfeld, John H.},
  year = 1980,
  journal = {J. Colloid Interface Sci.},
  volume = {78},
  number = {2},
  pages = {485--501},
  publisher = {Elsevier},
  url = {https://www.sciencedirect.com/science/article/pii/0021979780905871},
  urldate = {2026-04-09}
}

@article{lushnikovEvolutionCoagulatingSystems1976,
  title = {Evolution of Coagulating Systems: {{III}}. {{Coagulating}} Mixtures},
  shorttitle = {Evolution of Coagulating Systems},
  author = {Lushnikov, A. A.},
  year = 1976,
  journal = {J. Colloid Interface Sci.},
  volume = {54},
  number = {1},
  pages = {94--101},
  publisher = {Elsevier},
  url = {https://www.sciencedirect.com/science/article/pii/0021979776902885},
  urldate = {2026-04-09}
}

@book{knuthArtComputerProgramming1997,
  title = {The {{Art}} of {{Computer Programming}}: {{Fundamental Algorithms}}, {{Volume}} 1},
  shorttitle = {The {{Art}} of {{Computer Programming}}},
  author = {Knuth, Donald E.},
  year = 1997,
  publisher = {Addison-Wesley Professional},
  url = {https://books.google.com/books?hl=en&lr=&id=x9AsAwAAQBAJ&oi=fnd&pg=PA100&dq=Knuth+The+Art+of+Computer+Programming+Volume+1+Fundamental+Algorithms&ots=B041SaQ26z&sig=d9EzDliflP3gUMT-vcNt4CsclpE},
  urldate = {2026-04-21}
}

@article{kostoglouExtendedCellAverage2007,
  title = {Extended Cell Average Technique for the Solution of Coagulation Equation},
  author = {Kostoglou, Margaritis},
  year = 2007,
  journal = {J. Colloid Interface Sci.},
  volume = {306},
  number = {1},
  pages = {72--81},
  publisher = {Elsevier},
  url = {https://www.sciencedirect.com/science/article/pii/S002197970600960X},
  urldate = {2026-04-22}
}

@article{chaudhuryExtendedCellaverageTechnique2013,
  title = {An Extended Cell-Average Technique for a Multi-Dimensional Population Balance of Granulation Describing Aggregation and Breakage},
  author = {Chaudhury, Anwesha and Kapadia, Avi and Prakash, Anuj V. and Barrasso, Dana and Ramachandran, Rohit},
  year = 2013,
  journal = {Adv. Powder Technol.},
  volume = {24},
  number = {6},
  pages = {962--971},
  publisher = {Elsevier},
  url = {https://www.sciencedirect.com/science/article/pii/S0921883113000095},
  urldate = {2026-04-22}
}

@misc{gurrutxagaMonteCarloMethod2026,
  title = {A {{Monte Carlo}} Method for Tracking Dust Properties during Coagulation in Protoplanetary Disks},
  author = {Gurrutxaga, Nerea and Vaikundaraman, Vignesh and Drazkowska, Joanna},
  year = 2026,
  month = apr,
  publisher = {arXiv},
  doi = {10.48550/arXiv.2604.24857},
  url = {https://ui.adsabs.harvard.edu/abs/2026arXiv260424857G},
  urldate = {2026-05-12}
}

\begin{appendix}

\section{Pseudocodes}\label{sec-a-pseudocode}
\FloatBarrier

Here, we show the pseudocodes for the one-component direct method \ref{alg:direct-onecomponent}, the two-component direct method \ref{alg:direct-twocomponent}, and the one-component tree method \ref{alg:tree-onecomponent}. The two-component tree method is omitted since it can be naturally composed from the two-component direct method and the one-component tree method.

\begin{algorithm}
\caption{Direct method for one-component SCE (one time step)}\label{alg:direct-onecomponent}

\begin{algorithmic}
\ForAll {$i, j$}
    \If {$\omega[i] < 10^{-40} \vee \omega[j] < 10^{-40}$}
        \State \textbf{continue}
    \EndIf
    \State $m_\mathrm{I+II} \gets m[i] + m[j]$
    \State $k \gets n(m_\mathrm{I+II})$\quad (Eq. (\ref{eq-podolak-n}))
    \State $\epsilon \gets \epsilon(m_\mathrm{I+II}, k)$\quad (Eq. (\ref{eq-podolak-epsilon}))
    \State $Roo \gets R(m[i], m[j]) \times \omega[i] \times \omega[j]$
    \State $\Delta \omega[k] \gets \Delta \omega[k] + 0.5 \times Roo \times \epsilon$\quad (1st term)
    \State $\Delta \omega[k-1] \gets \Delta \omega[k-1] + 0.5 \times Roo \times (1-\epsilon)$\quad (1st term)
    \State $\Delta \omega[i] \gets \Delta \omega[i] - Roo$\quad (2nd term)
\EndFor
\end{algorithmic}

\end{algorithm}

\begin{algorithm}
\caption{Direct method for two-component SCE (one time step)}\label{alg:direct-twocomponent}

\begin{algorithmic}
\ForAll {$mi, vi, mj, vj$}
    \If {$\omega[mi, vi] < 10^{-40} \vee \omega[mj, vj] < 10^{-40}$}
        \State \textbf{continue}
    \EndIf
    \State $m_\mathrm{I+II} \gets m[mi] + m[mj]$
    \State $v_\mathrm{I+II} \gets v_\mathrm{I+II}(m[mi], v[vi], m[mj], v[vj])$
    \State $mk \gets n(m_\mathrm{I+II})$\quad (Eq. (\ref{eq-podolak-n}))
    \State $vk \gets n(v_\mathrm{I+II})$\quad (Eq. (\ref{eq-podolak-n}))
    \State $\epsilon_m \gets \epsilon(m_\mathrm{I+II}, mk)$\quad (Eq. (\ref{eq-podolak-epsilon}))
    \State $\epsilon_v \gets \epsilon(v_\mathrm{I+II}, vk)$\quad (Eq. (\ref{eq-podolak-epsilon}))
    \State $Roo \gets R(m[mi], v[vi], m[mj], v[vj]) \times \omega[mi, vi] \times \omega[mj, vj]$
    \State $\Delta \omega[mk, vk] \gets \Delta \omega[mk, vk] + 0.5 \times Roo \times \epsilon_m \epsilon_v$
    \State $\Delta \omega[mk, vk-1] \gets \Delta \omega[mk, vk-1] + 0.5 \times Roo \times \epsilon_m (1-\epsilon_v)$
    \State $\Delta \omega[mk-1, vk] \gets \Delta \omega[mk-1, vk] + 0.5 \times Roo \times (1-\epsilon_m) \epsilon_v$
    \State $\Delta \omega[mk-1, vk-1] \gets \Delta \omega[mk-1, vk-1] + 0.5 \times Roo \times (1-\epsilon_m) (1-\epsilon_v)$
    \State $\Delta \omega[mi, vi] \gets \Delta \omega[mi, vi] - Roo$
\EndFor
\end{algorithmic}

\end{algorithm}

\begin{algorithm}
\caption{Tree method for one-component SCE (one time step)}\label{alg:tree-onecomponent}

\begin{algorithmic}
\State update\_tree($root, \omega$)
\ForAll{$i$}
    \If {$\omega[i] < 10^{-40}$}
        \State \textbf{continue}
    \EndIf
    \State $Roo \gets R(m[i], m[i])\times \omega[i]\times \omega[i]$
    \State $k \gets n(2\times m[i])$\quad (Eq. (\ref{eq-podolak-n}))
    \State $\epsilon \gets \epsilon(2\times m[i], k)$\quad (Eq. (\ref{eq-podolak-epsilon}))
    \State $\Delta \omega[k] \gets \Delta \omega[k] + 0.5 \times Roo \times \epsilon$\quad (1st term)
    \State $\Delta \omega[k-1] \gets \Delta \omega[k-1] + 0.5 \times Roo \times (1-\epsilon)$\quad (1st term)
    \State $\Delta \omega[i] \gets \Delta \omega[i] - Roo$\quad (2nd term)
    \State $j \gets root$
    \While{$j \neq \texttt{null}$}
        \If {$j.\Omega < 10^{-40}$}
            \State $j \gets j.next$
            \State \textbf{continue}
        \EndIf
        \State $L \gets ||j.m_\mathrm{max} - j.m_\mathrm{min}||$\quad (Eq. (\ref{eq-tree-distance}))
        \State $D \gets ||m[i] - j.m_\mathrm{ave}||$\quad (Eq. (\ref{eq-tree-distance}))
        \State $morecond1 \gets (L/D \ge \theta_c)$\quad (Eq. (\ref{eq-tree-openingangle}))
        \State $k_\mathrm{max} \gets n(m[i] + j.m_\mathrm{max})$\quad (Eq. (\ref{eq-podolak-n}))
        \State $k_\mathrm{min} \gets n(m[i] + j.m_\mathrm{min})$\quad (Eq. (\ref{eq-podolak-n}))
        \State $morecond2 \gets (k_\mathrm{max} - k_\mathrm{min} \ge k_c)$\quad (Eq. (\ref{eq-tree-kc}))
        \State $morecond3 \gets (j.m_\mathrm{min} \le m[i] \le j.m_\mathrm{max})$
        \If {$!j.is\_leaf \wedge (morecond1 \vee morecond2 \vee morecond3)$}\quad 
            \State $j \gets j.more$
            \State \textbf{continue}
        \EndIf
        \State $Roo \gets R(m[i], j.m_\mathrm{ave})\times \omega[i]\times j.\Omega$
        \State $\Delta \omega[i] \gets \Delta \omega[i] - Roo$\quad (2nd term)
        \If {$m[i] < j.m_\mathrm{ave}$}
            \State $j \gets j.next$
            \State \textbf{continue}
        \EndIf
        \State $m_\mathrm{I+II} \gets m[i] + j.m_\mathrm{ave}$
        \State $k \gets n(m_\mathrm{I+II})$\quad (Eq. (\ref{eq-podolak-n}))
        \State $\epsilon \gets \epsilon(m_\mathrm{I+II}, k)$\quad (Eq. (\ref{eq-podolak-epsilon}))
        \State $\Delta \omega[k] \gets \Delta \omega[k] + 0.5 \times Roo \times \epsilon$\quad (1st term)
        \State $\Delta \omega[k-1] \gets \Delta \omega[k-1] + 0.5 \times Roo \times (1-\epsilon)$\quad (1st term)
        \State $j \gets j.next$
    \EndWhile
\EndFor
\end{algorithmic}

\end{algorithm}

\FloatBarrier
\section{Additional tests}\label{sec-a-additionalkernels}
\FloatBarrier

\begin{figure*}
  \centering
  \includegraphics[width=17cm]{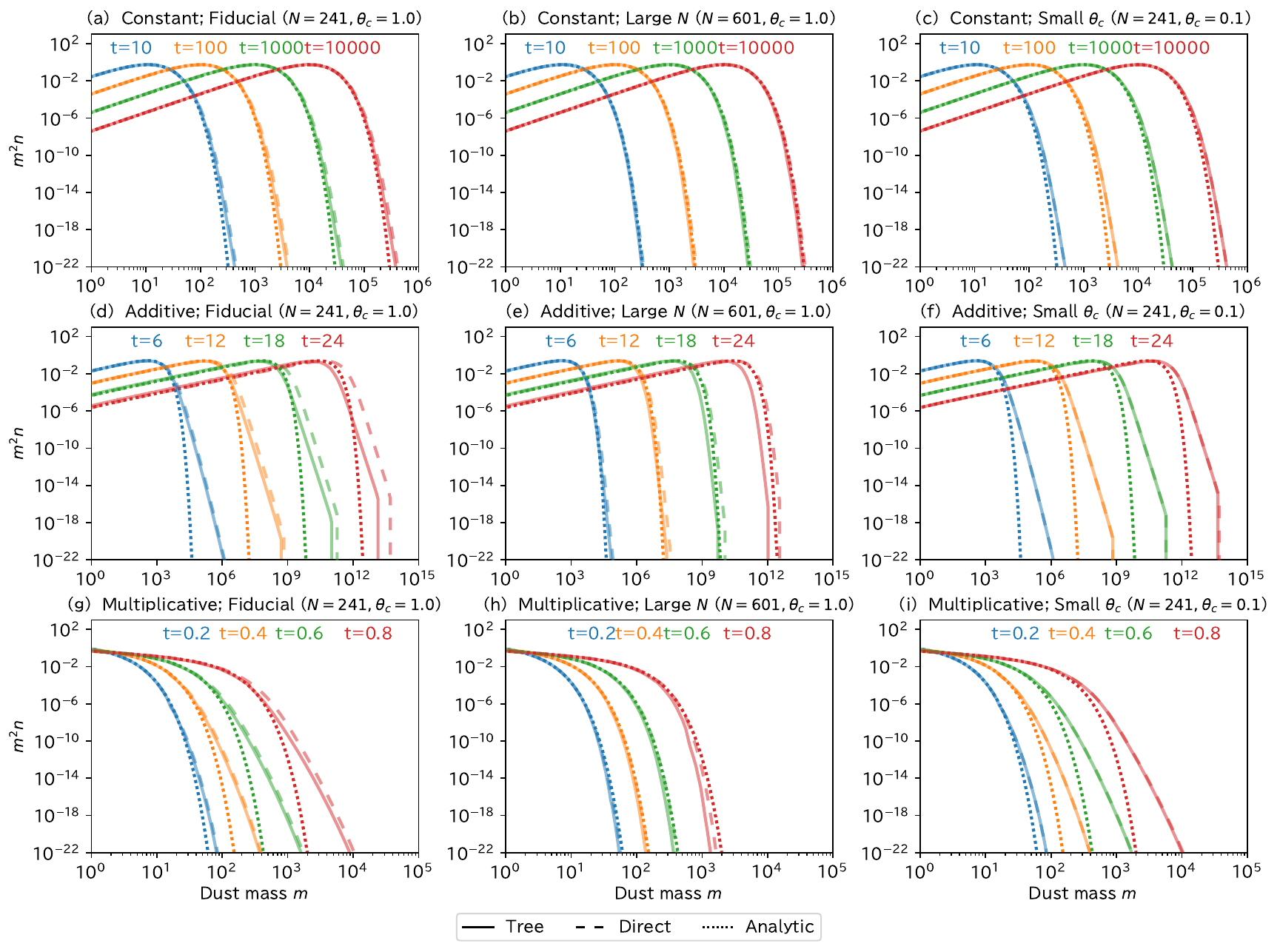}
  \caption{Analytic solutions and results of numerical calculation of one-component coagulation equation. The top panels (a), (b), and (c) are for the constant kernel, the middle-row panels (d), (e), and (f) are for the additive kernel, and the bottom panels (g), (h), and (i) are for the multiplicative kernel. The left panels (a), (d), and (g) are calculated with fiducial parameters $N_\mathrm{bd} = 16$ (i.e., $N=241$) and $\theta_c = 1$, the middle panels (b), (e), and (h) are calculated with a finer grid $N_\mathrm{bd} = 40$ (i.e., $N = 601$) and $\theta_c = 1$, and the right panels (c), (f), and (i) are calculated with a finer bin grouping $N_\mathrm{bd} = 16$ (i.e., $N=241$) and $\theta_c = 0.1$. In each plot, the sets of lines with four different colors show the snapshots at four different times. In each set, the lines show the tree method with the given $\theta_c$ and $k_c = 1000000$ (solid line), the direct method (dashed line), and the analytic solution (dotted line). All numerical results are calculated with adaptive time stepping.}
  \label{fig-appendix-1D-subex}
\end{figure*}

Here, additional tests for the three kernels with the one-component discrete SCE, or the one-component continuous SCE with the delta function as the initial condition, are shown.

For the one-component discrete SCE (Eq. \ref{eq-1D-discrete-SCE}), the following three analytic solutions corresponding to three main simple kernels are known. The three kernels are constant kernel $R (m_i, m_j)=\alpha$ (also known as “size-independent kernel” in chemical engineering), additive kernel $R (m_i, m_j) = \alpha (m_i + m_j)$ (also known as “size-dependent kernel” in chemical engineering), and multiplicative kernel $R (m_i, m_j) = \alpha m_i m_j$. Here, $\alpha$ is a constant. The analytic solution for the constant kernel is:
\begin{align}
  n(t, m) &= N_0 g(t)^2 (1-g(t))^{m-1}, \\
  g(t) &= \left(1 + \frac{\tau(t)}{2}\right)^{-1}, \\
  \tau(t) &= \alpha N_0 t.
\end{align}
The analytic solution for the additive kernel is:
\begin{align}
  n(t, m) &= N_0 \frac{m^{m-1}}{m!} g(t) (1 - g(t))^{m-1} e^{-m(1 - g(t))}, \\
  g(t) &= e^{-\tau (t)}, \\
  \tau(t) &= \alpha N_0 t.
\end{align}
And the analytic solution for the multiplicative kernel is:
\begin{align}
  n(t, m) &= N_0 \frac{(2m)^{m-1}}{m! m} \left(\frac{\tau (t)}{2}\right)^{m-1} e^{-m \tau (t)}, \\
  \tau (t) &= \alpha N_0 t
\end{align}
\citep{wetherillComparisonAnalyticalPhysical1990}. The initial condition is $n(0, m) = \delta(m, 1) N_0$ for all cases, where
\begin{align}
    \delta(x) &= \begin{cases}
        1 & \quad \mathrm{if}\ x=0,\\
        0 & \quad \mathrm{otherwise}
    \end{cases}
\end{align}
is the Kronecker delta function.

Figure \ref{fig-appendix-1D-subex} shows the exemplary results.

\end{appendix}
\end{document}